\documentclass[preprint,showpacs,preprintnumbers,amsmath,amssymb]{revtex4}%

\usepackage{graphicx}
\usepackage{dcolumn}
\usepackage{bm}
\usepackage{epsfig}
\usepackage{amsmath}
\usepackage{amsfonts}
\usepackage{amssymb}%
\usepackage{float}
\usepackage{subfigure}

\providecommand{\U}[1]{\protect\rule{.1in}{.1in}}

\newcommand{\be}{\begin{equation}}
\newcommand{\en}{\end{equation}}
\newcommand{\bea}{\begin{eqnarray}}
\newcommand{\ena}{\end{eqnarray}}

\begin{document}

\title{Reheating in small-field inflation on the brane: The Swampland Criteria and observational constraints 
in light of the PLANCK 2018 results}

\author{Constanza Osses}
\email{constanza.osses.g@mail.pucv.cl}

\author{Nelson Videla}
\email{nelson.videla@pucv.cl} 

\affiliation{ Instituto de
F\'{\i}sica, Pontificia Universidad Cat\'{o}lica de Valpara\'{\i}so,
Avenida Brasil 2950, Casilla 4059, Valpara\'{\i}so, Chile.}

\author{Grigoris Panotopoulos}
\email{grigorios.panotopoulos@tecnico.ulisboa.pt} \affiliation{ Centro de Astrof\'{\i}sica e Gravita{\c c}{\~a}o-CENTRA, Departamento de F{\'i}sica, Instituto Superior T\'ecnico-IST, Universidade de Lisboa-UL, Av. Rovisco Pais, 1049-001 Lisboa, Portugal}

\date{\today}

\begin{abstract}
We study cosmological inflation and its dynamics in the framework of the Randall-Sundrum II brane model. In particular, we analyze in detail four representative small-field inflationary potentials, namely Natural inflation, Hilltop inflation, Higgs-like inflation, and Exponential SUSY inflation, each characterized by two mass scales. We constrain the parameters for which a viable inflationary Universe emerges using the latest PLANCK results.
Furthermore, we investigate whether or not those models in brane cosmology are consistent with the recently proposed Swampland Criteria, and give predictions for the duration of reheating as well as for the reheating temperature after inflation. Our results show that (i) the distance conjecture is satisfied, (ii) the de Sitter conjecture and its refined version may be avoided, and (iii) the allowed range for the five-dimensional Planck mass, $M_5$,
is found to be $10^5\,\textup{TeV}\lesssim M_5\lesssim 10^{12}\,\textup{TeV}$. Our main findings indicate that non-thermal leptogenesis cannot work within the framework of RS-II brane cosmology, at least for the inflationary potentials considered here.
\end{abstract}


\maketitle

\section{Introduction}

Standard hot big-bang cosmology, based on four-dimensional General Relativity (GR) \cite{GR} combined with the cosmological principle, is supported by the three main pillars of modern cosmology. Those are i) the Hubble's law \cite{hubble},ii) the Primordial big-bang Nucleosynthesis (BBN) \cite{fowler}, and iii) the Cosmic Microwave Background (CMB) Radiation \cite{cmb}. The emerging cosmological model of the Universe seems to be overall quite successful, however some issues still remain regarding the initial conditions required for the big bang model, such as the horizon, the flatness, and the monopole problem. Cosmological inflation \cite{starobinsky1,inflation1,inflation2,inflation3} provides us with an elegant mechanism to solve those shortcomings all at once. Moreover, in the inflationary Universe, primordial curvature perturbations with an approximately scalar-invariant power spectrum, which seed CMB temperature anisotropies and the structure formation of the Universe, are generated from the vacuum fluctuations of a scalar field, the so called the inflaton \cite{Starobinsky:1979ty,R2,R202,R203,R204,R205,Abazajian:2013vfg}. Therefore, inflationary dynamics is currently widely accepted as the standard paradigm of the very early Universe, although we do not have a theory of inflation yet. For a classification of all single-field inflationary models based on a minimally coupled scalar field see \cite{kolb}, while for a large collection of inflationary models and their connection to Particle Physics see e.g. \cite{riotto,martin}. 

One can test the paradigm of cosmological inflation comparing its predictions on the $r-n_s$ plane with current cosmological and astronomical observations, and specially with those related to the CMB
temperature anisotropies from the PLANCK collaboration \cite{planck2,planck4} as well as the BICEP2/Keck-Array data \cite{bicep,Ade:2018gkx}. In particular, currently there only exists an
upper bound on the tensor-to-scalar ratio $r$, since the tensor power spectrum has not been measured yet. The PLANCK upper limit on the tensor-to-scalar-ratio, $r_{0.002}<0.10$ at 95$\%$ C.L., combined with the BICEP2/Keck Array (BK14) data, is further tightened, $r_{0.002}<0.064$. Besides, the tensor-to-scalar ratio can be related to the variation of $\Delta \phi$ of the field during inflation through the Lyth bound, assuming that $r$ is nearly constant \cite{Lyth:1996im}
\begin{equation}
\frac{\Delta \phi}{M_{\textup{pl}}}\simeq \sqrt{\dfrac{r}{8}}N_k,\label{lyth}
\end{equation}
where $N_k$ is the number of $e$-folds before the end of inflation, and $M_{\textup{pl}}$ is the reduced Planck mass associated with Newton's gravitational constant by $M_{\textup{pl}}=1/\sqrt{8\pi G}$. Models with $\Delta \phi<M_{\textup{pl}}$ and $\Delta \phi>M_{\textup{pl}}$ are called small-field and large-field models, respectively.
If next-generation CMB satellites, e.g. LiteBIRD \cite{Hazumi:2019lys}, COrE \cite{Finelli:2016cyd} and PIXIE \cite{Khatri:2013dha}, are not able to detect primordial B-modes, an upper limit of
$r<0.002\,(95\%\,\textup{C.L.})$ will be reached, implying that small-field inflation models will be favored, since a particular feature of these models is that tensor modes are much more suppressed
with respect to scalar modes than in the large-field models. In this type of models, the scalar field is rolling away from an unstable maximum of the potential, being a characteristic feature of spontaneous symmetry
breaking. Let us consider the inflaton potential of the form
\begin{equation}
V(\phi)=\Lambda^4\left[1-U(\phi)\right],\label{small-field} \end{equation}
where $\Lambda$ is a constant having a dimension of a mass and $U(\phi)$ is a function of $\phi$.

Natural Inflation (NI) with a pseudo-Nambu Goldstone boson (pNGB) as
the inflaton \cite{Freese:1990rb} arises in certain particle
physics model \cite{Adams:1992bn}. The scalar potential, which is flat due to shift
symmetries, has the form
\begin{equation}
V(\phi)=\Lambda^4\left[1-\cos\left(\frac{\phi}{f}\right)\right].\label{Vni}
\end{equation}
and it is characterized by two mass scales $f$ and $\Lambda$ with $f \gg
\Lambda$. It is assumed that a global symmetry is spontaneously broken at
some scale $f$, with a soft explicit symmetry breaking at a lower scale $\Lambda$.
Natural Inflation has been already studied in standard cosmology based on
GR \cite{Savage:2006tr,FK}. In particular, Natural Inflation is consistent with current data \cite{planck2,planck4}
for trans-Planckian values of the symmetry breaking scale $f$, for which it may be expected the low-energy effective theory, on which (\ref{Vni}) is based, to break down \cite{Banks:2003sx}. Another type of small-field models supported by Planck data are Hilltop inflation models, which are described by the potentials \cite{Boubekeur:2005zm,Kohri:2007gq}
\begin{equation}
V(\phi)=\Lambda^4\left[1-\left(\frac{\phi}{\mu}\right)^p\right],\label{hill}
\end{equation}
where $p$ is typically an integer power. In order to stabilize the potential from below, the former potentials are often written down as
\begin{equation}
V(\phi)=\Lambda^4\left[1-\left(\frac{\phi}{\mu}\right)^p\right]+... ,\label{hillm}  
\end{equation}
where higher order terms are included in the ellipsis. The fashionable models with $p=2$ and $p=4$ are ruled out by current observations for $\mu\lesssim M_{\textup{pl}}$ regardless of the omitted terms designated by the ellipsis. However, those models yield predictions favored by PLANCK 2018 when $\mu \gtrsim M_{\textup{pl}}$ for any value of the power $p$, which becomes indistinguishable from those of linear inflation, i.e. $V(\phi)\sim \phi$ \cite{Kallosh:2019jnl}. For numerical as well as analytic treatments of Hilltop inflation in the framework of GR, see Refs. \cite{Martin:2013tda,German:2020rpn,Antoniadis:2020bwi,Dimopoulos:2020kol}. A consistent modification of the quadratic Hilltop model ($p=2$) yields a Higgs-like potential \cite{Kallosh:2019jnl,Martin:2013tda}, which is used to describe dynamical symmetry breaking
\begin{equation}
V(\phi)=\Lambda^4\left[1-\left(\frac{\phi}{\mu}\right)^2\right]^2,\label{higgs}
\end{equation}
where the extra quartic term prevents the potential from becoming negative beyond the vacuum expectation value (VEV) $\mu$. It has been shown that such a potential remains favored by current data as long as the mass scales are high \cite{FK,Kallosh:2019jnl}. Another small-field model, derived
in the context of supergravity, corresponds to Exponential SUSY inflation, where the potential is given by \cite{Stewart:1994ts}
\begin{equation}
V(\phi)=\Lambda^4\left(1-e^{-\phi/f}\right)\label{susy}
\end{equation}
which is asymptotically flat in the limit
$\phi \rightarrow \infty$. This inflaton potential also appears in the context of D-brane inflation \cite{Dvali:1998pa} and it predicts a small value of the tensor-to-scalar for $f<M_{\textup{pl}}$ \cite{Tsujikawa:2013ila}, being inside the $(68\%\,\textup{C.L.})$ boundary constrained by PLANCK 2018 data \cite{Hirano:2019iie}.  Another supergravity-motivated model is 
K\"ahler moduli inflation \cite{Conlon:2005jm}
\begin{equation}
V(\phi)=\Lambda^4\left(1-c_1 \phi^{4/3}e^{-c_2 \phi^{4/3}}\right),
\end{equation}
which predicts a very small tensor-to-scalar ratio, well inside the $(68\%\,\textup{C.L.})$ contour \cite{Hirano:2019iie}.

The inflationary period ends when the equation-of-state parameter (EoS) becomes larger than $w=-1/3$, i.e. the slow-roll approximation breaks down, the expansion decelerates, and the Universe enters into the radiation era of standard Hot big-bang Cosmology \cite{Lyth:2009zz}. The transition era after the end of inflation, during which the inflaton is converted into the particles that populate the Universe later on is called reheating \cite{reh1,reh2} (for comprehensive reviews, see e.g. Refs. \cite{reh3,reh4,reh5}). As was shown Ref. \cite{Podolsky:2005bw}, the EoS parameter presents a sharp variation during the reheating phase due to the out-of-equilibrium nonlinear dynamics of fields. Unfortunately, the underlying physics of reheating is highly uncertain, complicated, and it cannot be directly probed by observations, although some bounds from BBN \cite{bound1,bound2}, the gravitino problem \cite{gravitino1,gravitino2,Okada:2004mh,gravitino3,gravitino4}, leptogenesis \cite{leptogenesis,buchmuller,davidson,lepto1,lepto2,lepto3,lepto4}, and the energy scale at the end of inflation do exist \cite{reh4,reh5}. There is, however, a strategy that allows us to obtain indirect constraints on reheating. First we parameterize our ignorance assuming for the fluid a constant equation-of-state $w_{re}$ during reheating. Next, we find certain relationships between the reheating temperature, $T_{re}$, and the duration of reheating, $N_{re}$, with $w_{re}$ and the inflationary observables \cite{Martin:2010kz,paper1,Martin:2014nya,paper2,Cai:2015soa,paper3,Rehagen:2015zma,Ueno:2016dim,Mishra:2021wkm}.

Considering that inflation opens up the window to probe physics in the very high energy regime, it is also tempting
to construct inflationary models in string theory \cite{Baumann:2014nda}. Although we do not have a full quantum gravity theory yet, string theory is believed to be a promising candidate, which possesses a space
of consistent low-energy effective field theories derived from it, called
the \emph{landscape} \cite{Bousso:2000xa,Giddings:2001yu,Kachru:2003aw}. The landscape consists of a vast amount of vacua described by different effective field theories (EFTs) at low energies. At the same time, there is another set of EFTs, dubbed the \emph{swampland}, which are not consistent with string theory. Accordingly, one can ask the question
what criteria a given low-energy EFT should satisfy in order to be contained in the string
landscape. In this direction, several criteria of this kind, dubbed \emph{swampland criteria} \cite{Ooguri:2006in,Obied:2018sgi} have been proposed so far, with the following implications for inflationary model-building
\begin{itemize}
\item {The distance conjecture:
\begin{equation}
 \frac{\Delta \phi}{M_{\textup{pl}}}< {\mathcal{O}}(1),\label{sw1}
 \end{equation}
}
\item {The de Sitter conjecture:
\begin{equation}
M_{\textup{pl}}\frac{\lvert V^{\prime}\rvert}{V} > c\sim {\mathcal{O}}(1). \label{sw2}
\end{equation}}
\end{itemize}
The distance conjecture implies that scalar fields cannot have
field excursions much larger than the Planck scale, since otherwise the validity of the EFT breaks down \cite{Ooguri:2016pdq}. As it can bee seen from Eq. (\ref{lyth}), in the context of inflation, field excursions are related to the tensor-to-scalar ratio. Accordingly, this conjecture limits the possibility of measuring tensor modes and hence primordial B-modes in the CMB. Specifically, for $N_k\gtrsim 50$, it is found $r\lesssim {\mathcal{O}}(10^{-3})$, which
lies on the edge of detectability for future experiments \cite{Hazumi:2019lys,Finelli:2016cyd,Khatri:2013dha}. In addition, the de Sitter conjecture states that slope of the scalar field potential satisfies a lower bound when $V>0$ \cite{Agrawal:2018own}. However, slow-roll inflation is in direct tension with those criteria \cite{Agrawal:2018own}, implying that  single-field models are ruled out as claimed in \cite{Kinney:2018nny,Garg:2018reu,Denef:2018etk}.
Nevertheless, those criteria are satisfied when studying inflation in non-standard, less conventional scenarios. For a representative list of related works, see  \cite{Achucarro:2018vey,Kehagias:2018uem,Matsui:2018bsy,Brahma:2018hrd,Motaharfar:2018zyb,Das:2018rpg,Dimopoulos:2018upl,Lin:2018kjm,Yi:2018dhl,Kamali:2019xnt,Brahma:2020cpy,Brandenberger:2020oav,Adhikari:2020xcg,Mohammadi:2020ake,Trivedi:2020wxf,Herrera:2020mjh}. Recently the \emph{refined de Sitter swampland
conjecture}, proposed in \cite{Garg:2018reu,Ooguri:2018wrx}, sates that:
\begin{itemize}
\item{Refined de Sitter conjecture:
\begin{equation}
 M_{\textup{pl}}\frac{\lvert V^{\prime}\rvert}{V} > c\sim {\mathcal{O}}(1)\quad \textup{or} \quad 
  M_{\textup{pl}}^2\frac{ V^{\prime \prime}}{V}<-c^{\prime}\sim {\mathcal{O}}(1).\label{swp3}
\end{equation}
}
\end{itemize}
With this refinement, which allows for a scalar field potential with maxima (hilltop) to exist, the conflicts with some small-field potentials, such as Higgs-like, QCD axion \cite{Denef:2018etk,Murayama:2018lie,Hamaguchi:2018vtv} and Hilltop \cite{Lin:2018rnx,Lin:2019fdk}, are resolved. 

Additionally, there is another Swampland conjecture proposed recently
in the literature, known as the Trans-Planckian Censorship Conjecture (TCC). Roughly speaking, the TCC claims that in a consistent quantum gravity theory, quantum fluctuations at sub-Planckian 
level are forbidden to exit the Hubble horizon during inflation. As a consequence, cosmic inflation is in direct conflict with this conjecture in regards to the upper bound on the tensor-to-scalar ratio, number of $e$-folds and energy scale of inflation \cite{Bedroya:2019snp,Bedroya:2019tba}.

A novel way to satisfy the refined swampland criteria is to consider 
inflation on the brane \cite{Brahma:2018hrd} and related works  \cite{Lin:2018kjm,Adhikari:2020xcg,Mohammadi:2020ake,Lin:2018rnx}. Furthermore, considering inflation in non-standard cosmologies is motivated by at least two facts, namely i) deviations from the standard Friedmann equation arise in higher-dimensional theories of gravity, and ii)
there is no observational test of the Friedmann equation before the BBN epoch. A well-studied example of
a novel higher-dimensional theory is the brane-world scenario, which inspired from M/superstring theory. Although 
brane models cannot be fully derived from the fundamental theory, they contain at least
the key ingredients found in M/superstring theory, such as extra dimensions, higher-dimensional objects
(branes), higher-curvature corrections to gravity (Gauss-Bonnet), etc. Since
superstring theory claims to give us a fundamental description of Nature, it is important to study what kind of 
cosmology it predicts.

Since there is a growing interest in studying inflationary models that meet both
observational data and Swampland Criteria, the main goal of the present work is to study the realization of some representative small-field inflation models, namely Natural inflation, Hilltop inflation, Higgs-like inflation  and Exponential SUSY inflation, in the framework of the RS-II brane model, in light of the recent PLANCK results and their consistency with the swampland criteria. Furthermore, we give predictions regarding the duration of reheating as well as the reheating temperature after inflation.

We organize our work as follows: After this introduction, in the next section we summarize the basics of the brane model as well as the dynamics of inflation and the basic formulas for determining the duration of reheating as well as for the reheating temperature after inflation. In sections from \ref{natbra} to \ref{SUSYbra}  we analyze 
each of the proposed small-field inflation models in the framework of RS-II model and present our results. Finally, in the last section we summarize our findings and exhibit our conclusions. We choose units so that $c=\hbar=1$.

\section{Basics of Braneworld inflation}\label{branerew}

\subsection{Braneworld cosmology}

In brane cosmology our four-dimensional world and the Standard Model (SM) of particle physics are confined to live on a 3-dimensional brane, whereas gravitons are allowed to propagate in the higher-dimensional bulk. Here we shall assume that only one additional spatial dimension, perpendicular to the brane, exists. Since the higher-dimensional Plank mass, $M_5$, is the fundamental mass scale instead of the usual four-dimensional Planck mass, $M_4$, the brane-world idea has been used to address the hierarchy problem of particle physics, first in the simple framework of a flat (4+$n$) space-time with 4 large dimensions and $n$ small compact dimensions \cite{Antoniadis:1997zg}, and later it was refined by Randall and Sundrum \cite{Randall:1999ee, Randall:1999vf}. For excellent introduction to brane cosmology see e.g. \cite{Langlois:2002bb}. In the RS-II model \cite{Randall:1999vf}, the four-dimensional effective field equations are computed to be \cite{Shiromizu:1999wj}
\begin{equation}
^{(4)}G_{\mu \nu}=-\Lambda_4g_{\mu \nu}+\frac{8\pi}{M_4^2} \tau_{\mu \nu}+\left(\frac{8\pi}{M_5^3}\right)^2\pi_{\mu \nu}-E_{\mu \nu},\label{4DEEQ}
\end{equation}
where $\Lambda_4$ is the four-dimensional cosmological constant, $\tau_{\mu \nu}$ is the matter stress-energy tensor on the brane, $\pi_{\mu \nu}=(1/12) \tau \tau_{\mu \nu}+(1/8) g_{\mu \nu} \tau_{\alpha \beta} \tau^{\alpha \beta}-(1/4) \tau_{\mu \alpha} \tau_\nu^\alpha-(1/24) \tau^2 g_{\mu \nu}$, and $E_{\mu \nu}=C_{\beta \rho \sigma}^\alpha n_\alpha n^\rho g_\mu^\beta g_\nu^\sigma$ is the projection of the five-dimensional Weyl tensor $C_{\alpha \beta \rho \sigma}$ on the brane, where $n^\alpha$ is the unit vector normal to the brane. $E_{\mu \nu}$ and $\pi_{\mu \nu}$ are the new terms, not present in standard four-dimensional Einstein's theory, and they encode the information about the bulk. The four-dimensional quantities are given in terms of the five-dimensional ones as follows \cite{Maartens:1999hf}
\begin{equation}
M_4=\sqrt{\frac{3}{4\pi}}\left(\frac{M_5^2}{\sqrt{\lambda}}\right)M_5,\label{M5M4}
\end{equation}
\begin{equation}
\Lambda_4 = \frac{4 \pi}{M_5^3} \left ( \Lambda_5+\frac{4 \pi \lambda^2}{3M_5^3} \right),\end{equation}
where $M_{\textup{pl}}=M_4/\sqrt{8\pi}\simeq 2.4\times 10^{18}\,\textup{GeV}$
is the reduced Planck mass, and $\lambda$ is the brane tension.

The Friedmann-like equation describing the backround evolution of a flat FRW Universe is found to be \cite{Binetruy:1999ut}
\begin{equation}
H^2=\frac{\Lambda_4}{3}+\frac{8\pi}{3M_4^2}\rho\left(1+\frac{\rho}{2\lambda}\right)+\frac{\mathcal{E}}{a^4}.\label{flatFE}
\end{equation}
where $a$ is the scale factor, $H$ is the Hubble parameter, $\rho$ is the total energy density of the cosmological fluid, and $\mathcal{E}$ is an integration constant coming from $E_{\mu \nu}$. The term $\frac{\mathcal{E}}{a^4}$ is known as the dark radiation, since it scales with $a$ the same way as radiation. However, during inflation this term will be rapidly diluted due to the quasi-exponential expansion, and therefore in the following we shall neglect it. The five-dimensional Planck mass is constrained by the standard big-bang nucleosynthesis to be $M_5 \gtrsim 10\,\textup{TeV}$ \cite{Cline:1999ts}, implying that $\lambda \gtrsim (1\,\textup{MeV})^4\sim (10^{-21}\,M_{\textup{pl}})^4$. A stronger constraint on $M_5$, namely $M_5 \gtrsim 10^5\,\textup{TeV}$, results from current tests for deviations from Newton’s gravitational law on scales larger than 1 mm \cite{Clifton:2011jh}. 

In the discussion to follow we shall set the four-dimensional cosmological constant $\Lambda_4$ to zero, i.e. we shall adopt the RS fine tuning $\Lambda_5=-4 \pi \lambda^2/(3 M_5^3)$, so that the model can explain the current cosmic acceleration without a cosmological constant. Finally, neglecting the term $\frac{\mathcal{E}}{a^4}$  the Friedmann-like equation (\ref{flatFE}) takes the final form
\begin{equation}
H^2=\frac{8\pi}{3M_4^2}\rho\left(1+\frac{\rho}{2\lambda}\right),\label{finalflatFE}
\end{equation}
upon which our study on brane inflation will be based.

\subsection{Inflationary dynamics}

At low energies, i.e., when $\rho\ll \lambda$, inflation in the brane-world scenario behaves in exactly the same way as standard inflation, but at higher energies we expect inflationary dynamics to be modified.

We consider slow-roll inflation driven by a scalar field $\phi$, for which the energy density $\rho$ and the pressure $P$ are given by
$\rho=\frac{\dot{\phi}^2}{2}+V(\phi)$ and $P=\frac{\dot{\phi}^2}{2}-V(\phi)$, respectively, where $V(\phi)$ is the scalar potential. Assuming that the scalar field is confined to the brane, the usual four-dimensional Klein-Gordon (KG) equation still holds
\begin{equation}
\ddot{\phi}+3H\dot{\phi}+V^{\prime}=0,\label{KG}
\end{equation}
where a prime denotes differentiation with respect to $\phi$, while an over dot denotes differentiation with respect to the cosmic time. In the slow-roll approximation the cosmological equations take the form  (\ref{finalflatFE}) and (\ref{KG})
\begin{equation}
H^2\simeq\frac{8\pi}{3M_4^2}V\left(1+\frac{V}{2\lambda}\right),\label{finalflatFESR}
\end{equation}
and
\begin{equation}
\dot{\phi}\simeq -\frac{V^{\prime}}{3H}.\label{KGSR}
\end{equation}
The brane-world correction term $V/\lambda$ in Eq. (\ref{finalflatFESR}) enhances the Hubble rate for a given potential. Thus there is an enhanced Hubble friction term in Eq. (\ref{KGSR}), as compared to GR, and brane-world effects will reinforce slow-roll for the same potential.

That way, using those two equations, it is possible to write down the expression for the slow-roll parameters on the brane as \cite{Maartens:1999hf}
\begin{eqnarray}
\epsilon & \equiv & \epsilon_V \frac{1+V/\lambda}{\left(1+V/2\lambda\right)^2},\label{ep}\\
\eta & \equiv & \eta_V \frac{1}{1+V/2\lambda},\label{et}
\end{eqnarray}
where $\epsilon_V=\frac{M_4^2}{16 \pi}\left(\frac{V^{\prime}}{V}\right)^2$
and $\eta_V=\frac{M_4^2}{8 \pi}\,\frac{V''}{V}$ 
are the usual slow-roll parameters of standard cosmology for a canonical scalar field.
Considering the definition of $\epsilon_V$ for standard inflation, the de Sitter swampland conjecture Eq. (\ref{sw2}) and the first equation of its refined version in (\ref{swp3}) imply 
\begin{equation}
\epsilon_V\sim c^2/2\sim {\mathcal{O}}(1),
\end{equation}
which rules out slow-roll inflation, since the former is in conflict with $\epsilon_V\ll1$. Slow-roll inflation on the brane implies that $\epsilon\ll1$ and $\left|\eta\right|\ll1$, which can be achieved in  the high-energy regime, i.e., $V\gg \lambda$, despite the fact that both $\epsilon_V$ and $\eta_V$ are large due to the large slope of the potential. This feature is crucial for avoiding the refined swampland criteria \cite{Brahma:2018hrd}. In the high-energy limit, Eqs. (\ref{ep}) and (\ref{et}) become
\begin{eqnarray}
\epsilon&\simeq & \epsilon_V \left(\frac{4 \lambda}{V}\right),\label{epHE}\\
\eta &\simeq & \eta_V \left(\frac{2 \lambda}{V}\right),\label{etHE}
\end{eqnarray}
while in the low-energy limit $V\ll \lambda$, Eqs. (\ref{ep}) and (\ref{et})
are reduced to the usual slow-roll parameters of standard cosmology. 
Clearly, the deviations from standard slow-roll inflation can be seen in the high-energy regime, as both parameters are suppressed by a factor $V/\lambda$.

The number of $e$-folds in the slow-roll approximation, using  (\ref{finalflatFE}) and (\ref{KG}), yields
\begin{equation}
N_k \simeq -\frac{8\pi}{M_4^2}\int_{\phi_{k}}^{\phi_{end}}\frac{V}{V^{\prime}}\left(1+\frac{V}{2\lambda}\right)\, d\phi,\label{Nfolds}
\end{equation}
where $\phi_{k}$ and $\phi_{end}$ are the values of the scalar field when the cosmological scales cross the Hubble-radius and at the end of inflation, respectively. As it can be seen, the number of $e$-folds is increased due to an extra term
of $V/\lambda$. This implies a more amount of inflation, between these two values of the field, compared to standard
inflation.

\subsection{Perturbations}

In the following we shall briefly review cosmological perturbations in brane-world inflation. We consider the gauge invariant quantity $\zeta=-\psi-H\frac{\delta \rho}{\dot{\rho}}$. Here, $\zeta$ is defined on slices of uniform density and reduces to the curvature perturbation at super-horizon scales. A fundamental
feature of $\zeta$ is that it is nearly constant on super-horizon scales \cite{Riotto:2002yw}, and in fact this property does not depend on the gravitational field equations \cite{Wands:2000dp}. Therefore, for the spatially flat gauge, we 
have $\zeta=H\frac{\delta \phi}{\dot{\phi}}$, where $\left|\delta \phi\right|=H/2\pi$. That way, using the slow-roll approximation, the amplitude of scalar perturbations is given by \cite{Maartens:1999hf}
\begin{equation}
P_S=\frac{H^2}{\dot{\phi}^2}\left(\frac{H}{2\pi}\right)^2 \simeq \frac{128\pi}{3 M_4^6}\frac{V^3}{V^{\prime\, 2}}\left(1+\frac{V}{2\lambda}\right)^3.\label{AS}
\end{equation}
On the other hand, the tensor perturbations are more involved since the gravitons can propagate into the bulk. The major uncertainty comes from the tensor $E_{\mu \nu}$, which describes the impact on the four-dimensional cosmology from the five-dimensional bulk, and whose evolution is not determined by the four-dimensional effective theory alone. In our work we make an approximation neglecting back-reaction due to metric perturbations in the fifth dimension, and setting $E_{\mu \nu}$ = 0. To discover whether back-reaction will have a significant effect or not, a more rigorous treatment is required. The amplitude of tensor perturbations is given by \cite{Maartens:1999hf}
\begin{equation}
P_T=\frac{64\pi}{M^2_4}\left(\frac{H}{2\pi}\right)^2F^2(x),\label{TS}
\end{equation}
where
\begin{eqnarray}
F(x) &=& \left[\sqrt{1+x^2}-x^2\ln\left(\frac{1}{x}+\sqrt{1+\frac{1}{x^2}}\,\right)\,\right]^2\nonumber\\
&=& \left[\sqrt{1+x^2}-x^2 \sinh^{-1}\left(\frac{1}{x}\right)\right]^{-1/2},\label{FT}
\end{eqnarray}
and $x$ is given by
\begin{equation}
x=HM_4\sqrt{\frac{3}{4\pi \lambda}}.\label{xFT}
\end{equation}

The expressions for the spectra are, as always, to be evaluated at the Hubble radius crossing $k = aH$.
As expected, in the the low-energy limit the expressions for the spectra become the same as those derived without considering the brane effects. However, in the high-energy limit, these expressions become
\begin{eqnarray}
P_S &\simeq & \frac{16\pi}{3 M_4^6 \lambda^3}\frac{V^6}{V^{\prime \,2}},\label{ASHE}\\
P_T  &\simeq & \frac{32\, V^3}{M_4^4 \lambda^2}.\label{ATHE}
\end{eqnarray}

The scale dependence of the scalar power spectra is determined by the scalar spectral index, which in the slow-roll approximation obeys the usual relation
\begin{eqnarray}
n_s &=&1+\frac{d \ln \mathcal{P}_{S}}{d \ln k}\nonumber,\\
n_s &\simeq & 1-6\epsilon+2\eta. \label{ns}
\end{eqnarray}

The amplitude of tensor perturbations can be parameterized by the tensor-to-scalar ratio, defined to be \cite{Lyth:2009zz}
\begin{equation}
r\equiv \frac{P_T}{P_S},\label{rr}
\end{equation}
which implies that in the low-energy limit this expression becomes $r\simeq 16 \epsilon_V$, where $\epsilon_V$ is
the standard slow-roll parameter, whereas in the high-energy limit we have \cite{Bassett:2005xm}
\begin{equation}
r \simeq 24 \epsilon,\label{rHE}
\end{equation}
with $\epsilon$ corresponding to Eq. (\ref{epHE}).

As we have seen, at late times the brane-world cosmology is identical to the standard one. During the early Universe, particularly during inflation, there may be changes to the perturbations predicted by the standard cosmology, if the energy density is sufficiently high compared to the brane tension. 



\subsection{Reheating}

Here we shall briefly describe how to compute the number of $e$-folds of reheating $N_{re}$ as well as the reheating temperature $T_{re}$ in terms of the scalar spectral index for single-field inflation in the high-energy regime of RS-II brane-world scenario. For the derivation of the main formulas, we  mainly follow Refs. \cite{paper1,paper2,paper3}.

Reheating after inflation is important for itself as a mechanism achieving what we know as the hot big-bang Universe. The energy of the inflaton field becomes in thermal radiation during the process of reheating through particle creation while the inflaton field oscillates around the minimum of its potential. If one considers that during reheating phase the main contribution to the energy density of the Universe comes from a component having an effective equation-of-state parameter (EoS) $w_{re}$, and its energy density can be related to the scale factor through $\rho\propto a^{-3(1+w_{re})}$, we can write down the following relation
\begin{equation}
    \frac{\rho_{end}}{\rho_{re}}=\left(\frac{a_{end}}{a_{re}}\right)^{-3(1+w_{re})},\label{rhoendre}
\end{equation}
where the subscripts $end$ and $re$ denote the end of inflation and the end of reheating phase, respectively.

The number of $e$-folds of reheating are related to the scale factor both at the end of inflation and reheating according to
\begin{equation}
    e^{-N_{re}}=\frac{a_{end}}{a_{re}}\label{enre}.
\end{equation}

Then, by combining Eqs. \eqref{rhoendre} and \eqref{enre}, we can write the number
of $e$-folds of reheating as
\begin{equation}
    N_{re}=\frac{1}{3(1+w_{re})}\ln\left(\frac{\rho_{end}}{\rho_{re}}\right).\label{nrerho}
\end{equation}

On the other hand, we consider the Friedmann-like equation \eqref{finalflatFE}
in the high-energy limit $\rho \gg \lambda$ 
\begin{equation}
H^2 \simeq \frac{4\pi}{3M_4^2 \lambda}\rho^2,
\end{equation}
and the slow-roll parameter $\epsilon$, defined as
\begin{equation}
\epsilon=-\frac{\dot{H}}{H^2}.
\end{equation}
By combining the time derivative of Eq. \eqref{finalflatFE} with the continuity equation for the scalar field $\dot{\rho}=-3H(\rho+P)$, $\epsilon$ is expressed as follows
\begin{eqnarray}
\epsilon&=&\frac{6\dot{\phi}^2 /2}{\dot{\phi}^2 /2 + V}.
\end{eqnarray}
From the last equation, we solve for the kinetic term $\frac{\dot{\phi}^2}{2}$, yielding
\begin{equation}
\frac{\dot{\phi}^2}{2}=\frac{V\epsilon}{6-\epsilon}.   
\end{equation}
So, we can write down the expression for the energy density of the scalar field $\rho=\frac{\dot{\phi}^2}{2}+V$ in terms of the slow-roll parameter $\epsilon$ and the scalar field potential $V$ as follows
\begin{eqnarray}
\rho&=&V\left(\frac{\dot{\phi}^2}{2V}+1\right),\\
\rho&=&V\left(\frac{\epsilon}{6-\epsilon}+1\right).
\end{eqnarray}
Accordingly, the relationship between the energy density and the potential at
the end of inflation ($\epsilon(\phi_{end})=1$) is given by
\begin{equation}
  \rho_{end}=\frac{6}{5}\,V(\phi_{end})=\frac{6}{5}\,V_{end},\label{rhoend}
\end{equation}
which is slightly different from those already obtained in \cite{Bhattacharya:2019ryo}, where 
$\rho_{end}=\frac{7}{6}V_{end}$. Otherwise, in GR it is found that $\rho_{end}=\frac{3}{2}V_{end}$ \cite{paper1,paper2,paper3}.

Replacing (\ref{rhoend}) in Eq. \eqref{nrerho} we obtain
\begin{equation}
N_{re}=\frac{1}{3(1+w_{re})}\ln\left(\frac{6}{5}\frac{V_{end}}{\rho_{re}}\right).\label{nrein}
\end{equation}
At the end of reheating phase, the energy density of the universe is assumed to be
\begin{equation}
\rho_{re}=\frac{\pi^2}{30}g_{re}T_{re}^4,\label{rho_re}
\end{equation}
where $g_{re}$ is the number of internal degrees of freedom of relativistic particles
at the end of reheating. Assuming that the degrees of freedom come from the particles
in the Standard Model, $g_{re}={\mathcal{O}}(100)$ for $\gtrsim 175$\,GeV \cite{reh4,reh5}, while for a Minimal Supersymmetric Standard Model ()MSSM, $g_{re}={\mathcal{O}}(200)$ \cite{Adhikari:2020xcg,Adhikari:2019uaw}.

On the other hand, the entropy is defined as 
\begin{equation}
    s=\frac{2\pi^2}{45}gT^3 ,\label{s}
\end{equation}
where the temperature is inversely proportional to the scale factor for radiation ($T\propto a^{-1}$). Then, by replacing the temperature in Eq. \eqref{s}, we have that $s\propto a^{-3}$. Assuming the conservation
of entropy, it yields $gT^3a^3=const.$ Now, if we apply the entropy
conservation between reheating and today
\begin{equation}
    g_{re}T_{re}^3a_{re}^3=g_0T_0^3 a_0^3,\label{re0}
\end{equation}
where $g_0$ denotes the number of internal degrees of freedom of relativistic particles
today, which comes from photons and neutrinos. Then, Eq. (\ref{re0}) becomes
\begin{equation}
    g_{re}T_{re}^3=\left(\frac{a_0}{a_{re}}\right)^3\left[2T_0^3 + \frac{21}{4}T_{\nu 0}^3\right].\label{conservation_s}
\end{equation}
For the contribution coming from neutrinos at the right-hand side of (\ref{conservation_s}), we use $T_{\nu 0}=\left(\frac{4}{11}\right)^{1/3} T_0$, where $T_0=2.725$ K is the temperature of the
universe today. The ratio $\frac{a_0}{a_{re}}$ can be written as 
\begin{equation}
\frac{a_0}{a_{re}}=\frac{a_0}{a_{eq}}\frac{a_{eq}}{a_{re}},
\end{equation}
where we introduce $e^{-N_{RD}}=\frac{a_{re}}{a_{eq}}$, with $N_{RD}$ being
the duration in $e$-folds of the radiation dominated epoch. Accordingly, Eq. (\ref{conservation_s}) is rewritten
as
\begin{equation}
    T_{re}=T_0\left(\frac{a_0}{a_{eq}}\right)e^{N_{RD}}\left(\frac{43}{11g_{re}}\right).\label{T}
\end{equation}

Furthermore, we may compare the wavelength ($\lambda_0 \simeq \frac{a_0}{k}$) with the Hubble radius ($d_H \simeq\frac{1}{H_0}$) today , so
\begin{eqnarray}
\frac{d_H}{\lambda_0}&=&\frac{k}{a_0 H_0}\\
&=&\frac{a_k H_k}{a_0 H_0},
\end{eqnarray}
where the subscript $k$ denotes when the scale crosses the Hubble radius. Incorporating the intermediates eras, for the ratio $\frac{a_0}{a_{eq}}$ we have (see, e.g. \cite{paper3})
\begin{equation}
\frac{a_0}{a_{eq}}=\frac{a_0 H_k}{k}e^{-N_k}e^{-N_{re}}e^{-N_{RD}}.
\end{equation}
Using this result in \eqref{T} we find
\begin{equation}
    T_{re}=\left(\frac{43}{11g_{re}}\right)^{1/3}\left(\frac{a_0 T_0}{k}\right) H_k e^{-N_k} e^{-N_{re}}.\label{trein}
\end{equation}
Upon replacement of Eq. (\ref{trein}) in Eq. \eqref{nrein}, one solves for $N_{re}$ giving
\begin{eqnarray}
N_{re}=\frac{4}{1-3\,w_{re}}\left[-\frac{1}{4}\ln\left(\frac{36}{\pi^2 g_{re}}\right) - \frac{1}{3}\ln\left(\frac{11 g_{re}}{43}\right) - \ln\left(\frac{k}{a_0 T_0}\right) - \ln \left(\frac{V_{end}^{1/4}}{H_k}\right) - N_k \right].
\end{eqnarray}
 By assuming $g_{re}\sim {\mathcal{O}}(100)$ and using the pivot scale $\frac{k}{a_0}=0.05\,\textup{Mpc}^{-1}$ from PLANCK, we arrive to the final expression for the number of $e$-folds of reheating
\begin{equation}
N_{re}=\frac{4}{1-3\,w_{re}}\left[61.6 - N_k - \ln \left(\frac{V_{end}^{1/4}}{H_k}\right)\right],\label{nre}
\end{equation}
where $H_k$ can be written down using the definition of the tensor-to-scalar ratio $r=P_T/P_S$. Taking $P_S$ at the 
pivot scale and using the expression for $P_T$ in the high-energy limit (\ref{ATHE}), one finds
\begin{equation}
H_k=\left(\frac{\pi^2}{6}r P_S M_4\sqrt{\frac{\lambda}{12 \pi}}\right)^{1/3}.
\end{equation}

Finally, combining Eqs. (\ref{rho_re}) and \eqref{nrein} the reheating temperature is computed as follows
\begin{equation}
T_{re}=\left(\frac{36}{100\,\pi^2}\right)^{1/4}\,\left(\frac{6}{5}\,V_{end}\right)^{1/4}\,e^{-\frac{3}{4}(1+w_{re})N_{re}}.\label{tre}
\end{equation}

Here, the model-dependent expressions are the Hubble rate at the instant when the cosmological scale crosses the Hubble radius, $H_k^{-1}$, the number of $e$-folds $N_k$, and the inflaton potential at the end of the inflationary expansion, $V_{end}$. Thus, it is implicit that $N_{re}, T_{re}$ depend on the observables $P_{s}$, $n_{s}$ and $r$ that we have already discussed. It is also remarkable the dependence of $N_{re}$ and $T_{re}$ on the 5-dimensional Planck mass, which enters into $V_{end}$ and $H_k$.

\section{Natural inflation on the brane}\label{natbra}

\subsection{Dynamics of inflation}

The Natural inflation potential is given by Eq. (\ref{Vni})
\begin{equation}
V(\phi)=\Lambda^4\left[1-\cos\left(\frac{\phi}{f}\right)\right].\label{Vnat}
\end{equation}
Applying Eqs. \eqref{epHE} and \eqref{etHE} to this potential, we obtain the slow-roll parameters in the high-energy regime
\begin{eqnarray}
\epsilon =\alpha\,\frac{(1+\cos(y))}{(1-\cos(y))^2},\label{epni}\\
\eta =\alpha\, \frac{\cos(y)}{(1-\cos(y))^2},\label{etni}
\end{eqnarray}
where $y$ and $\alpha$ are dimensionless parameters, which by definition are given by
\begin{eqnarray}
y &\equiv& \frac{\phi}{f},\\
\alpha &\equiv& \frac{M_4^2 \lambda}{4\pi f^2 \Lambda^4},\label{alphani}
\end{eqnarray}
respectively.
An important quantity to be computed is the field at Hubble horizon
crossing $\phi_k$, at which observables, such as the scalar power spectrum, the spectral index and the tensor-to-scalar ratio, are evaluated. In doing so, we first impose the condition $\epsilon \equiv 1$, which allows us to compute the value
of the inflaton field at the end of inflation
\begin{equation}
\cos(y_{end})=\cos \left(\frac{\phi_{end}}{f}\right)=\frac{1}{2}\left(2+\alpha-\sqrt{\alpha}\sqrt{8+\alpha}\right).\label{yend}
\end{equation}

Replacing this value of the field and the potential in Eq. \eqref{Nfolds} and performing the integral, the number of $e$-folds $N_k$ is computed to be
\begin{equation}
N_k=\frac{1}{\alpha}\left[\cos(y_{k})-\cos{(y_{end})}-2\ln\left(\frac{1+\cos(y_{k})}{1+\cos(y_{end})}\right)\right]\label{efoldsni}.
\end{equation}

Then, we solve for $y_k=\frac{\phi_k}{f}$, yielding
\begin{equation}
\cos(y_{k})=\cos\left( \frac{\phi_{k}}{f}\right)=-1-2\, W_{-1}\left[z(N_k,\alpha)\right],\label{ycmb}
\end{equation}
where $W_{-1}$ denotes the negative branch of the Lambert function \cite{Veberic:2012ax}, and its argument is given by
\begin{equation}
z(N_k,\alpha)\equiv \frac{1}{4}\sqrt{e^{-2-\frac{\sqrt{\alpha}}{2}\left(\sqrt{\alpha}+2N_k\sqrt{\alpha}-\sqrt{8+\alpha}\right)}\left(4+\alpha-\sqrt{\alpha}\sqrt{8+\alpha}\right)^2}.\label{zN}
\end{equation}

\subsection{Cosmological perturbations}

Using the potential (\ref{Vnat}) in the expression for the scalar power spectrum (Eq. \eqref{ASHE}), it leads to
\begin{equation}
P_S=\frac{1}{12 \pi^2 \alpha^3}\gamma^4\,\frac{(1-\cos(y))^5}{(1+\cos(y))},\label{asv}
\end{equation}
where $\gamma=\frac{\Lambda}{f}$ is a new dimensionless parameter. If we replace Eqs. \eqref{epni} and \eqref{etni} into \eqref{ns}, we obtain the expression for the spectral index
\begin{equation}
n_s=1-2\alpha\,\frac{(3+2\cos(y))}{(1-\cos(y))^2}.\label{nsv}
\end{equation}

The tensor-to-scalar ratio as a function of the scalar field is obtained after replacing (\ref{Vnat}) in Eq. \eqref{rr}
\begin{equation}
r=24\alpha\,\frac{(1+\cos(y))}{(1-\cos(y))^2}.\label{rrV}
\end{equation}

After evaluating those observables at the Hubble radius crossing with \eqref{ycmb}, we find
\begin{eqnarray}
\label{asni}
P_S &=& \frac{4}{3 \pi \alpha^3}\gamma^4\frac{(1+W_{-1}\left[z(N_k,\alpha)\right])^5}{\left(-W_{-1}\left[z(N_k,\alpha)\right]\right)},\\ 
\label{nsni}
n_s &=&  1-\frac{\alpha}{2}\,\frac{\left(1-4W_{-1}\left[z(N_k,\alpha)\right]\right)}{(1+W_{-1}\left[z(N_k,\alpha)\right])^2},\\ 
\label{rrni}
 r &=& 12\,\alpha\,\frac{\left(-W_{-1}\left[z(N_k,\alpha)\right]\right)}{(1+W_{-1}\left[z(N_k,\alpha)\right])^2}.
\end{eqnarray}

\begin{figure}[ht!]
\centering
\includegraphics[scale=0.5]{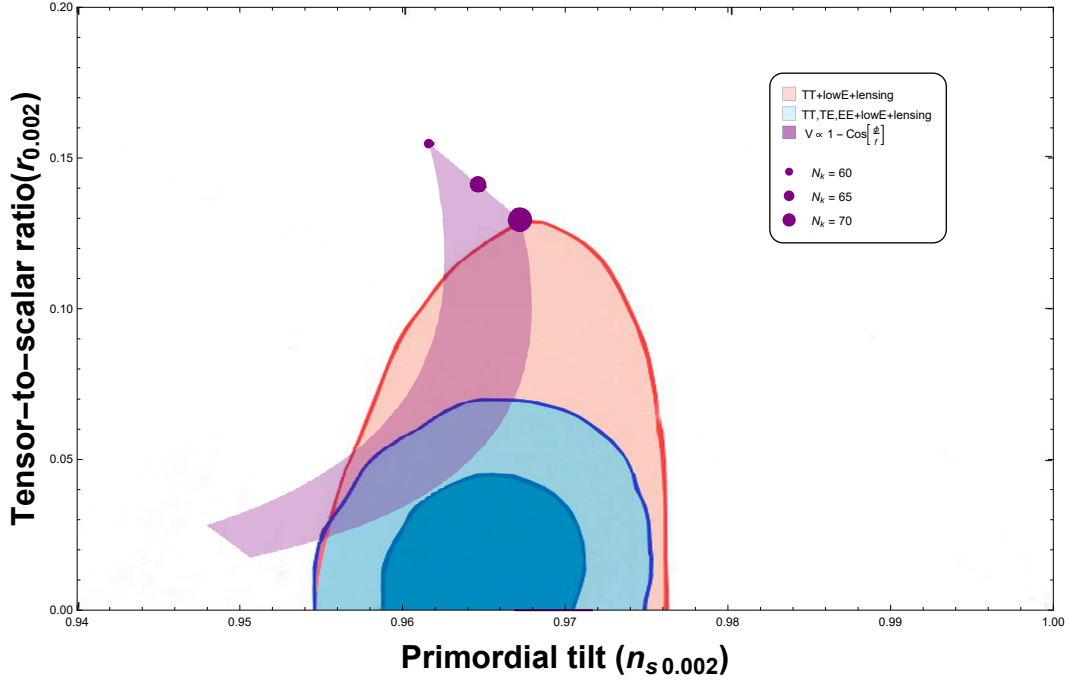}
\caption{We show the plot of the tensor-to-scalar ratio $r$ versus the scalar spectral index $n_s$ for Natural inflation on the brane along with the two-dimensional marginalized joint confidence contours for ($n_s,r$) at the 68$\%$ (blue region) and 95$\%$ (light blue region) C.L., from the latest PLANCK 2018 results.}\label{contornoNI}
\end{figure}

The predictions for Natural Inflation regarding the $n_s-r$ plane may be generated plotting Eqs. (\ref{nsni}) and (\ref{rrni}) parametrically, varying simultaneously the dimensionless parameter $\alpha$ in a wide range and the number $e$-folds $N_k$ within the range $N_k=60-70$. In Fig. \ref{contornoNI}, we have considered the two-dimensional marginalized joint confidence contours for ($n_s,r$) at the 68$\%$ (blue region) and 95$\%$ (light blue region) C.L., from the latest PLANCK 2018 results.


The allowed values for $\alpha$ are found when a given curve, for a fixed number of $e$-folds, enters (from above) and leaves (from below) the 2$\sigma$ region. One obtains that for $N_k = 65$, the predictions of the model are within the 
95$\%$ C.L. region from PLANCK data, for $\alpha$ being in the range
\begin{eqnarray}
\,3.93\times 10^{-2} \lesssim \alpha \lesssim 7.02\times 10^{-2}.
\end{eqnarray}

In that case, the prediction for the tensor to scalar ratio is the following
\begin{equation}
0.068 \gtrsim r \gtrsim 0.035.
\end{equation}


Accordingly, for $N_k=70$, the predictions are within the 95$\%$ C.L. for the following range of $\alpha$
\begin{equation}
3.25\times 10^{-2} \lesssim \alpha \lesssim 7.85\times 10^{-2},    
\end{equation}
while $r$ is found to be in the range
\begin{equation}
0.069 \gtrsim r \gtrsim 0.023.
\end{equation}

Thus, combining the previous constraints on $\alpha$ with Eq.(\ref{asni}) and the amplitude of the scalar spectrum $P_S\simeq 2.2\times 10^{-9}$, we obtain the corresponding allowed ranges for the dimensionless parameter $\gamma$
\begin{eqnarray}
6.58\times 10^{-4} \lesssim \gamma \lesssim 6.99\times 10^{-4}, \\
6.11\times 10^{-4} \lesssim \gamma \lesssim 6.61\times 10^{-4},
\end{eqnarray}
for $N_k=65$ and $N_k=70$, respectively.
The allowed ranges for $\alpha$ and $\gamma$ are summarized in Table \ref{constalphani}.


\begin{table}[H]
\begin{center}
\begin{tabular}{| c | c | c |}\hline
 $N_k$   & Constraint on $\alpha$  & Constraint on $\gamma$ \\\hline
   \,65\,  & \,$0.0393 \lesssim \alpha \lesssim 0.0702$\,  & \,$6.58\times 10^{-4} \lesssim \gamma \lesssim 6.99\times 10^{-4}$\,  \\
  \,70\,  & \,$0.0325 \lesssim \alpha \lesssim 0.0785$\,  & \,$6.11\times 10^{-4} \lesssim \gamma \lesssim 6.61\times 10^{-4}$\,  \\ \hline
\end{tabular}
\caption{Results for the constraints on the parameters $\alpha$ and $\gamma$ for Natural inflation in the high-energy of Randall-Sundrum brane model, using the last data of PLANCK.} \label{constalphani}
\end{center}
\end{table}


After replacing the relation between the 4-dimensional and 5-dimensional Planck masses (Eq. \eqref{M5M4}) into the definition of $\alpha$ (Eq.(\ref{alphani})) and using the fact that $ \Lambda=\gamma f$, the following expressions for the mass scales $f$ and $\Lambda$ are derived



\begin{equation}
f = \left ( \frac{3}{16 \pi^2 \alpha \gamma^4} \right )^{1/6} M_5, \label{fni}
\end{equation}
\begin{equation}
\Lambda = \gamma f=\gamma \left ( \frac{3}{16 \pi^2 \alpha \gamma^4} \right )^{1/6} M_5. \label{Lni}
\end{equation}

Evaluating those expressions at the several values for $\alpha$ and $\gamma$ (Table \ref{constalphani}), we may obtain a value for the brane tension $\lambda$ as well as the allowed
ranges for the mass scales $f$ and $\Lambda$ for any given value of $M_5$. 
If we consider the lower limit for the five-dimensional Planck mass, $M_5=10^{5}$\,TeV \cite{Clifton:2011jh}, it yields 
$\lambda=1.60\times 10^{-3}$\,TeV$^4$, while the corresponding constraints on the mass scales (in units of TeV) are shown 
in Table \ref{Tflni5}.


\begin{table}[H]
\begin{center}
\begin{tabular}{| c | c | c |}\hline
 $N_k$   & Constraint on $f$\,[TeV]  & Constraint on $\Lambda$\,[TeV] \\\hline
  \,65\,  & \,$1.17\times 10^7 \gtrsim f \gtrsim 1.02\times 10^7$\,  & \,$7.70\times 10^{3} \gtrsim \Lambda \gtrsim 7.14\times 10^{3}$\,  \\
  \,70\,  & \,$1.27\times 10^7 \gtrsim f \gtrsim 1.04\times 10^7$\,  & \,$7.76\times 10^{3} \gtrsim \Lambda \gtrsim 6.88\times 10^{3}$\,  \\ \hline
\end{tabular}
\caption{Results for the constraints on the mass scales $f$ and $\Lambda$ for Natural inflation in the high-energy of Randall-Sundrum brane model $M_5=10^{5}$\,TeV, using the last data of PLANCK.} \label{Tflni5}
\end{center}
\end{table}


In order to obtain an upper bound for the 5-dimensional Planck mass, we take into account that the inflationary dynamics takes places in the high-energy regime, $V \gg \lambda$. In doing so, we realize that
during inflation $V\simeq \Lambda^4$, then if we solve Eq. \eqref{M5M4} for the brane tension $\lambda$, the condition
for the high-energy regime imposes the following constraint on the amplitude of the potential
\begin{equation}
\Lambda^4 \gg \frac{3M_5^6}{4\pi M_4^2}.\label{Lggni}
\end{equation} 

Combining Eqs. (\ref{Lni}) and \eqref{Lggni}, one finds the following upper bound for the 5-dimensional Planck mass
\begin{equation}
M_5 \ll \sqrt{\frac{4\pi}{3}} \gamma^2 \left(\frac{3}{16\pi^2 \alpha \gamma^4}\right)^{1/3} M_4^2.
\end{equation}

If we replace the allowed values for $\alpha$ and $\gamma$ in the last equation, we find that the 5-dimensional Planck mass is such  $M_5 \ll 10^{14}$ TeV. So, if we assume that the maximum allowed value for $M_5$ is two orders of magnitude less, i.e. $M_5=10^{12}$ TeV, the brane tension is computed to be $\lambda=1.60\times 10^{39}$\,TeV$^4$, while the constraints on $f$ and $\Lambda$ are displayed in Table \ref{Tflni11}.


\begin{table}[H]
\begin{center}
\begin{tabular}{| c | c | c |}\hline
 $N_k$   & Constraint on $f$\,[TeV]  & Constraint on $\Lambda$\,[TeV] \\\hline
  \,65\,  & \,$1.17\times 10^{14} \gtrsim f \gtrsim 1.02\times 10^{14}$\,  & \,$7.70\times 10^{10} \gtrsim \Lambda \gtrsim 7.14\times 10^{10}$\,  \\
  \,70\,  & \,$1.27\times 10^{14} \gtrsim f \gtrsim 1.04\times 10^{14}$\,  & \,$7.76\times 10^{10} \gtrsim \Lambda \gtrsim 6.88\times 10^{10}$\,  \\ \hline
\end{tabular}
\caption{Results for the constraints on the mass scales $f$ and $\Lambda$ for Natural inflation in the high-energy of Randall-Sundrum brane model $M_5=10^{12}$\,TeV, using the last data of PLANCK.} \label{Tflni11}
\end{center}
\end{table}


From Tables \ref{Tflni5} and \ref{Tflni11}, the mass scales $f$ and $\Lambda$ take sub-Planckian values and there is a hierarchy between them consistent with $f \gg \Lambda$, achieving an almost flat potential. Moreover, the constraints already found on $\alpha$  and Eqs. (\ref{yend}) and (\ref{ycmb}) imply that during inflation
the dynamics is such that $\phi \sim f$, therefore Natural Inflation in the high-energy regime of the RS-II brane-model takes place at sub-Planckian values of the scalar field. It is worth mentioingn that our results for the mass scales differ almost by one order or magnitude in comparison to those already found in Ref. \cite{Videla:2016nlt} when using $M_5=10^{5}$\,TeV. In addition, our results with the upper limit $M_5=10^{12}$\,TeV are similar to those found in Ref. \cite{Mohammadi:2020ake} so far, where the authors used $M_5=5\times 10^{12}$\,TeV.

\begin{figure}[ht!]
\centering
\includegraphics[width=0.47\textwidth]{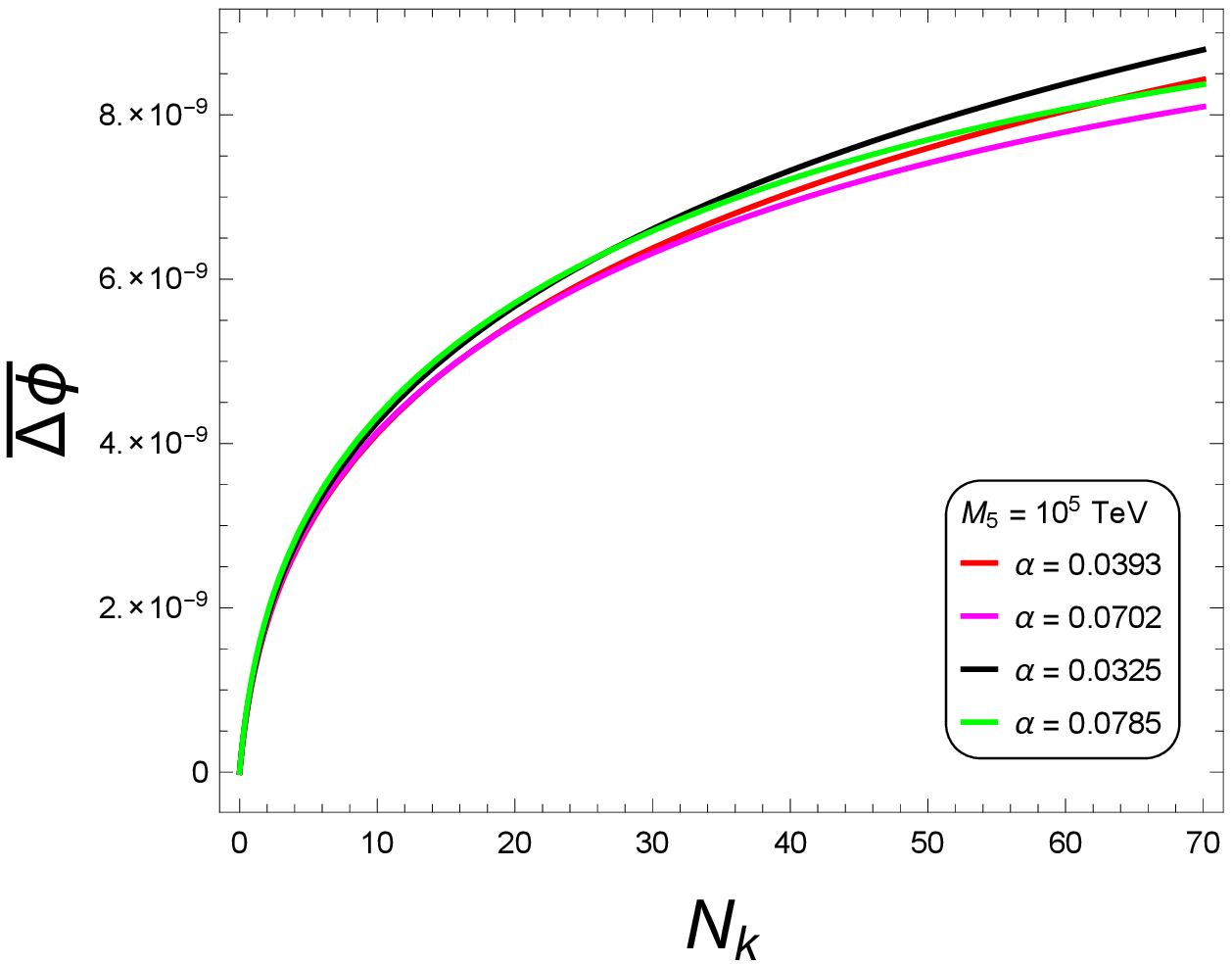}
\includegraphics[width=0.45\textwidth]{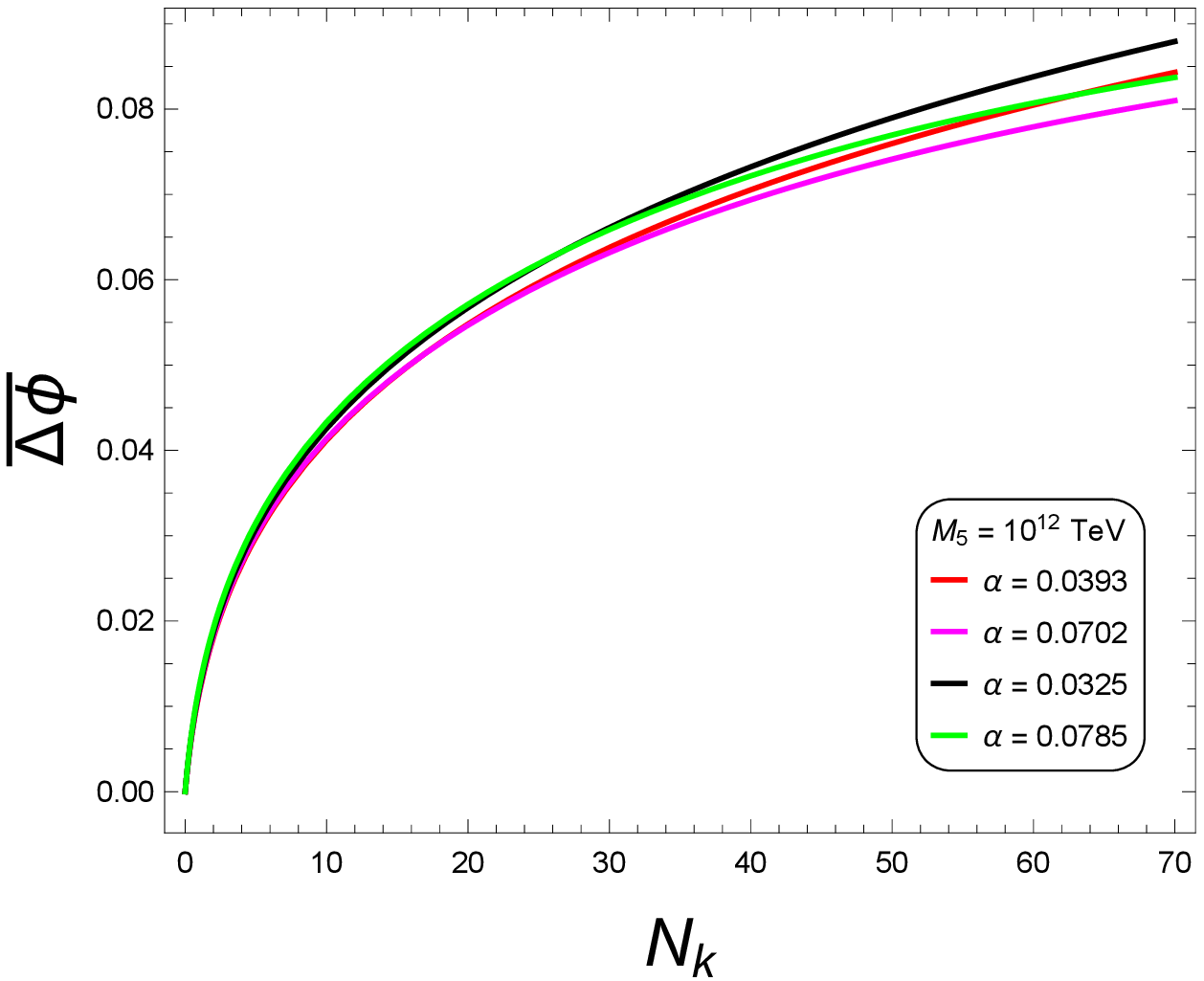}
\includegraphics[width=0.47\textwidth]{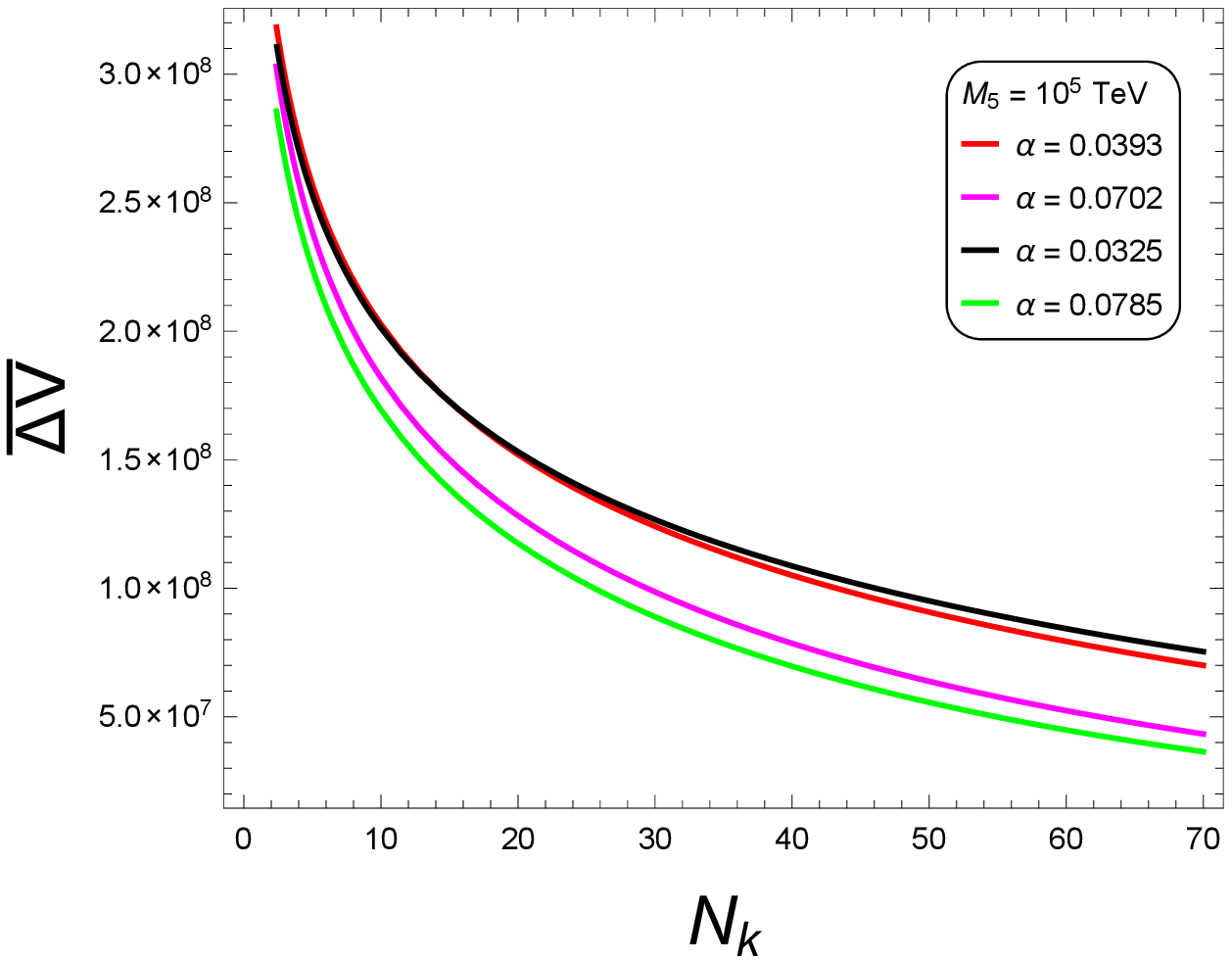}
\includegraphics[width=0.45\textwidth]{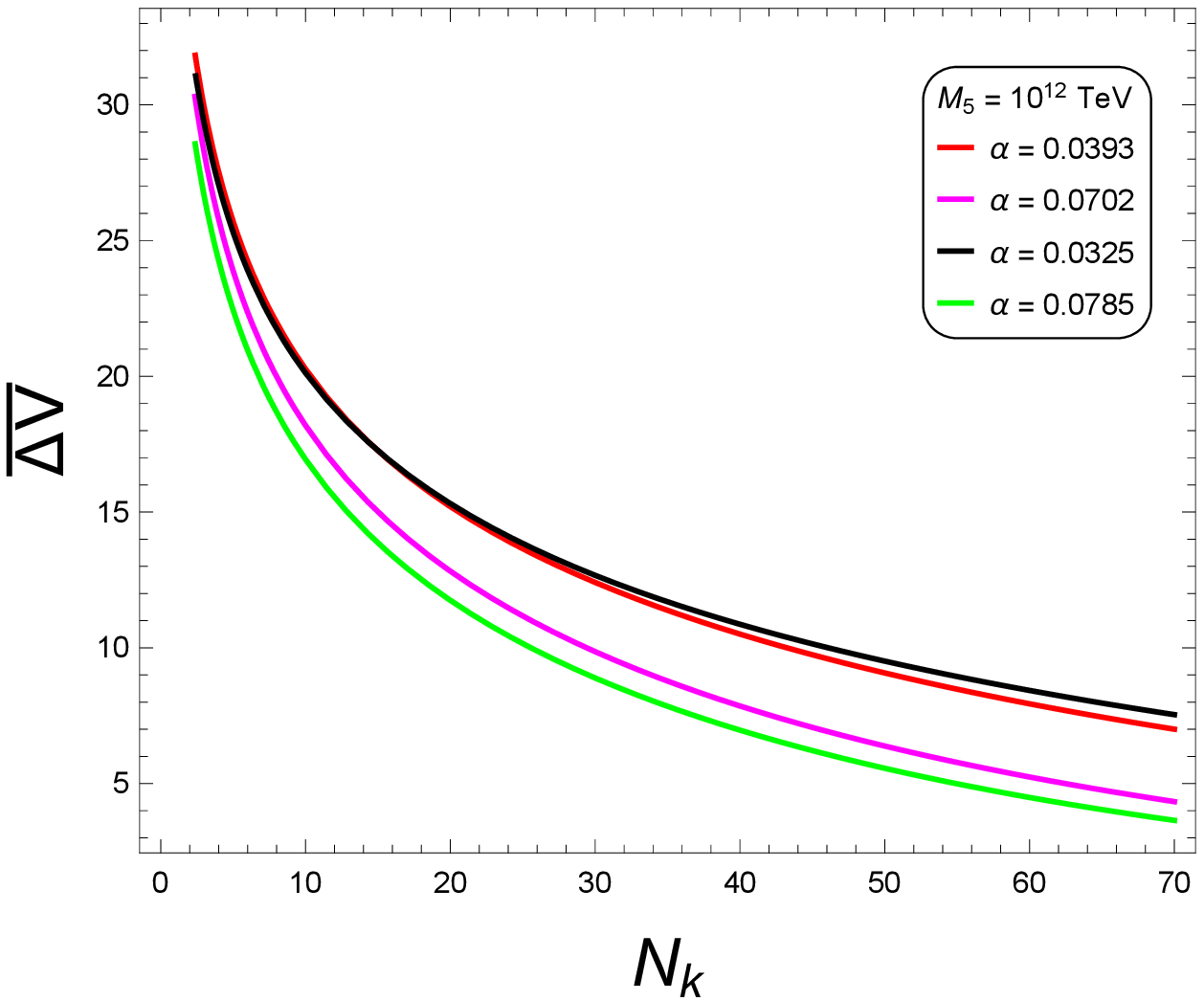}
\caption{Plots of the Swampland criteria for $M_5=10^5$\,TeV (left) and $M_5=10^{12}$\,TeV (right) in terms of the number of $e$-folds for the different values of $\alpha$. Top panels show the behaviour of the distance conjecture where $\Delta \phi/M_{pl}\equiv\overline{\Delta \phi}$, while the bottom panels show the de Sitter conjecture where $M_{pl}\,|V'|/V\equiv\overline{\Delta V}$.} \label{fig:NIswampland}
\end{figure}

After obtaining the allowed parameter space where Natural Inflation in the high-energy limit of Randall-Sundrum brane model is viable, we want to see if the Swampland Criteria are met in this model. Fig. \ref{fig:NIswampland} shows the distance
conjecture (\ref{sw1}) and the de Sitter 
conjecture (\ref{sw2}) of the Swampland Criteria: the top and bottom panels shows the behaviour of $\Delta \phi/M_{pl}\equiv\overline{\Delta \phi}$ and $M_{pl}\,|V'|/V\equiv\overline{\Delta V}$, respectively, against the number of $e$-folds $N_k$ for some values of $\alpha$ and the lower (left) and upper (right) limits of $M_5$. We note that for the distance conjecture $\overline{\Delta \phi}$, it increases as both $N_k$ and the 5-dimensional Planck mass increase, but the curves are always less than the unity since the scale mass $f$ is always sub-Planckian, so the distance conjecture is fulfilled. For the de Sitter conjecture, we note that $\overline{\Delta V}$ decreases with $N_k$, but it increases
as $M_5$ increases. In this case, we must be careful because $\overline{\Delta V}$ is related to the slow-roll parameter $\epsilon_V$ in General Relativity, yielding values much larger than this conjecture requires. Nevertheless, as we discussed in Section \ref{branerew}, slow-roll inflation on the brane implies that $\epsilon\ll1$ and $\left|\eta\right|\ll1$, which can be achieved in  the high-energy limit, i.e., $V\gg \lambda$ despite the fact that both $\epsilon_V$ and $\eta_V$ are large. In this way, the de Sitter conjecture and its refined version are avoiding. Additionally, our results for the distance conjecture are similar to those found in Ref. \cite{Mohammadi:2020ake} while although our plots have the same behavior for the de Sitter conjecture, the values of $\overline{\Delta V}$ differ by several order of magnitude when we use $M_5=10^{12}$\,TeV.

\subsection{Reheating}

\begin{figure}[ht!]
\centering
\includegraphics[width=0.45\textwidth]{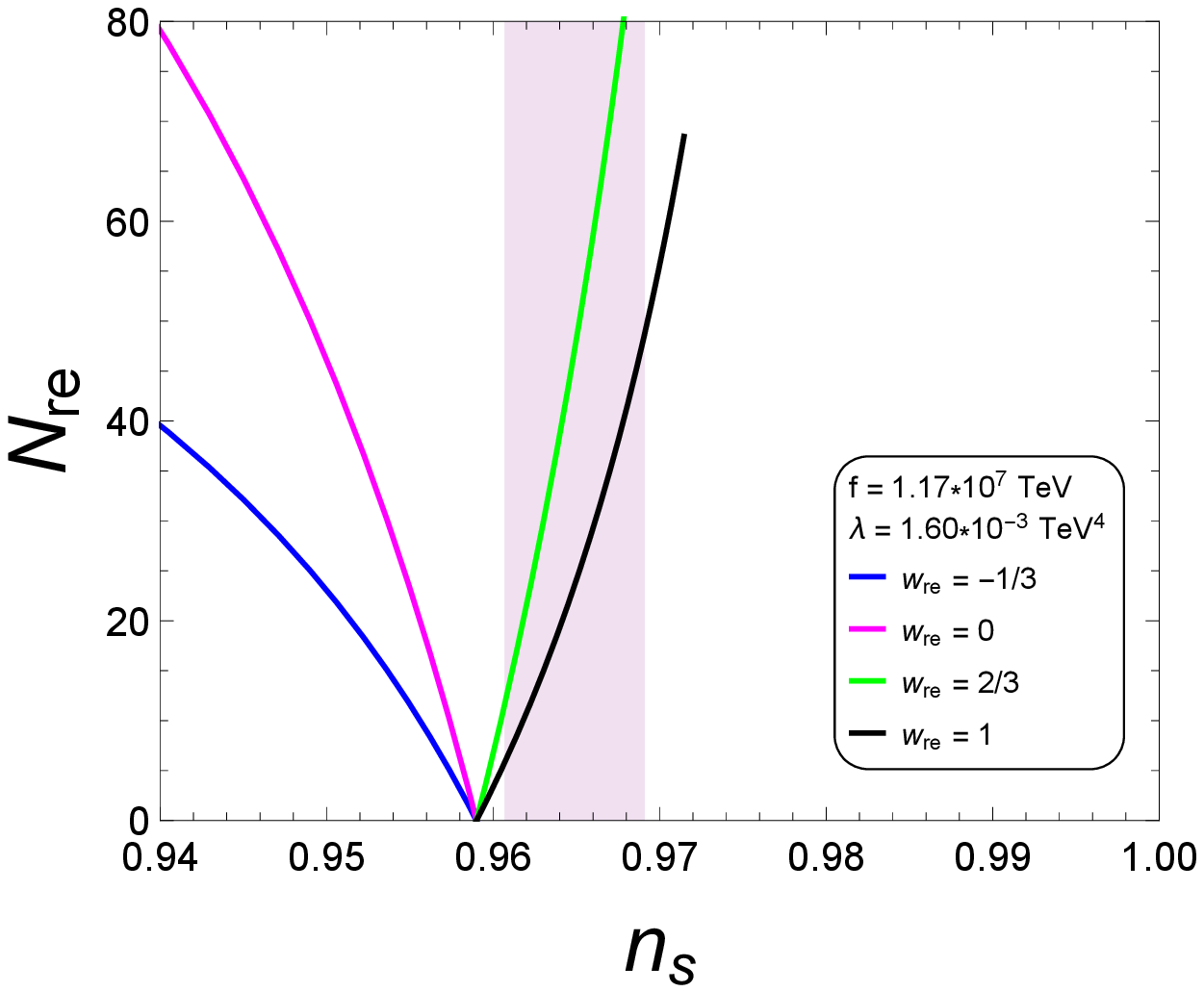}
\includegraphics[width=0.45\textwidth]{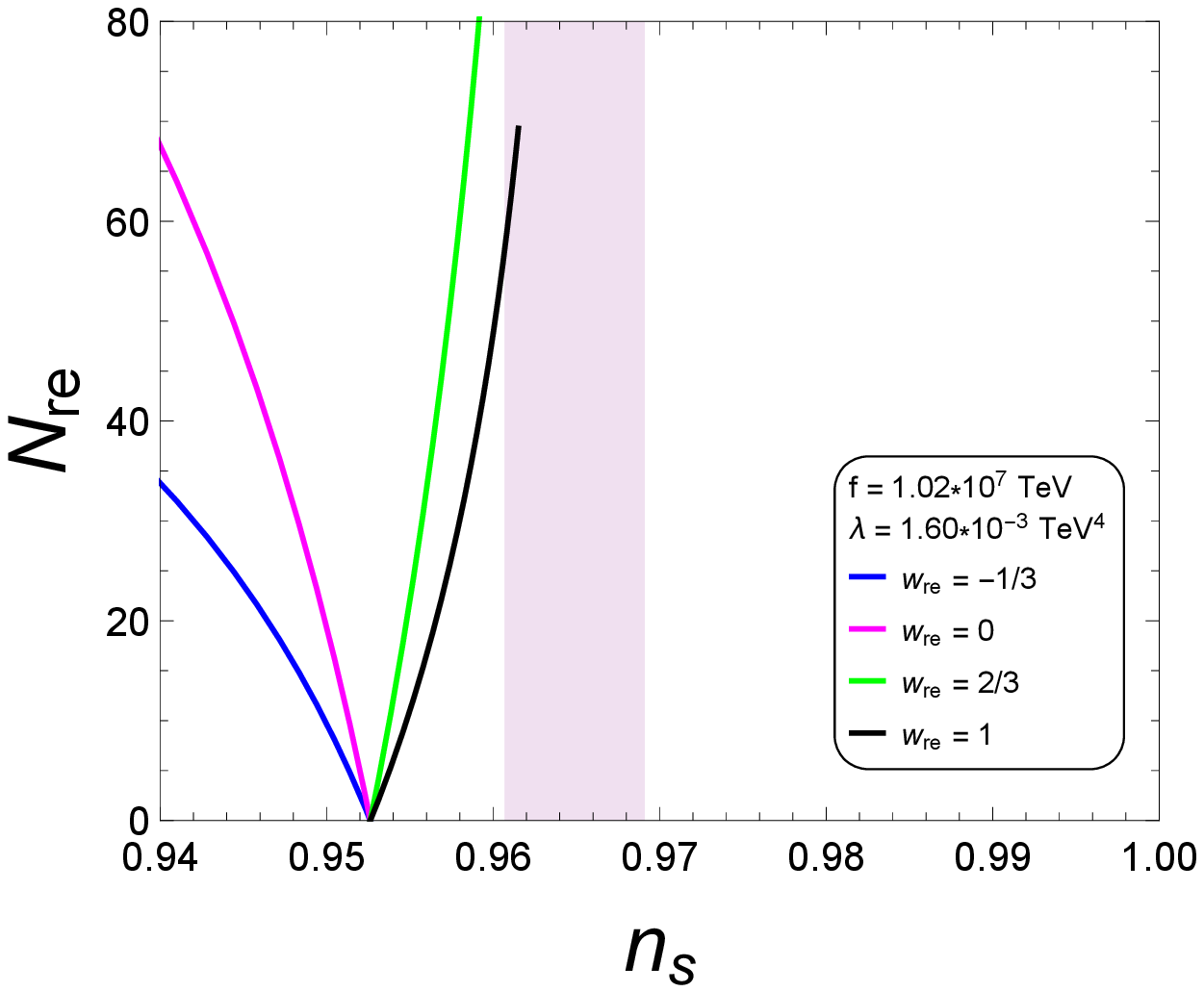}
\includegraphics[width=0.45\textwidth]{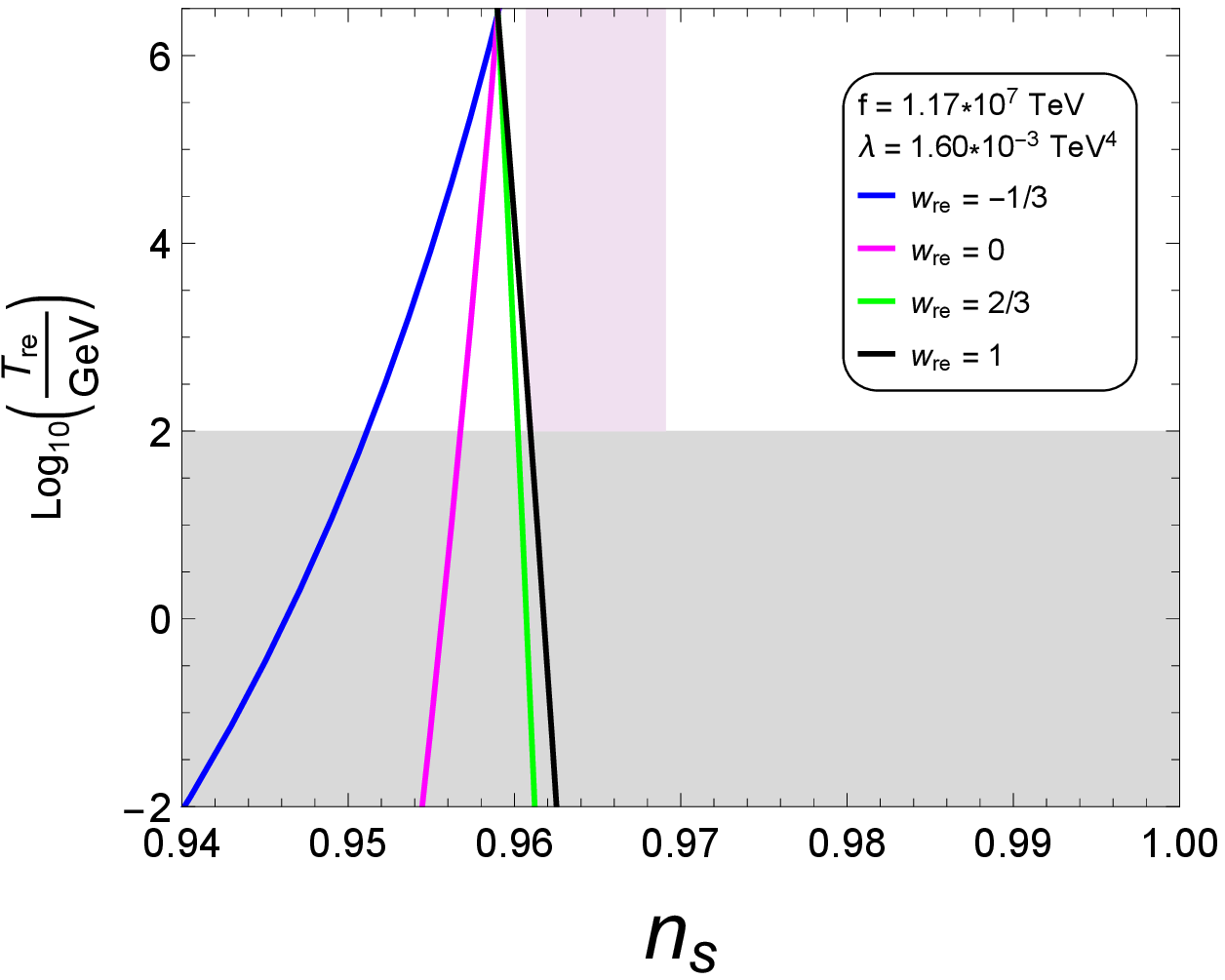}
\includegraphics[width=0.45\textwidth]{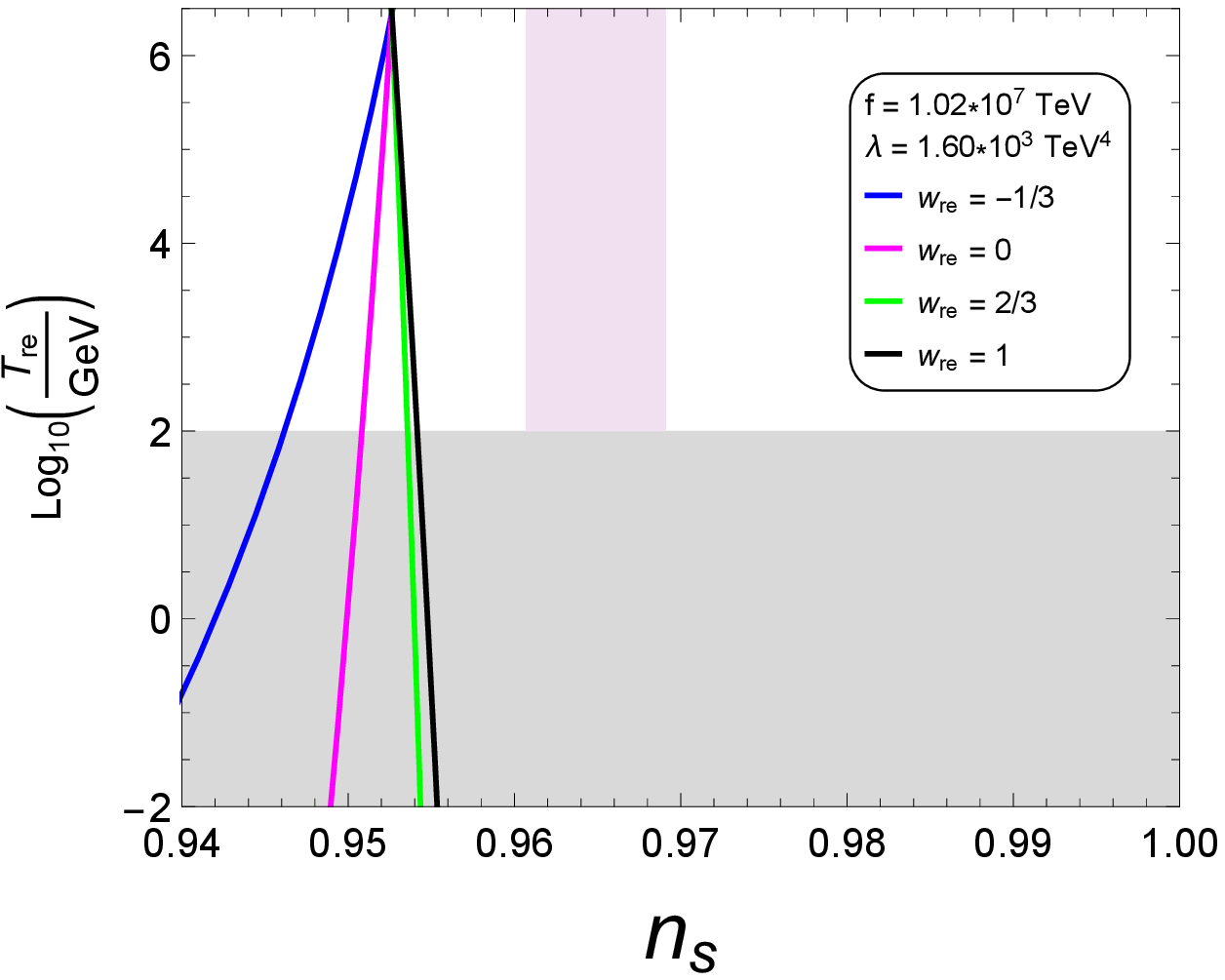}
\caption{Plots of $N_{re}$ and $T_{re}$ as functions of $n_s$ using the lower limit of the five-dimensional Planck mass ($M_5=10^5$\,TeV) for Natural inflation. The blue, pink, green and black curves corresponds to the following values of the EoS parameter: $w_{re} = -1/3$, $0$, $2/3$ and $1$, respectively. The purple region indicates the observational constraints on the spectral index, $n_s=0.965\pm 0.004$. The gray region show temperatures being below the electroweak scale, $T<10^2$\,GeV. Also, in
order to be consistent with BBN, it is required $T_{re}\gtrsim 10$\,MeV. The two plots on the left and on the right correspond to $\alpha=0.0393$ and $\alpha=0.0702$, respectively.} \label{fig:NI5}
\end{figure}

\begin{figure}[ht!]
\centering
\includegraphics[width=0.45\textwidth]{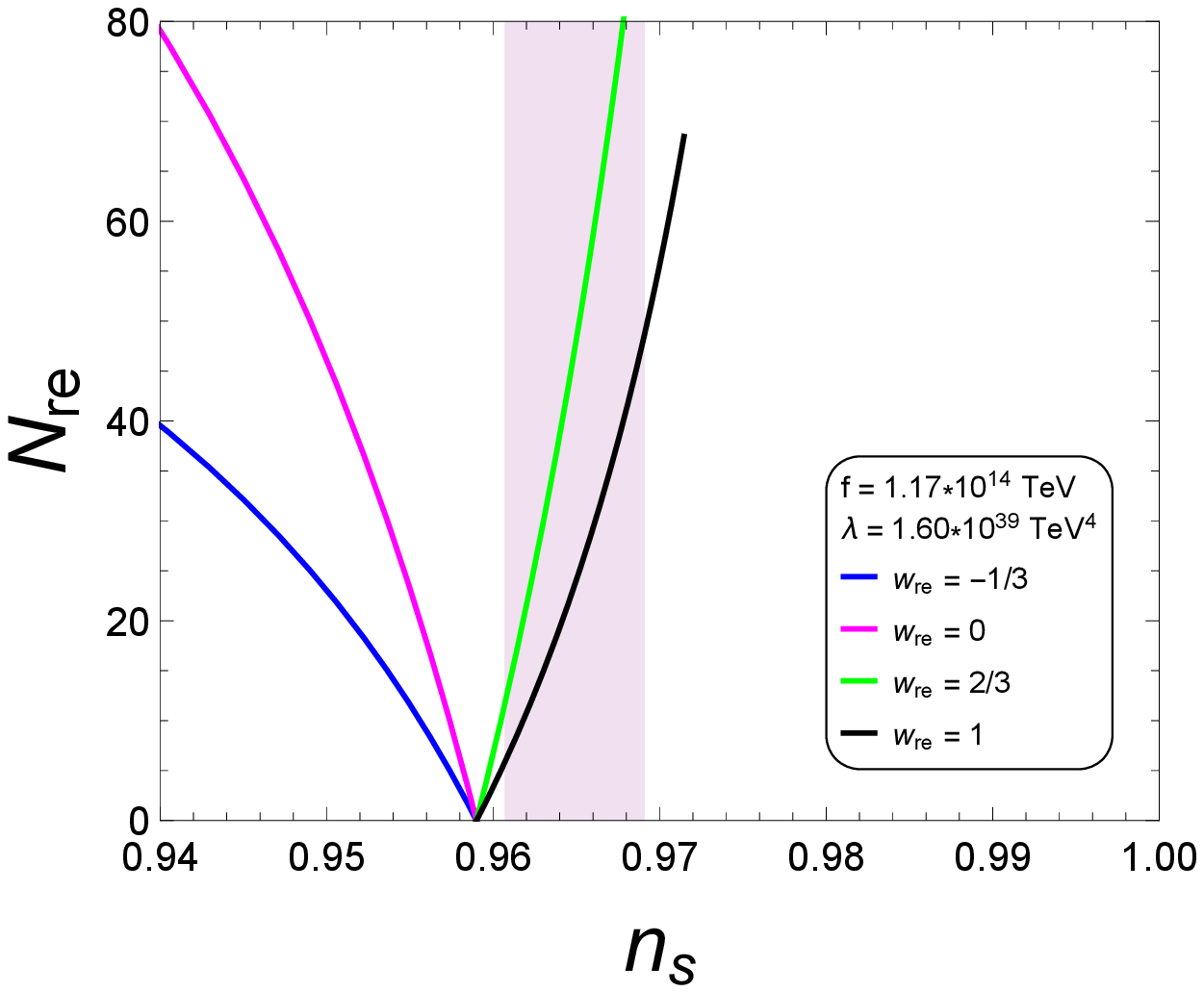}
\includegraphics[width=0.45\textwidth]{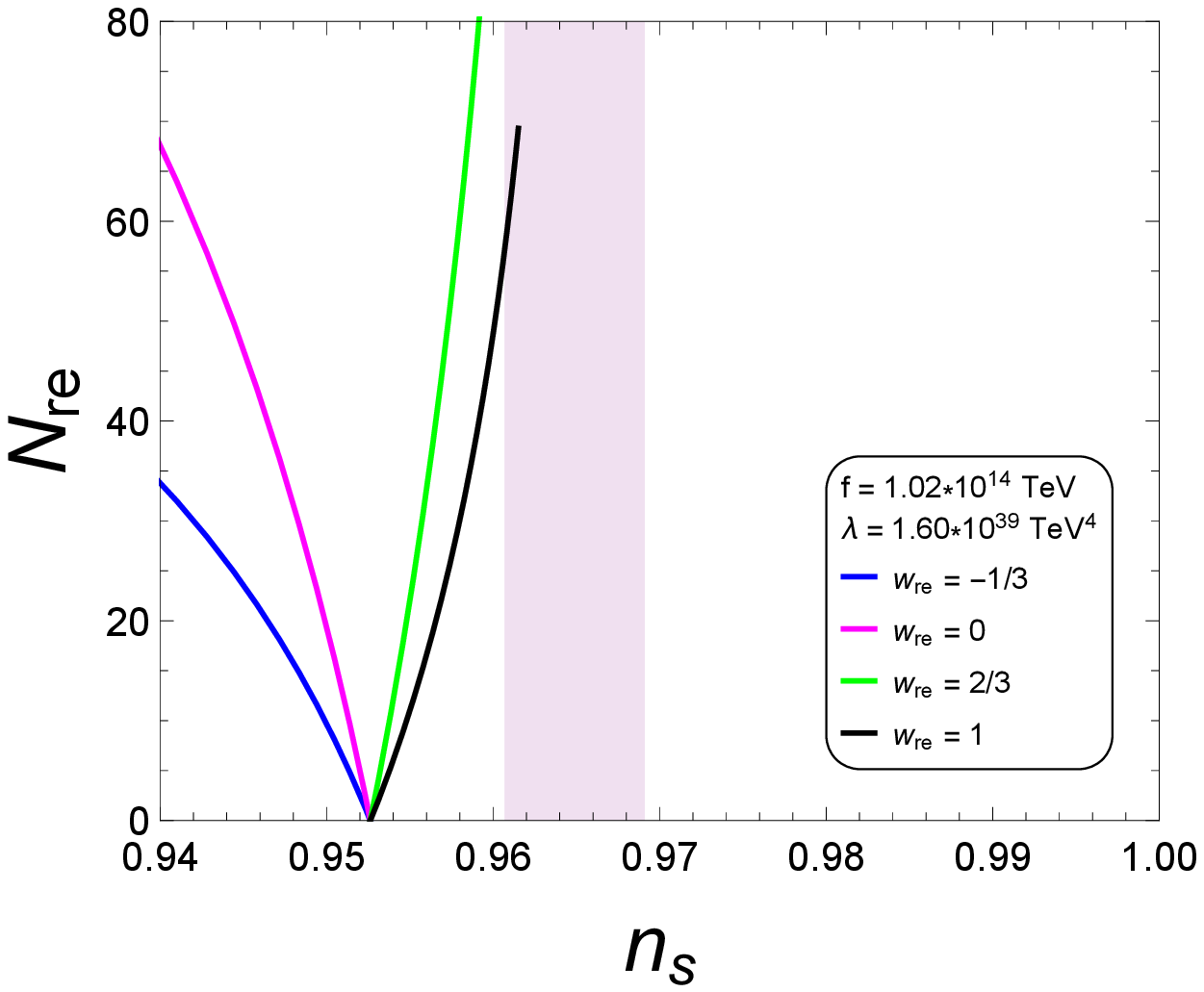}
\includegraphics[width=0.45\textwidth]{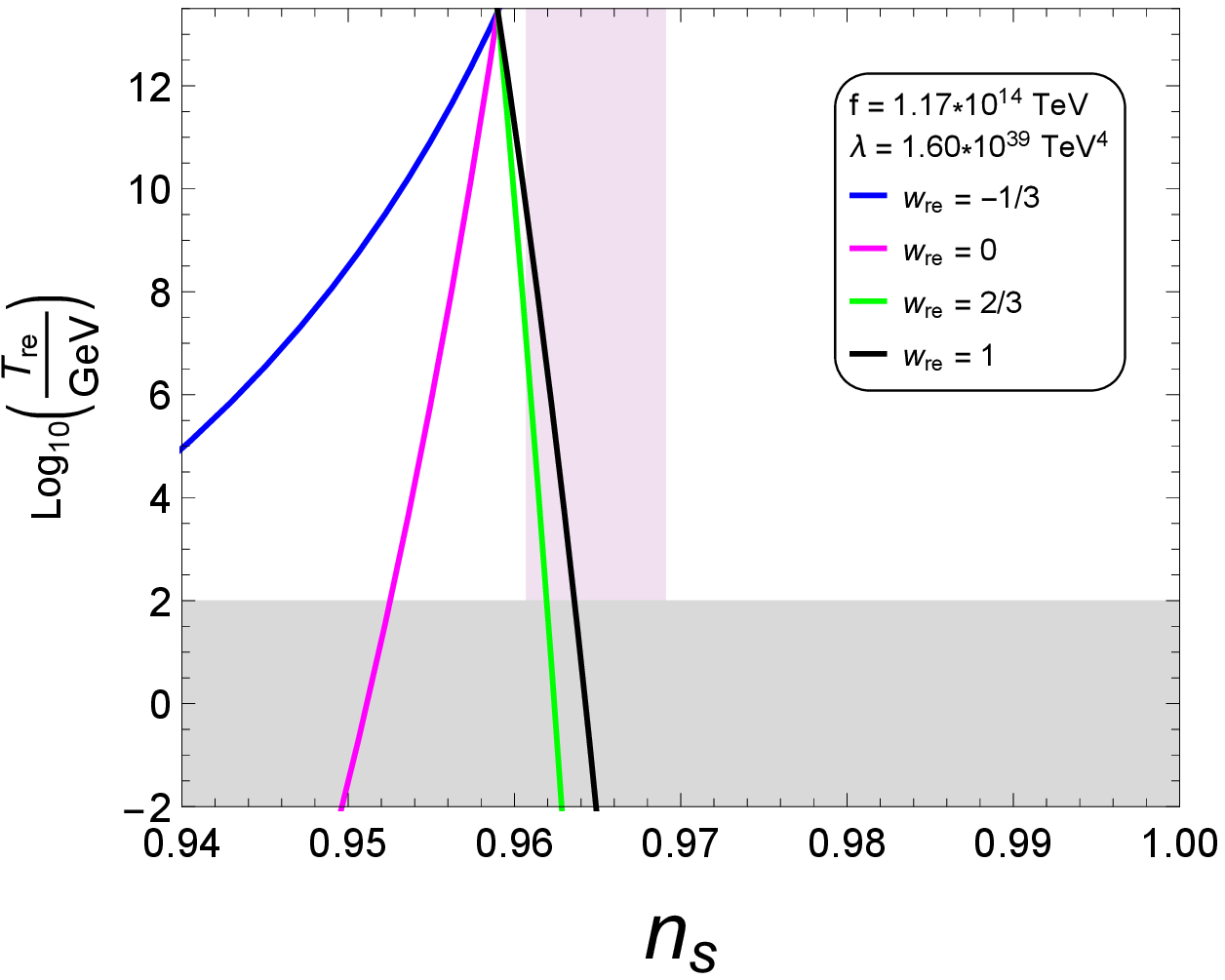}
\includegraphics[width=0.45\textwidth]{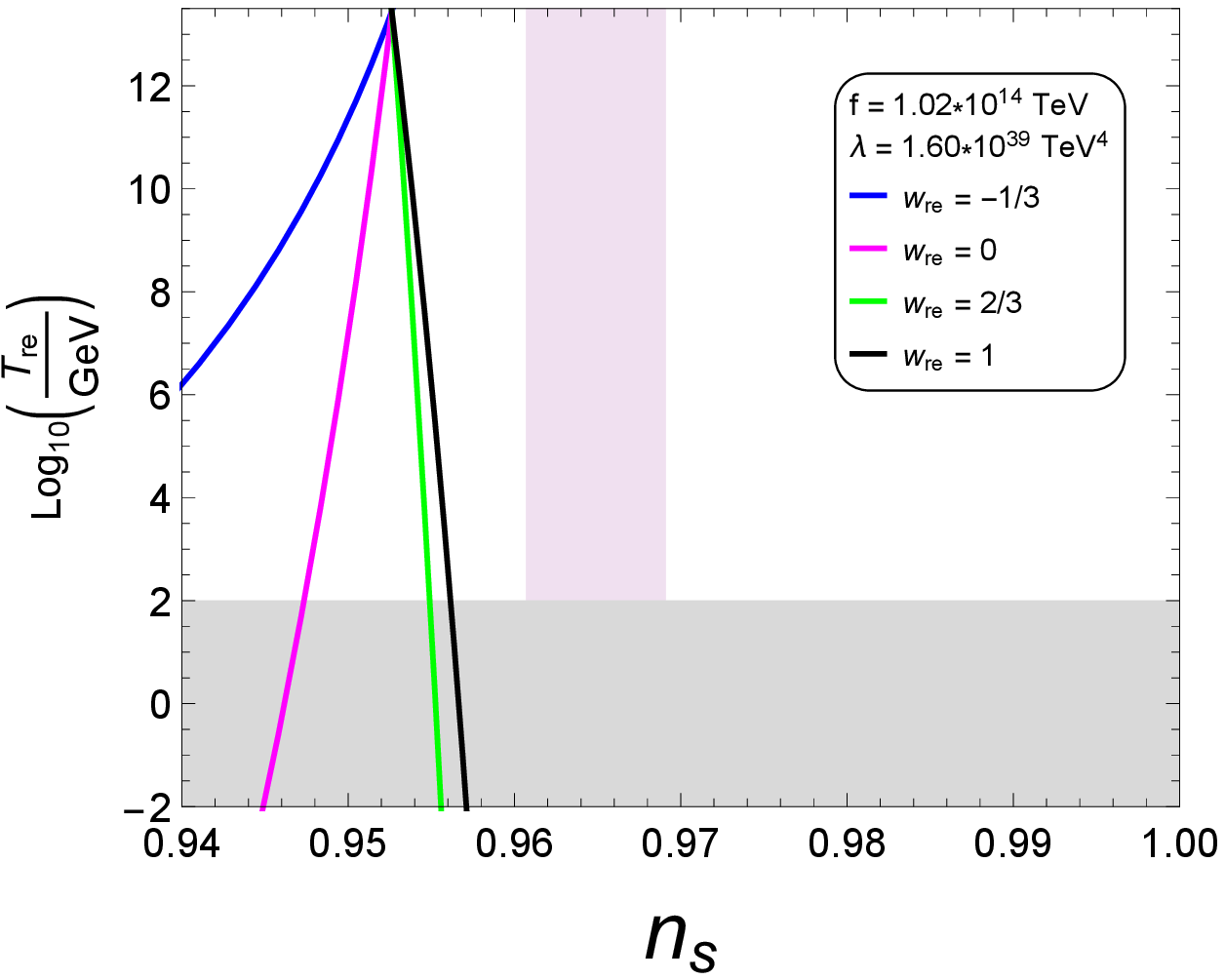}
\caption{Same as Fig. \ref{fig:NI5} but for the upper limit of the five-dimensional Planck mass, $M_5=10^{12}$\,TeV.}\label{fig:NI12}
\end{figure}

We now investigate the predictions regarding the number of $e$-folds as well as the temperature associated with the reheating epoch $N_{re}$ and $T_{re}$, respectively. In doing so, we plot
parametrically Eqs. (\ref{nsni}), \eqref{nre}, and \eqref{tre} with respect to the number of $e$-folds $N_k$ for several values of the effective equation-of-state parameter $w_{re}$ over the range $-\frac{1}{3}\leq w_{re}\leq1$, as well as 
$\alpha$, which encodes the information about the mass scales $f$ and $\Lambda$, and the brane tension $\lambda$. In Fig. \ref{fig:NI5}, we show the plots for reheating when using the lower limit of the 5-dimensional Planck mass, namely $M_5=10^5$\,TeV and two allowed values of $\alpha$ at $N_k = 65$. On the other hand, in Fig. \ref{fig:NI12} we use the upper limit on $M_5$, $M_5=10^{12}$\,TeV for the same values of $\alpha$. For the other values of $\alpha$, the prediction of reheating has the same behaviour, however we will show the plots that fit better with current observational data. Firstly, we must note that, for the two values of $M_5$, the point at which the curves converge (implying instantaneous reheating, i.e. $N_{re}\rightarrow 0$) is gradually shifted to the left when we increasing the dimensionless parameter $\alpha$. Another important finding is that the temperature at which all curves intersect, i.e. the maximum reheating temperature, increases as the 5-dimensional Planck mass $M_5$ increases. In particular, for $M_5=10^5$\,TeV the maximum reheating temperature is about $10^7$\,GeV, while for $M_5=10^{12}$\,TeV, it is about $10^{14}$\,GeV. Then, a new phenomenology arises in comparison to the former analysis within the standard scenario, in which the maximum reheating temperature (if reheating is instantaneous) is $T_{re}\lesssim 7\times 10^{15}$\,GeV. This dependence reheating temperature on five-dimensional Planck mass has been realised in Ref. \cite{Bhattacharya:2019ryo} so far, where the authors reconstructed the inflationary potential in the RS-II brane-world. Furthermore, our reheating temperature, which depends strongly on the five-dimensional Planck mass, for $M_5=10^{12}$\,TeV is at least two orders of magnitude greater than those found in \cite{Mohammadi:2020ake} in which case the temperature is more sensitive to the number of $e$-folds.




If we assume that during reheating, the universe is governed by an effective equation-of-state of the form $P = w_{re} \rho$, where $P$ and $\rho$ denote the pressure and the energy density, respectively, of the fluid in which the inflaton decays. Then, it becomes important to find what EoS parameter 
$w_{re}$ is preferred by current observational bounds. For doing that, we analyze when each curve of the reheating temperature plots against the scalar spectral index enters to the purple region (at 1$\sigma$ of $n_s$) and meets
the point at which all curves converges.
Therefore, an allowed range for the scalar spectral index $n_s$ as well as $N_k$ is found when fixing $\alpha$. For consistency, we display the results for the plots of Fig. \ref{fig:NI12} in Table \ref{wni}. It is worth noting that these values for the duration of reheating differ from those obtained in \cite{Mohammadi:2020ake} which are found to be $N_{re}\approx 20$ for $w_{re} = 1$ and $N_k=65$ when they use $M_5=5\times10^{12}$\,TeV. On the other hand, it should be noted that for $\alpha=0.0702$ and $\alpha=0.0785$ (plots not shown), none of the curves enter to the purple region, while for $\alpha=0.0393$ and $M_5=10^5$\,TeV only one curve, corresponding to $w_{re}=1$ is inside but for $M_5=10^{12}$\,TeV two curves ($w_{re}=2/3$ and $w_{re}=1$) are inside. For $\alpha=0.0325$ (plots not shown) the same two curves are inside.



\begin{table}[H]
\begin{center}
\begin{tabular}{ cc }   
\begin{tabular}{ |c|c|c| } \hline
$w_{re}$       &  $N_k$\\\hline
   \,2/3\,   & \,59 - 62\,  \\ 
   \,1\,     & \,59 - 65\,  \\ \hline
\end{tabular} \\
\end{tabular}
\caption{Summary of the allowed
range for the number of $e$-folds for each EoS parameter $w_{re}$ for $M_5=10^{12}$\,TeV when the dimensionless parameter $\alpha$ is fixed to be $\alpha=0.0393$.}\label{wni}
\end{center}
\end{table}


\begin{figure}[ht!]
 \centering
  \includegraphics[width=0.45\textwidth]{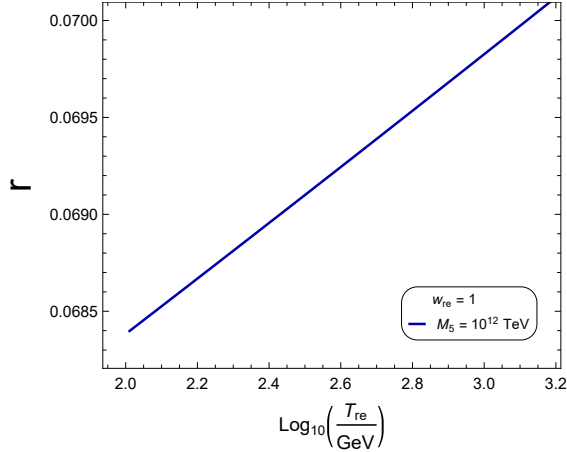}\caption{Plot of the tensor-to-scalar ratio $r$ against the reheating temperature for Natural inflation for $w_{re}=1$ and $\alpha=0.0393$ when using $M_5=10^{12}$\,TeV.} \label{fig:rNI}
\end{figure}


We also want to know what are the allowed values for the tensor-to-scalar ratio in terms of the reheating temperature for some values belonging the allowed range of the 5-dimensional Planck mass. In doing so, we plot parametrically Eqs. (\ref{rrni}) and (\ref{tre}) with respect to the number of $e$-folds, which varies according to the available found so far with $M_5$ and $\alpha$ fixed (see Table \ref{wni}). The values obtained for $r$ must be consistent with the upper limit by PLANCK 2018 data in combination with the BICEP2/Keck Array (BK14) data. In this way we can discard those values of $\alpha$, $T_{re}$ and $w_{re}$ for which $r$  does not meet this bound. We emphasize that this method must be consistent with the previous analysis. The only values of $\alpha$ and $w_{re}$ in consistency with the former, correspond to $\alpha=0.0393$ and $w_{re}=1$, as it is shown in Fig. \ref{fig:rNI}. Firstly, we observe that the curves starts at $T_{re}\approx 10^2$\,GeV which is consistent with the previous plots of reheating. In principle, temperatures within the range $10\,\textup{MeV}\lesssim T_{re}\lesssim 10^2$\,GeV (gray region from Figs. \ref{fig:NI5} and \ref{fig:NI12} may not be discarded, but would be interesting for baryogenesis \cite{Davidson:2000dw}. Next, we note that the curve for the lower limit of $M_5$ is well outside the upper limit on $r$, so we can discard it in principle. Consequently, it is found that for $w_{re}=1$, the reheating temperature must be in the range of
\begin{eqnarray}
10^2\,\textup{GeV}\lesssim T_{re}\lesssim 10^{3}\,\textup{GeV},
\end{eqnarray}
when $M_5=10^{12}$\,TeV.

\section{Hilltop Inflation on the brane}\label{hillbra}

\subsection{Dynamics of inflation}

Quadratic Hilltop inflation is driven by the potential \eqref{hill}
\begin{equation}
V(\phi)=\Lambda^4 \left[1-\left(\frac{\phi}{\mu}\right)^2 \right]\label{hpot}
\end{equation}

In this case, the slow-roll parameters in the high-energy limit are given by
\begin{eqnarray}
\epsilon &=&\frac{\alpha \,x^2}{(1-x^2)^3},\label{epsilonh}\\
\eta &=&-\frac{\alpha}{2\,(1-x^2)^2},\label{etah}
\end{eqnarray}
where the dimensionless parameters are defined as follows
\begin{eqnarray}
x &\equiv& \frac{\phi}{\mu},\\
\alpha &\equiv& \frac{M_4^2 \lambda}{\pi^2\, \mu^2 \, \Lambda^4}.\label{alphah}
\end{eqnarray}

Unlike Natural Inflation, for our quadratic Hilltop potential, the scalar field at Hubble-radius crossing $\phi_k$ is found by means numerically. In that case, we start with the definition of the number of $e$-folds in terms of the Hubble rate
\begin{equation}
dN = H dt.
\end{equation}

Then, using the slow-roll approximation $3H\dot{\phi} + V \approx 0$ within the
high-energy limit and the relation between $x$ and $\phi$, we have a differential expression which gives
us $x(N)$
\begin{equation}
    dN\simeq -\frac{4\pi\mu V^2}{\lambda M_4^2 V'}\,dx.
\end{equation}

We obtain the numerical solution for $x_k$ by means introducing the initial condition $x(N=0)=x_{end}$, where $x_{end}=\phi_{end}/\mu$ is obtained from the condition at the end of inflation, i.e. $\epsilon(x_{end})=1$ from Eq. (\ref{epsilonh}).

\subsection{Cosmological perturbations}

\begin{figure}[ht!]
\centering
\includegraphics[scale=0.5]{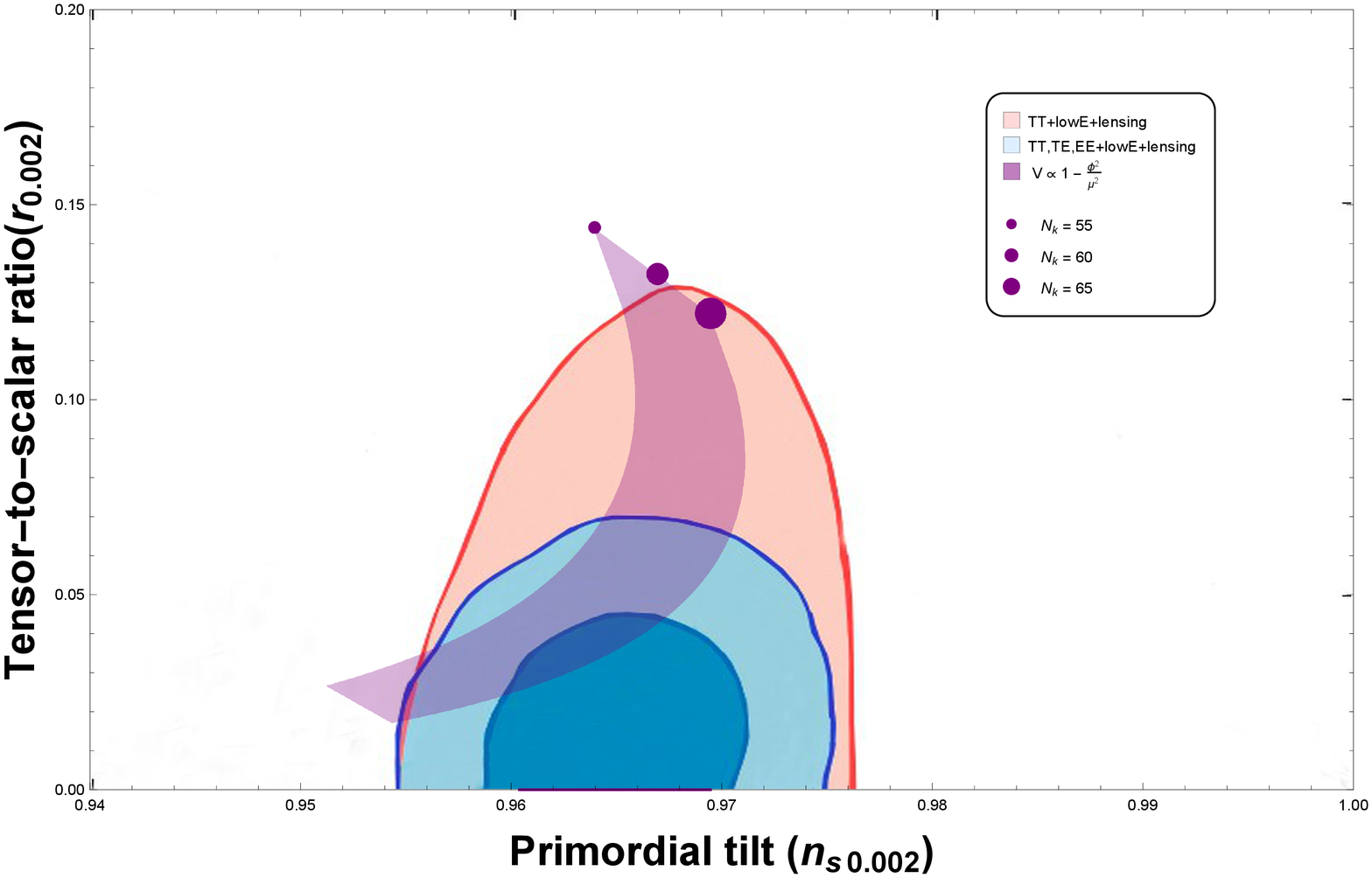}
\caption{Plot of the tensor-to-scalar ratio $r$ versus the scalar spectral index $n_s$ for quadratic Hilltop inflation on the brane along with the two-dimensional marginalized joint confidence contours for ($n_s,r$) at the 68$\%$ (blue region) and 95$\%$ (light blue region) C.L., from the latest PLANCK 2018 results.}\label{contornoH}
\end{figure}

Replacing the potential \eqref{hpot} into Eq. \eqref{ASHE} we found the following expression for the scalar power spectrum
as a function of the scalar field
\begin{equation}
P_S=\frac{4\,\gamma^4\,(1-x^2)^6}{3\,\pi^2 \,\alpha^3 \,x^2},\label{perth}
\end{equation}
where $\gamma=\frac{\Lambda}{\mu}$ is a dimensionless parameter.
To obtain the scalar spectral index and 
the tensor to scalar ratio, both evaluated
at the Hubble-radius crossing, one first
replaces the solution for $x_k$ in 
$\epsilon$ and $\eta$ and uses Eqs. \eqref{ns} and \eqref{rr}. Next, we plot
parametrically $n_s$ and $r$, varying simultaneously $\alpha$ in a wide range and $N_k$ within the range $N_k=55-65$. Fig. \ref{contornoH} shows the tensor-to-scalar ratio against the spectral index plot using the two-dimensional marginalized joint confidence contours for ($n_s,r$) at the 68$\%$ (blue region) and 95$\%$ (light blue region) C.L., from the latest PLANCK 2018 results.
Using the same method as in Natural inflation to found the allowed values of $\alpha$, one obtains that the predictions of the model are within the 95\% C.L. region from PLANCK data if, for $N_k=55$, $\alpha$ lies in the range $1.35\times 10^{-2}\lesssim \alpha \lesssim 3.19\times 10^{-2}$, and the corresponding prediction for the tensor-to-scalar ratio is $0.070 \gtrsim r \gtrsim 0.036$.
Accordingly, for $N_k=60$, the allowed range of the dimensionless parameter 
$\alpha$ is $1.04\times 10^{-2}\lesssim \alpha \lesssim 3.72\times 10^{-2}$, while $r$ is found to be in the range $0.069 \gtrsim r \gtrsim 0.024$. In the same fashion, for $N_k=65$, $\alpha$ is
found to be in range range $9.10\times 10^{-3}\lesssim \alpha \lesssim 3.96\times 10^{-2}$, while the prediction for $r$ is $0.065 \gtrsim r \gtrsim 0.017$.

By combining Eq. \eqref{perth} with the constraints on $\alpha$, and the amplitude of the scalar spectrum $P_S\simeq 2.2\times 10^{-9}$, the allowed values
for $\gamma$ are found to be in the range
\begin{eqnarray}
3.33\times 10^{-4} \lesssim \gamma \lesssim 4.00\times 10^{-4},\\
2.98\times 10^{-4} \lesssim \gamma \lesssim 3.85\times 10^{-4},\\
2.77\times 10^{-4} \lesssim \gamma \lesssim 3.65\times 10^{-4},
\end{eqnarray}
for $N_k=55$, $N_k=60$ and $N_k=65$, respectively.
The allowed ranges for $\alpha$ and $\gamma$ are summarized in Table \ref{constalphah}.

 

\begin{table}[H]
\begin{center}
\begin{tabular}{| c | c | c |}\hline
 $N_k$   & Constraint on $\alpha$  & Constraint on $\gamma$ \\\hline
   \,55\,  & \,$0.0135 \lesssim \alpha \lesssim 0.0319$\,  & \,$3.33\times 10^{-4} \lesssim \gamma \lesssim 4.00\times 10^{-4}$\,  \\
   \,60\,  & \,$0.0104 \lesssim \alpha \lesssim 0.0372$\,  & \,$2.98\times 10^{-4} \lesssim \gamma \lesssim 3.85\times 10^{-4}$\,  \\
  \,65\,  & \,$0.0091 \lesssim \alpha \lesssim 0.0396$\,  & \,$2.77\times 10^{-4} \lesssim \gamma \lesssim 3.65\times 10^{-4}$\,  \\ \hline
\end{tabular}
\caption{Results of the constraints on the parameters $\alpha$ and $\gamma$ for Hilltop inflation in the high-energy of Randall-Sundrum brane model, using the last data of PLANCK.} \label{constalphah}
\end{center}
\end{table}


The expressions for the mass scales $\mu$ and $\Lambda$ are obtained after replacing Eq. \eqref{M5M4} into the definition of $\alpha$ (Eq. \eqref{alphah}), yielding
\begin{equation} 
\mu = \left ( \frac{3}{4 \pi^2 \alpha \gamma^4} \right )^{1/6} M_5, \label{muh}
\end{equation}
\begin{equation}
\Lambda = \gamma \mu=\gamma \left ( \frac{3}{4 \pi^2 \alpha \gamma^4} \right )^{1/6} M_5. \label{Lh}
\end{equation}

After evaluating those expressions at the several values for $\alpha$ and $\gamma$ (Table \ref{constalphah}) and considering the lower limit for the five-dimensional Planck mass, $M_5=10^5$\,TeV, the brane tension is found to be $\lambda=1.60\times 10^{-3}$\,TeV$^4$ while the corresponding constraints on the mass scales (in units of TeV) are shown in the top panel of Table \ref{Tmulh}. Using the same method to found an upper limit of the five-dimensional Planck mass as in Natural inflation, we obtain that $M_5\ll 10^{14}$\,TeV. Assuming as maximum allowed value $M_5=10^{12}$ TeV, we obtain $\lambda=1.60\times 10^{39}$\,TeV$^4$ and the corresponding values for mass scales are shown in the bottom panel of Table \ref{Tmulh}.


\begin{table}[H]
\centering
\begin{tabular}{| c | c | c |}\hline
 $N_k$   & Constraint on $\mu$\,[TeV]  & Constraint on $\Lambda$\,[TeV] \\\hline
  \,55\,  & \,$2.78\times 10^7 \gtrsim \mu \gtrsim 2.13\times 10^7$\,  & \,$9.24\times 10^{3} \gtrsim \Lambda \gtrsim 8.51\times 10^{3}$\,  \\
  \,60\,  & \,$3.12\times 10^7 \lesssim \mu \gtrsim 2.13\times 10^7$\,  & \,$9.30\times 10^{3} \gtrsim \Lambda \gtrsim 8.19\times 10^{3}$\,  \\
  \,65\,  & \,$3.35\times 10^7 \gtrsim \mu \gtrsim 2.18\times 10^7$\,  & \,$9.28\times 10^{3} \gtrsim \Lambda \gtrsim 7.97\times 10^{3}$\,  \\ \hline
\end{tabular}\vspace{1cm}
\begin{tabular}{| c | c | c |}\hline
 $N_k$   & Constraint on $\mu$\,[TeV]  & Constraint on $\Lambda$\,[TeV] \\\hline
  \,55\,  & \,$2.78\times 10^{14} \gtrsim \mu \gtrsim 2.13\times 10^{14}$\,  & \,$9.24\times 10^{10} \gtrsim \Lambda \gtrsim 8.51\times 10^{10}$\,  \\
  \,60\,  & \,$3.12\times 10^{14} \gtrsim \mu \gtrsim 2.13\times 10^{14}$\,  & \,$9.30\times 10^{10} \gtrsim \Lambda \gtrsim 8.19\times 10^{10}$\,  \\
  \,65\,  & \,$3.35\times 10^{14} \gtrsim \mu \gtrsim 2.18\times 10^{14}$\,  & \,$9.28\times 10^{10} \gtrsim \Lambda \gtrsim 7.97\times 10^{10}$\,  \\ \hline
\end{tabular}
\caption{Results for the constraints on the mass scales $\mu$ and $\Lambda$ for quadratic Hilltop inflation in the high-energy limit of Randall-Sundrum brane model using the last data of PLANCK. The top table shows the results using $M_5=10^{5}$\,TeV while the bottom table shows the results using $M_5=10^{12}$\,TeV.} \label{Tmulh}
\end{table}

For this model, the plots for the Swampland criteria, which are not shown, but these present the same behavior that those shown in FIG. \ref{fig:NIswampland}. For the distance conjecture, $\overline{\Delta \phi}$ increases with the number of $e$-folds but also increases as the 5-dimensional Planck mass grows, so this conjecture is fulfilled. On the other hand, for the de Sitter conjecture, $\overline{\Delta V}$ decreases with both the number of $e$-folds and $M_5$ which, having in mind the discussion in Section \ref{natbra}, it is avoided.

\subsection{Reheating}

\begin{figure}[ht!]
\centering
\includegraphics[width=0.45\textwidth]{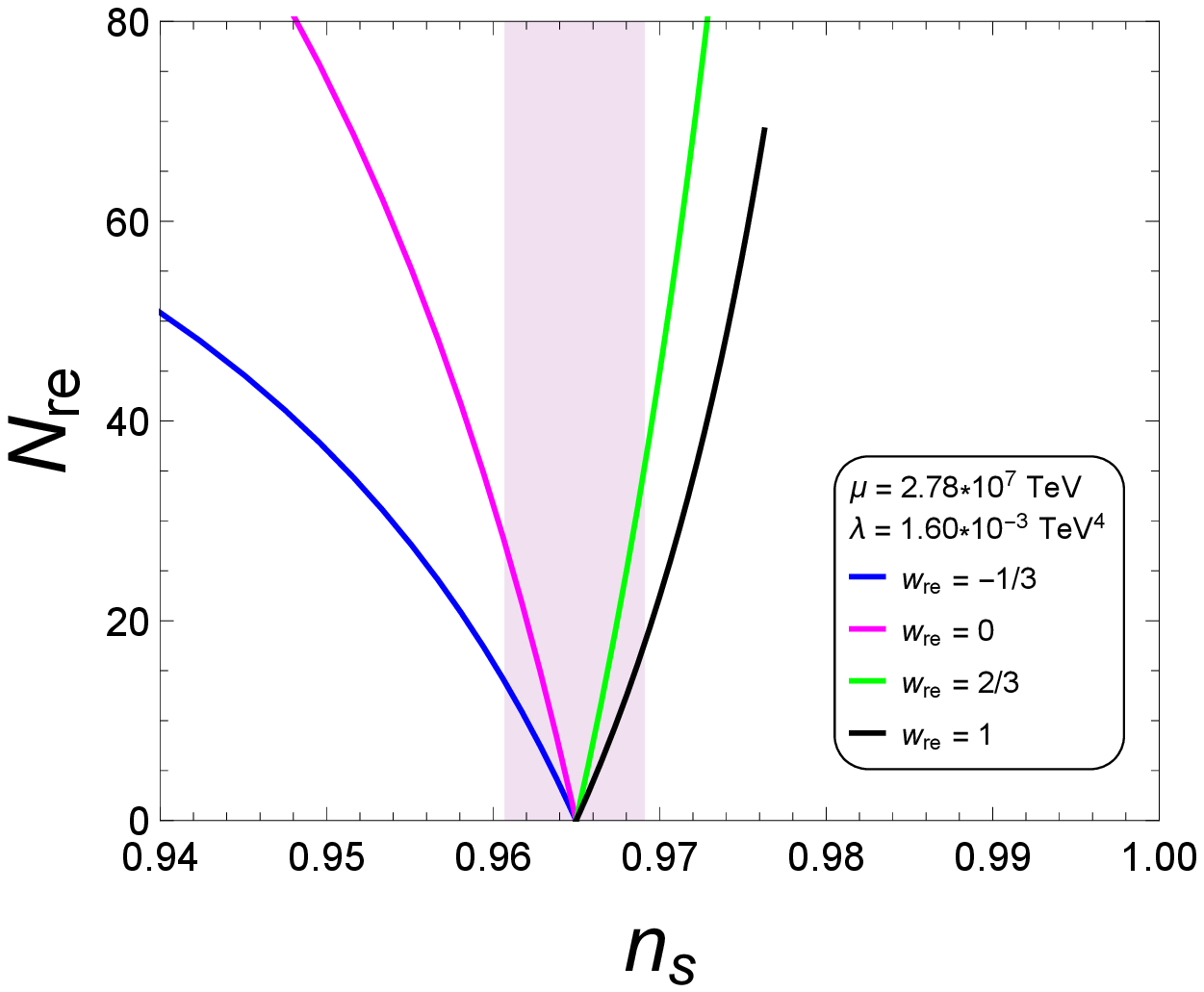}
\includegraphics[width=0.45\textwidth]{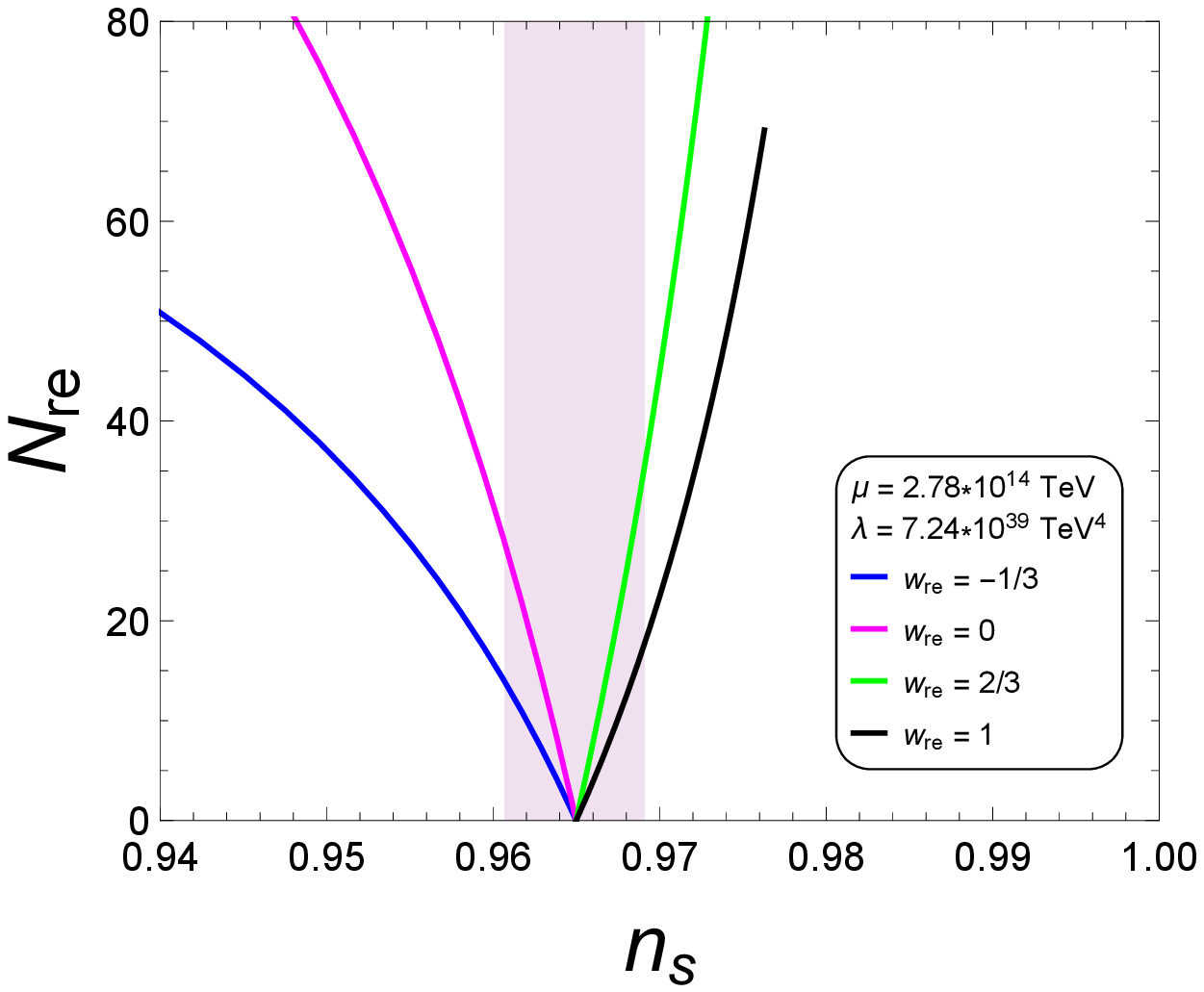}
\includegraphics[width=0.45\textwidth]{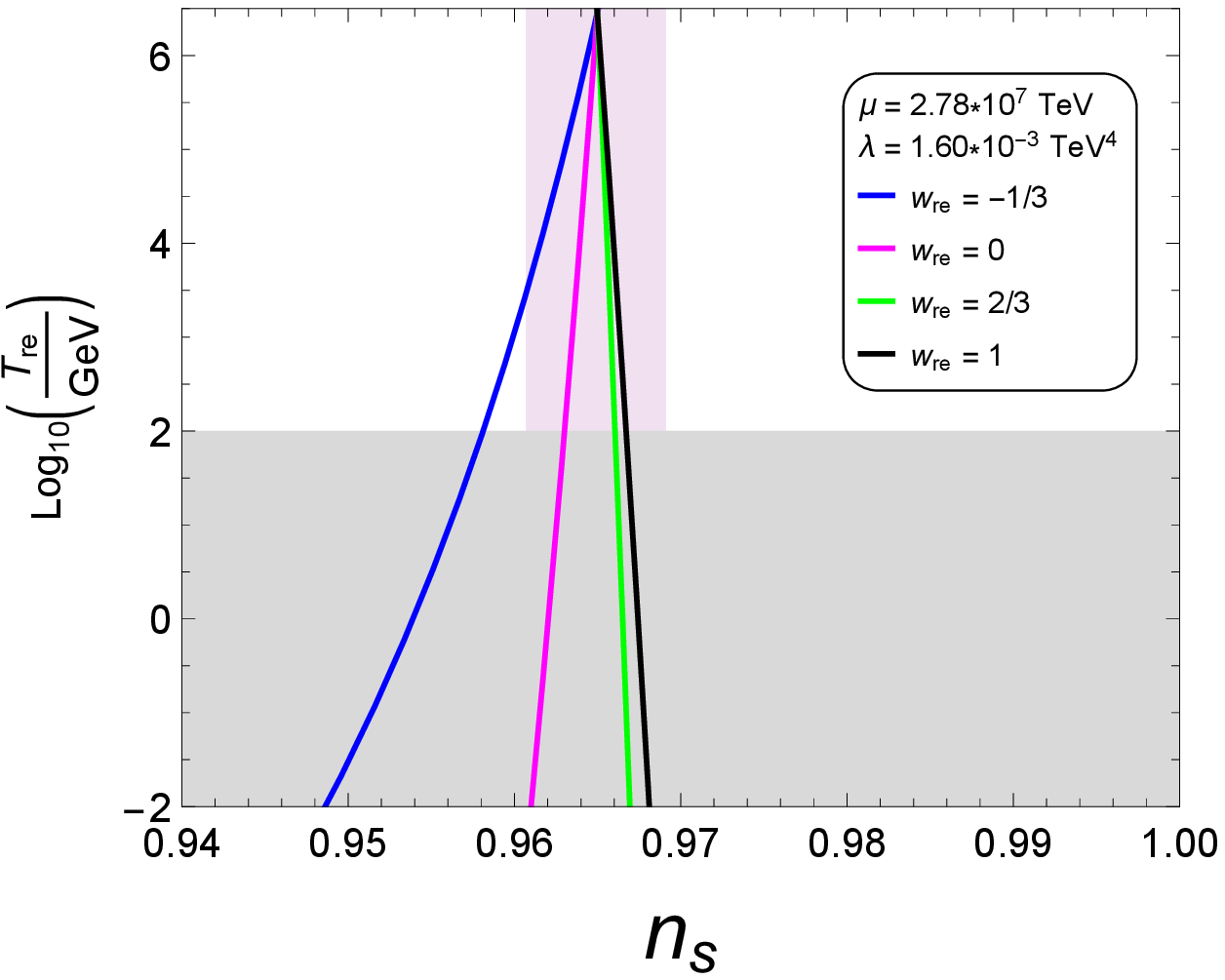}
\includegraphics[width=0.45\textwidth]{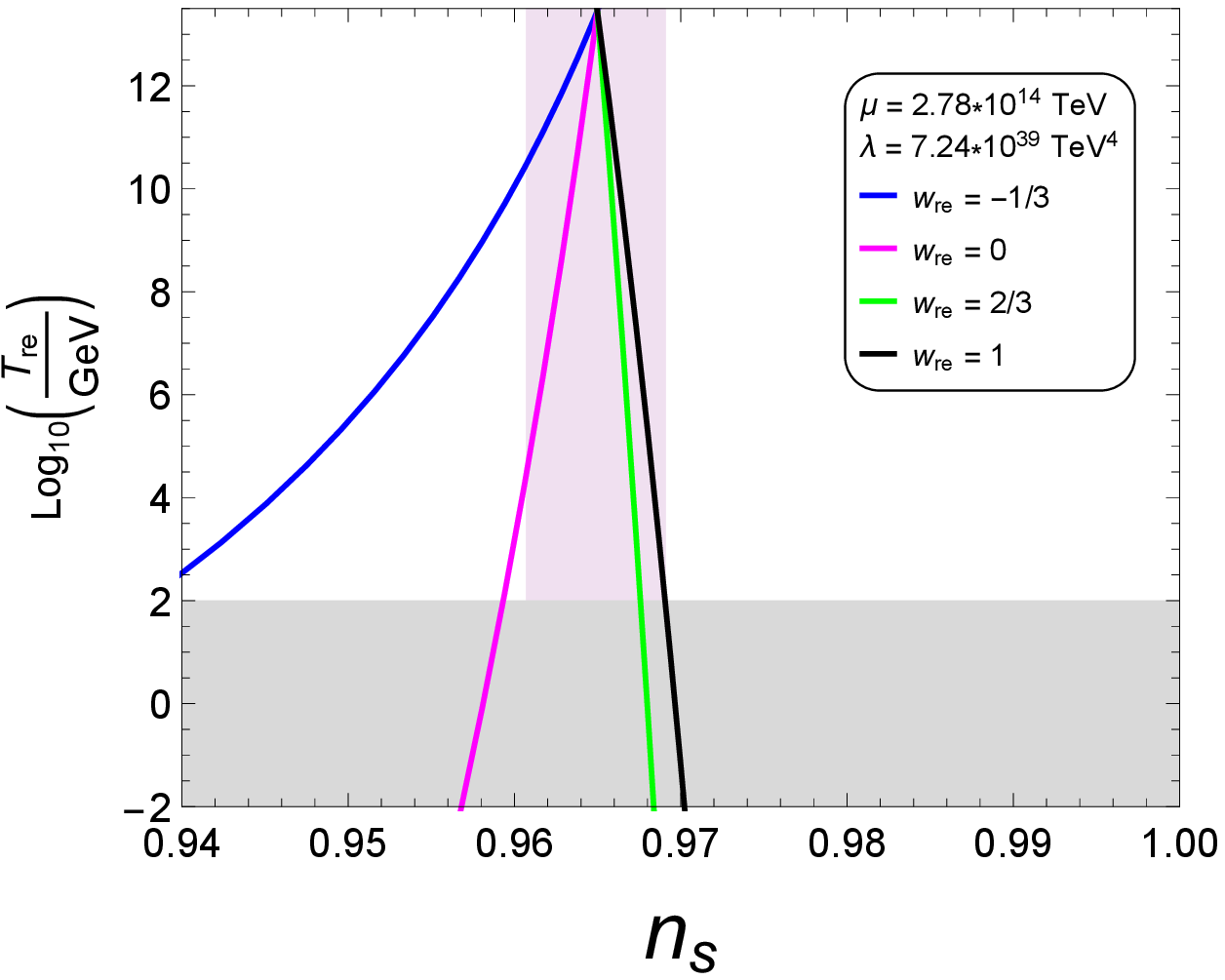}
\caption{Plots of $N_{re}$ and $T_{re}$ as functions of $n_s$ for Hilltop inflation. The left panels shows the plots for $M_5=10^{5}$\,TeV while the right panels shows the plot for $M_5=10^{12}$\,TeV. The curves and the shading regions are the same as FIG. \ref{fig:NI5} and all plots corresponds to $\alpha=0.0135$.} \label{fig:H}
\end{figure}

In the same way as Natural inflation, we can give predictions for reheating plotting parametrically Eqs. \eqref{nre} and \eqref{tre} with respect to $\alpha$ and $N_k$ over the range of the effective EoS $-\frac{1}{3}\leq w_{re}\leq1$. In despite this type of potential is unbounded from below, i.e. does not present a minimum around which the inflaton oscillates and reheating is achieved, we may assume that 
the details of reheating are encoded in the effective EoS parameter $w_{re}$. Yet another possibility to achieve reheating is by adding extra terms as those in Eq. (\ref{hillm}) for $p=2$, which stabilizes the potential and prevent it becomes negative. In FIG. \ref{fig:H}, we show the plots for reheating using $M_5=10^5$\,TeV (left panels) and $M_5=10^{12}$\,TeV (right panels) for $\alpha=0.0135$ that corresponds to the constraints at $N_k=55$. Even though it is not shown in the plots, the behavior of the convergence point is the same as in Natural inflation, i.e., the point at which the curves converges shifts to the left when $\alpha$ increases. As it can be seen, the maximum temperature of reheating also increases with the five-dimensional Planck mass, giving $T_{re}\approx10^7$\,GeV for $M_5=10^5$\,TeV and $T_{re}\approx10^{14}$\,GeV for $M_5=10^{12}$\,TeV.

Analyzing when each curve of the reheating temperature plots enters to the purple region and meets the converge point
of instantaneous reheating, an allowed range for $N_k$ is found when fixing $\alpha$. For consistency, we display the results for the plots of Fig. \ref{fig:H} in Table \ref{wh}. It should be noted that for $\alpha=0.0104$ and $\alpha=0.0091$ (plots not shown), all of the four curves enter to the purple region, while for $\alpha=0.0319$, $\alpha=0.0372$ and  $\alpha=0.0396$ (plots not shown), none of the curves enter. 

\begin{table}[H]
\begin{center}
\begin{tabular}{ cc }   
\begin{tabular}{ |c|c|c| } \hline
 $w_{re}$       &  $N_k$\\\hline
   \,-1/3\,  & \,49 - 56\,  \\
   \,0\,     & \,53 - 56\,  \\
   \,2/3\,   & \,56 - 58\,  \\ 
   \,1\,     & \,56 - 59\,  \\ \hline
\end{tabular} &  
\hspace{1cm}
\begin{tabular}{ |c|c|c| } \hline
$w_{re}$       &  $N_k$\\\hline
   \,-1/3\,  & \,49 - 56\,  \\
   \,0\,     & \,49 - 56\,  \\
   \,2/3\,   & \,56 - 61\,  \\ 
   \,1\,     & \,56 - 65\,  \\ \hline
\end{tabular} \\
\end{tabular}
\caption{Summary of the allowed range for the number of $e$-folds for each EoS parameter $w_{re}$  when the dimensionless parameter $\alpha$ is fixed to  $\alpha=0.0135$. The left and right tables corresponds to a 5-dimensional Planck mass of $M_5=10^5$\,TeV and $M_5=10^{12}$\,TeV respectively.}\label{wh}
\end{center}
\end{table}

\begin{figure}[ht!]
\centering
\includegraphics[width=0.45\textwidth]{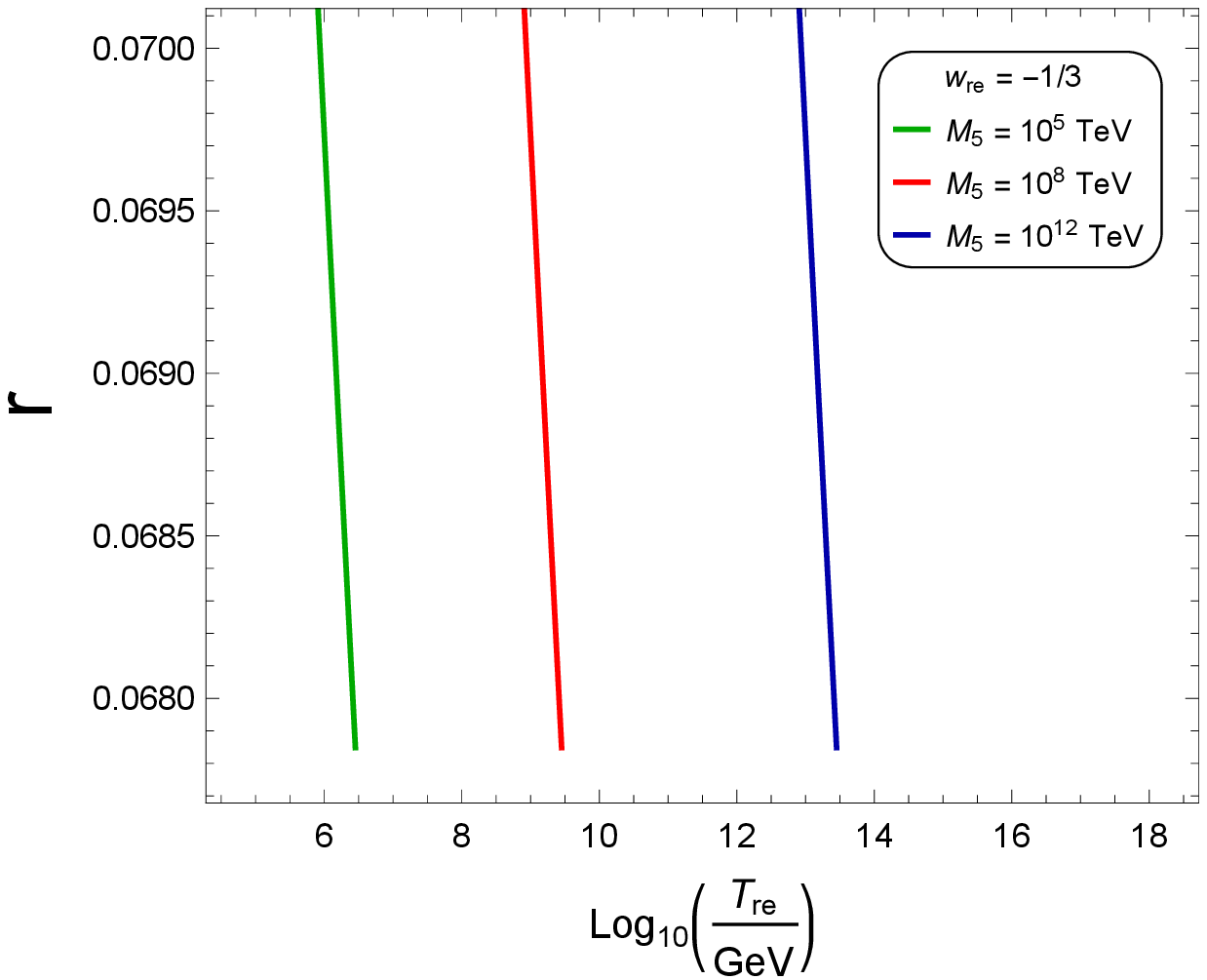}
\includegraphics[width=0.45\textwidth]{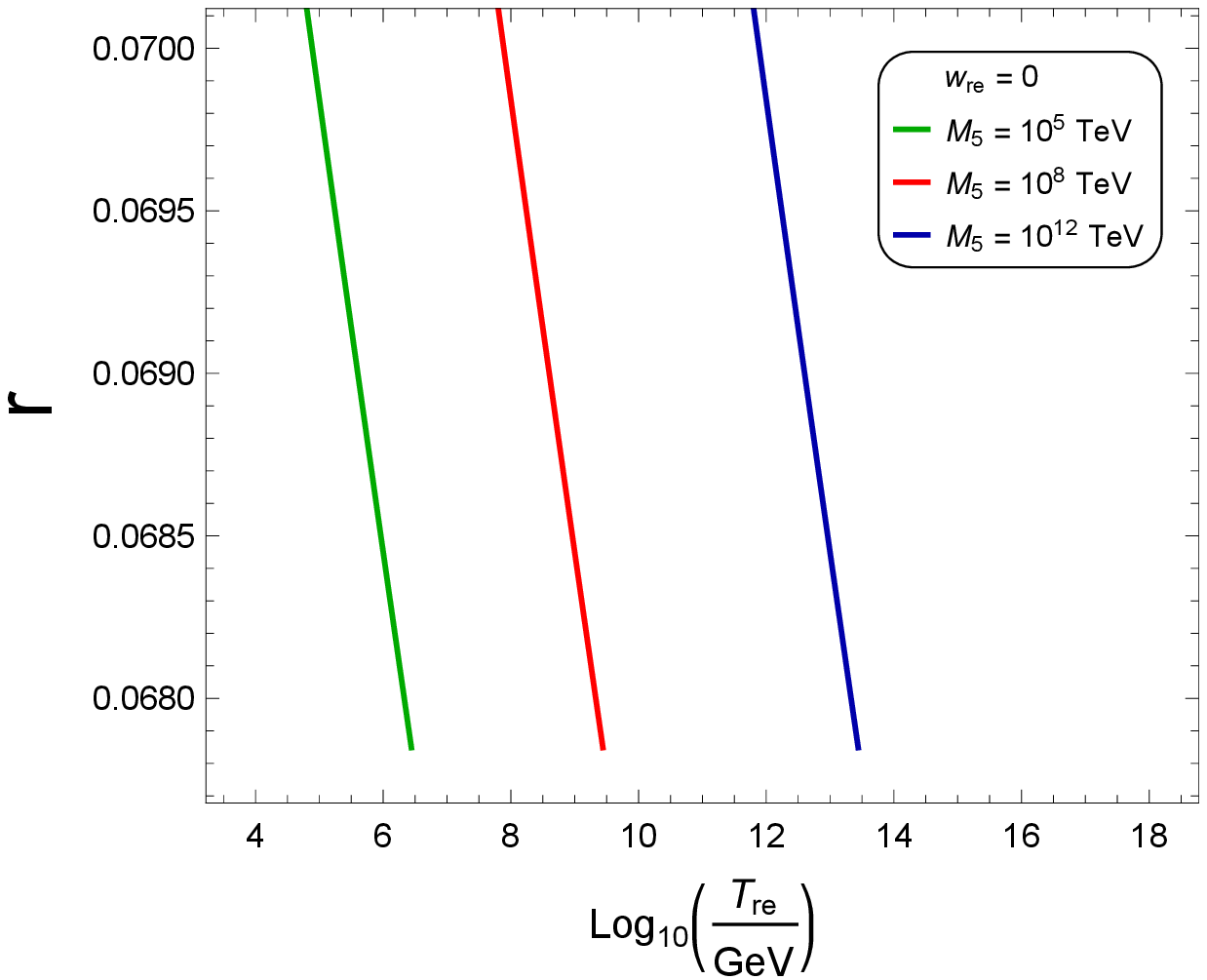}
\includegraphics[width=0.45\textwidth]{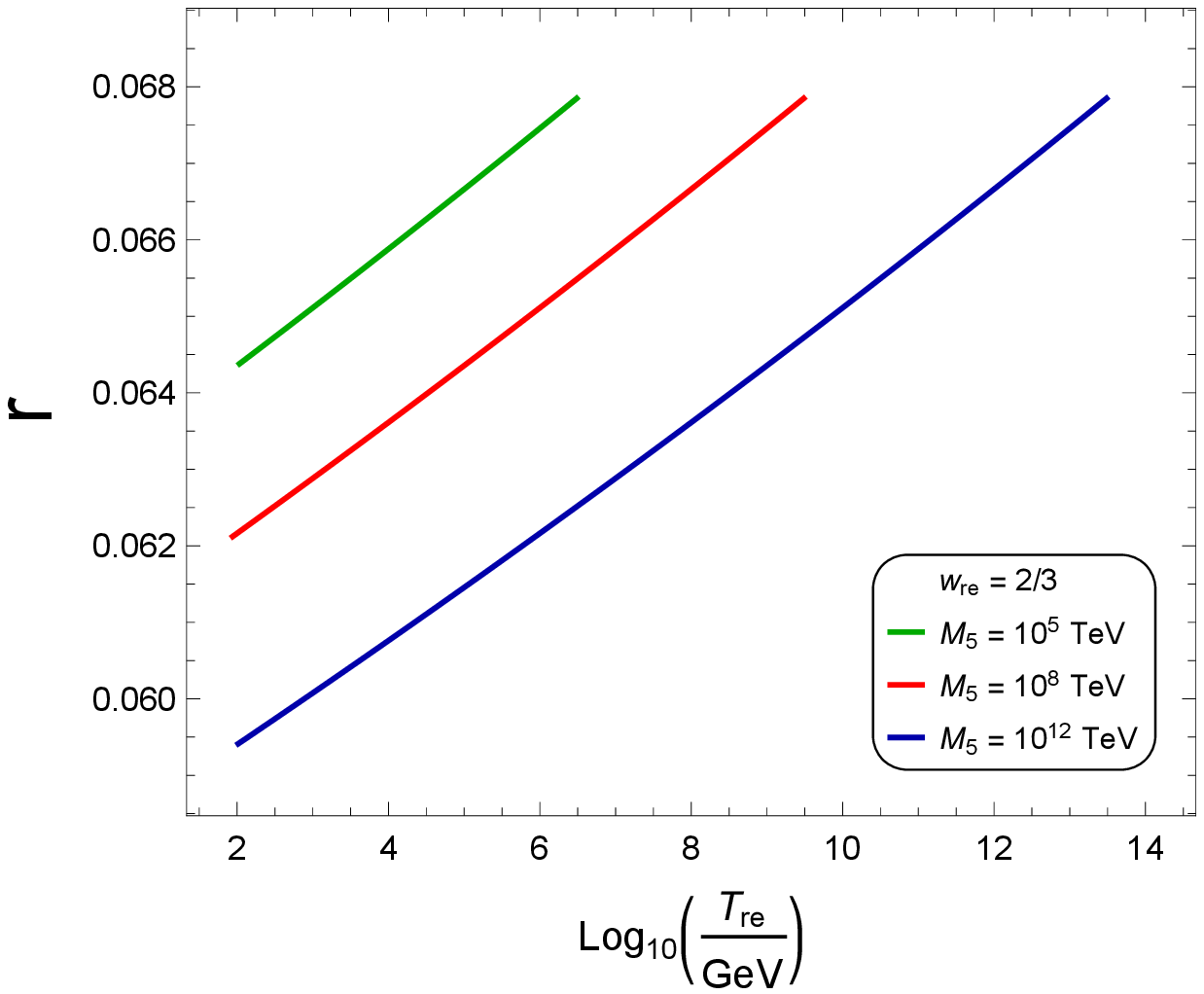}
\includegraphics[width=0.45\textwidth]{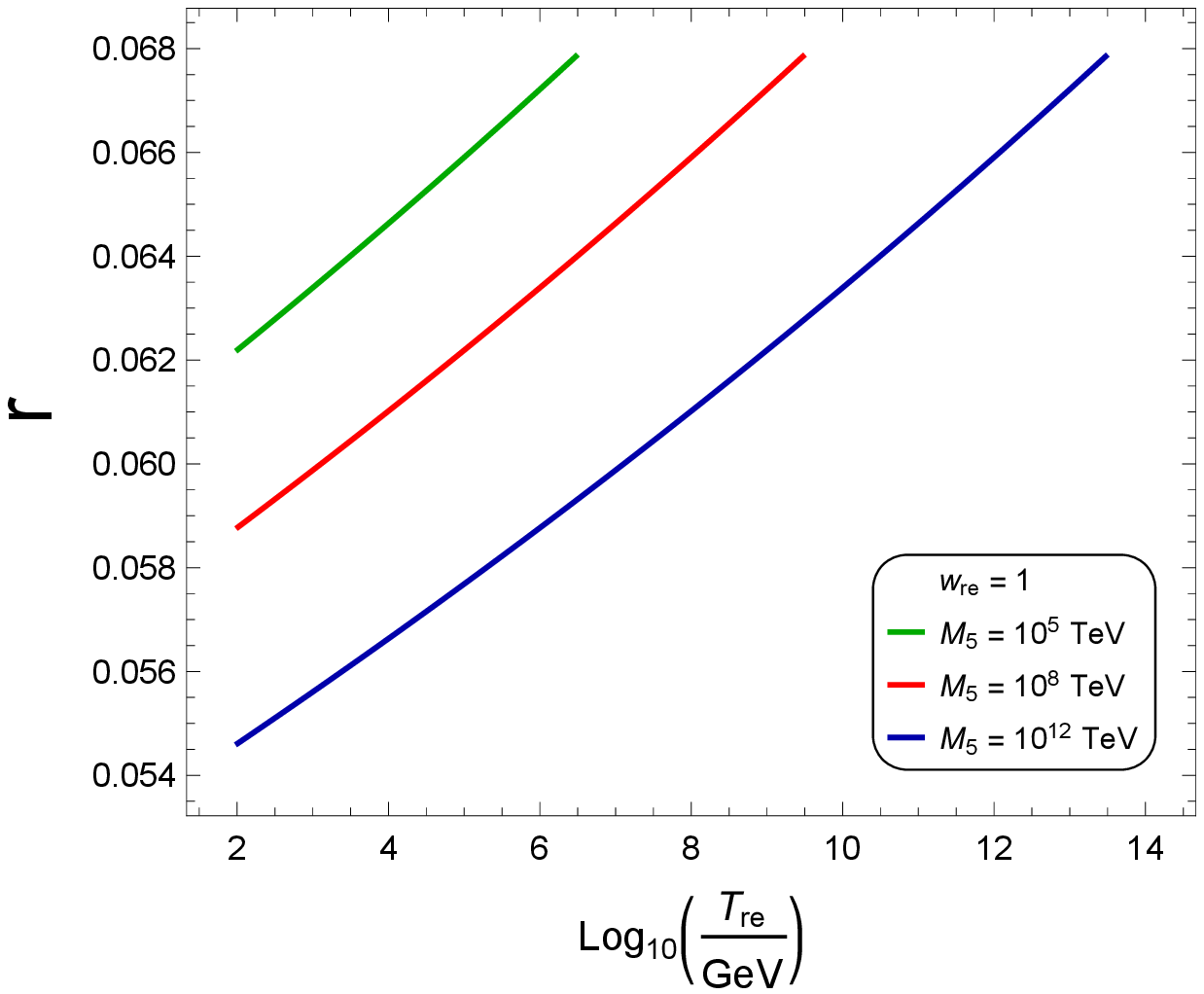}
\caption{Plots for the tensor-to-scalar ratio against the reheating temperature for quadratic Hilltop inflation for $w_{re}=-1/3,\,0,\,2/3,\,1$ and $\alpha=0.0135$. The green, red and blue lines corresponds to a mass of $M_5=10^{5}$\,TeV, $M_5=10^8$\,TeV and $M_5=10^{12}$\,TeV respectively.} \label{fig:rH1}
\end{figure}
Plotting parametrically Eqs. \eqref{rHE} and \eqref{tre}, both evaluated at the Hubble radius crossing, with respect to the number of $e$-folds, it is possible to express the tensor-to-scalar ratio, $r$, in terms of $T_{re}$. Then, one constrains simultaneously $r$ and $T_{re}$ when $\alpha$ is fixed, and for certain values of the EoS parameter and the 5-dimensional Planck mass. From Fig. \ref{fig:rH1}, it is found that for
$w_{re}=-1/3$ and $w_{re}=0$, the available values of $T_{re}$ are $10^6$ GeV, $10^9$ GeV, and $10^{14}$ GeV, when $M_5$ is fixed to $10^5$ TeV, $10^8$ TeV, and $10^{12}$ TeV, respectively. Consequently, the allowed ranges
for $T_{re}$ when $w_{re}$ is set to 2/3 and 1, read
\begin{eqnarray}
10^2\,\textup{GeV}\lesssim T_{re}\lesssim 10^{6}\,\textup{GeV},\\
10^2\,\textup{GeV}\lesssim T_{re}\lesssim 10^{9}\,\textup{GeV},\\
10^2\,\textup{GeV}\lesssim T_{re}\lesssim 10^{14}\,\textup{GeV},
\end{eqnarray}
when $M_5$ is set to $10^5$ TeV, $10^8$ TeV, and $10^{12}$ TeV, respectively.

\section{Higgs-like Inflation on the brane}\label{doubbra}

\subsection{Dynamics of inflation}

The potential for Higgs-like inflation is given by Eq. \eqref{higgs}
\begin{equation}
V(\phi)=\Lambda^4\,\left[1-\left(\frac{\phi}{\mu}\right)^2\right]^2\label{higgspot}
\end{equation}

The slow-roll parameters in the high-energy limit are computed to be
\begin{eqnarray}
\epsilon &=&\frac{4\, \alpha \,x^2}{(1-x^2)^4},\label{epsilonhiggs}\\
\eta &=&\frac{2\, \alpha \,x^2}{(1-x^2)^4} - \frac{\alpha}{(1-x^2)^3},\label{etahiggs}
\end{eqnarray}
where the dimensionless parameter are defined as
\begin{eqnarray}
x &\equiv& \frac{\phi}{\mu},\\
\alpha &\equiv& \frac{M_4^2 \lambda}{\pi \mu^2 \Lambda^4}.\label{alphahiggs}
\end{eqnarray}

Similarly to Hilltop inflation, we solve numerically the expression for the scalar field at the Hubble-radius crossing. 
Using the definition of the number of $e$-folds and the KG equation in the slow-roll approximation, a first order differential equation for $x_k=\phi/f$ is obtained. The former is solved by using using as initial condition $x(N=0)=x_{end}$, where $x_{end}=\phi_{end}/\mu$ is obtained from the condition at the end of inflation, i.e. $\epsilon(x_{end})=1$.

\subsection{Cosmological perturbations}

\begin{figure}[ht!]
\centering
\includegraphics[scale=0.5]{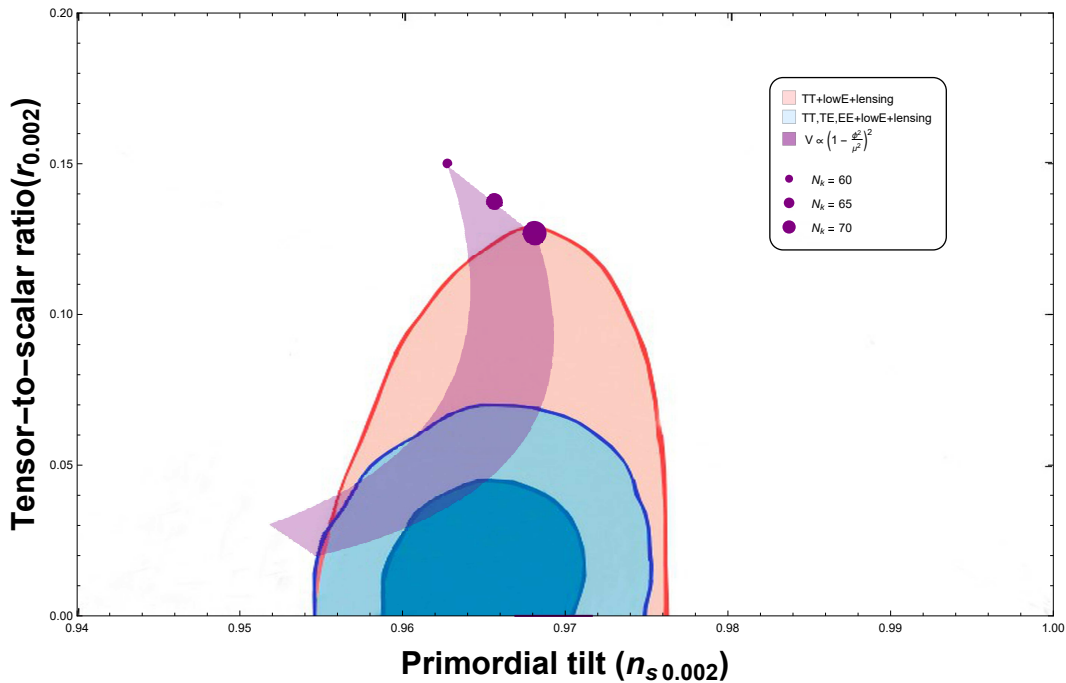}
\caption{Plot of the tensor-to-scalar ratio $r$ versus the scalar spectral index $n_s$ for Higgs-like inflation on the brane along with the two-dimensional marginalized joint confidence contours for ($n_s,r$) at the 68$\%$ (blue region) and 95$\%$ (light blue region) C.L., from the latest PLANCK 2018 results.}\label{contornohiggs}
\end{figure}

The scalar power spectrum is found replacing the potential \eqref{higgspot} into Eq. \eqref{ASHE}, which yields
\begin{equation}
P_S=\frac{\gamma^4\,(1-x^2)^{10}}{2\,\pi^2 \,\alpha^3 \,x^2},\label{perthiggs}
\end{equation}
where $\gamma=\frac{\Lambda}{\mu}$.
Evaluating the slow-roll parameters $\epsilon$ and $\eta$ at the solution for $x_k$, and using Eqs. \eqref{ns} and \eqref{rr}, we may obtain both the scalar spectral index and the tensor-to-scalar ratio, and generate the $n_s - r$ plane. Here, we vary simultaneously the number $e$-folds $N_k$ within the range $N_k=60-70$, and $\alpha$ in a wide range. Fig. \ref{contornohiggs} shows the tensor-to-scalar ratio against the scalar spectral index plot using the two-dimensional marginalized joint confidence contours for ($n_s,r$) at the 68$\%$ (blue region) and 95$\%$ (light blue region) C.L., from the latest PLANCK 2018 results.



Following the same procedure as before to find the allowed values of $\alpha$, one obtains the predictions of the model within the 95\% C.L. region from PLANCK data. For $N_k=60$, $\alpha$ must be within the range $1.15\times 10^{-2}\lesssim \alpha \lesssim 1.55\times 10^{-2}$. Consequently, the tensor-to-scalar ratio lies in the interval $0.061 \gtrsim r \gtrsim 0.044$. For $N_k=65$, the predictions are found to be $8.00\times 10^{-3}\lesssim \alpha \lesssim 1.88\times 10^{-2}$ and $0.070 \gtrsim r \gtrsim 0.027$. Finally, for $N_k=70$, $\alpha$ is found within
the range $6.80\times 10^{-3}\lesssim \alpha \lesssim 2.00\times 10^{-2}$, while the corresponding values of the tensor-to-scalar ratio are given by $0.069 \gtrsim r \gtrsim 0.020$.\\

Combining the constraints on $\alpha$ with Eq. \eqref{perthiggs} and the amplitude of the scalar spectrum $P_S\simeq 2.2\times 10^{-9}$ we obtain the corresponding allowed ranges for $\gamma$
\begin{eqnarray}
2.57\times 10^{-4} \lesssim \gamma \lesssim 2.67\times 10^{-4},\\
2.30\times 10^{-4} \lesssim \gamma \lesssim 2.56\times 10^{-4},\\
2.14\times 10^{-4} \lesssim \gamma \lesssim 2.42\times 10^{-4},
\end{eqnarray}
for $N_k=60$, $N_k=65$ and $N_k=70$, respectively.
The allowed ranges for $\alpha$ and $\gamma$ are summarized in Table \ref{constalphahiggs}.


\begin{table}[H]
\begin{center}
\begin{tabular}{| c | c | c |}\hline
 $N_k$   & Constraint on $\alpha$  & Constraint on $\gamma$ \\\hline
   \,60\,  & \,$0.0115 \lesssim \alpha \lesssim 0.0155$\,  & \,$2.57\times 10^{-4} \lesssim \gamma \lesssim 2.67\times 10^{-4}$\,  \\
   \,65\,  & \,$0.0080 \lesssim \alpha \lesssim 0.0188$\,  & \,$2.30\times 10^{-4} \lesssim \gamma \lesssim 2.56\times 10^{-4}$\,  \\
  \,70\,  & \,$0.0068 \lesssim \alpha \lesssim 0.0200$\,  & \,$2.14\times 10^{-4} \lesssim \gamma \lesssim 2.42\times 10^{-4}$\,  \\ \hline
\end{tabular}
\caption{Results of the constraints on the parameters $\alpha$ and $\gamma$ for Higgs-like inflation in the high-energy limit of the Randall-Sundrum brane model, using the last data of PLANCK.} \label{constalphahiggs}
\end{center}
\end{table}


After replacing Eq. \eqref{M5M4} into the definition of $\alpha$ (Eq. \eqref{alphahiggs}), and using the fact that $\Lambda=\gamma\,\mu$, the following expressions for $\mu$ and $\Lambda$ in terms of $M_5$ are derived
\begin{equation} 
\mu = \left ( \frac{3}{4 \pi^2 \alpha \gamma^4} \right )^{1/6} M_5, \label{fd}
\end{equation}
\begin{equation}
\Lambda = \gamma \mu=\gamma \left ( \frac{3}{4 \pi^2 \alpha \gamma^4} \right )^{1/6} M_5. \label{Ld}
\end{equation}

Evaluating those expressions at several values for $\alpha$ and $\gamma$ (Table \ref{constalphahiggs}) and considering the lower and upper limit for the five-dimensional Planck mass, the brane tension is found to be the same as in the previous models, i.e.  $\lambda=1.60 \times 10^{-3}$\,TeV$^4$ for $M_5=10^5$\,TeV and $\lambda=1.60 \times 10^{39}$\,TeV$^4$ for $M_5=10^{12}$\,TeV. The top panels of Table \ref{Tmulhiggs} show the corresponding values of the mass scales for the lower limit of $M_5$, while the bottom panels show the values of the mass scales for the upper limit.



\begin{table}[H]
\centering
\begin{tabular}{| c | c | c |}\hline
 $N_k$   & Constraint on $\mu$\,[TeV]  & Constraint on $\Lambda$\,[TeV] \\\hline
  \,60\,  & \,$3.39\times 10^7 \gtrsim \mu \gtrsim 3.14\times 10^7$\,  & \,$8.71\times 10^{3} \gtrsim \Lambda \gtrsim 8.39\times 10^{3}$\,  \\
  \,65\,  & \,$3.88\times 10^7 \gtrsim \mu \gtrsim 3.13\times 10^7$\,  & \,$8.92\times 10^{3} \gtrsim \Lambda \gtrsim 8.01\times 10^{3}$\,  \\
  \,70\,  & \,$4.18\times 10^7 \gtrsim \mu \gtrsim 3.22\times 10^7$\,  & \,$8.94\times 10^{3} \gtrsim \Lambda \gtrsim 7.78\times 10^{3}$\,  \\ \hline
\end{tabular}\vspace{1cm}
\begin{tabular}{| c | c | c |}\hline
 $N_k$   & Constraint on $\mu$\,[TeV]  & Constraint on $\Lambda$\,[TeV] \\\hline
  \,60\,  & \,$3.39\times 10^{14} \gtrsim \mu \gtrsim 3.14\times 10^{14}$\,  & \,$8.71\times 10^{10} \gtrsim \Lambda \gtrsim 8.39\times 10^{10}$\,  \\
  \,65\,  & \,$3.88\times 10^{14} \gtrsim \mu \gtrsim 3.13\times 10^{14}$\,  & \,$8.92\times 10^{10} \gtrsim \Lambda \gtrsim 8.01\times 10^{10}$\,  \\
  \,70\,  & \,$4.18\times 10^{14} \gtrsim \mu \gtrsim 3.22\times 10^{14}$\,  & \,$8.94\times 10^{10} \gtrsim \Lambda \gtrsim 7.78\times 10^{10}$\,  \\ \hline
\end{tabular}
\caption{Results for the constraints on the mass scales $\mu$ and $\Lambda$ for Higgs-like inflation in the high-energy limit of the Randall-Sundrum brane model using the last data of PLANCK. The top table shows the results using $M_5=10^{5}$\,TeV while the bottom table shows the results using $M_5=10^{12}$\,TeV.} \label{Tmulhiggs}
\end{table}

Following the analysis performed for Natural Inflation, it can be shown that the plots for the two conjectures of the Swampland Criteria, which are not shown, exhibit a similar behavior with those shown in Fig. \ref{fig:NIswampland}. For the distance conjecture, $\overline{\Delta \phi}$ increases with both the number of $e$-folds and the 5-dimensional Planck mass, so this conjecture is fulfilled. On the other hand, for the de Sitter conjecture, $\overline{\Delta V}$ decreases with the number of $e$-folds and $M_5$. For the same arguments given before, the de Sitter Swampland criteria and its refined version are avoided.

\subsection{Reheating}

\begin{figure}[ht!]
\centering
\includegraphics[width=0.45\textwidth]{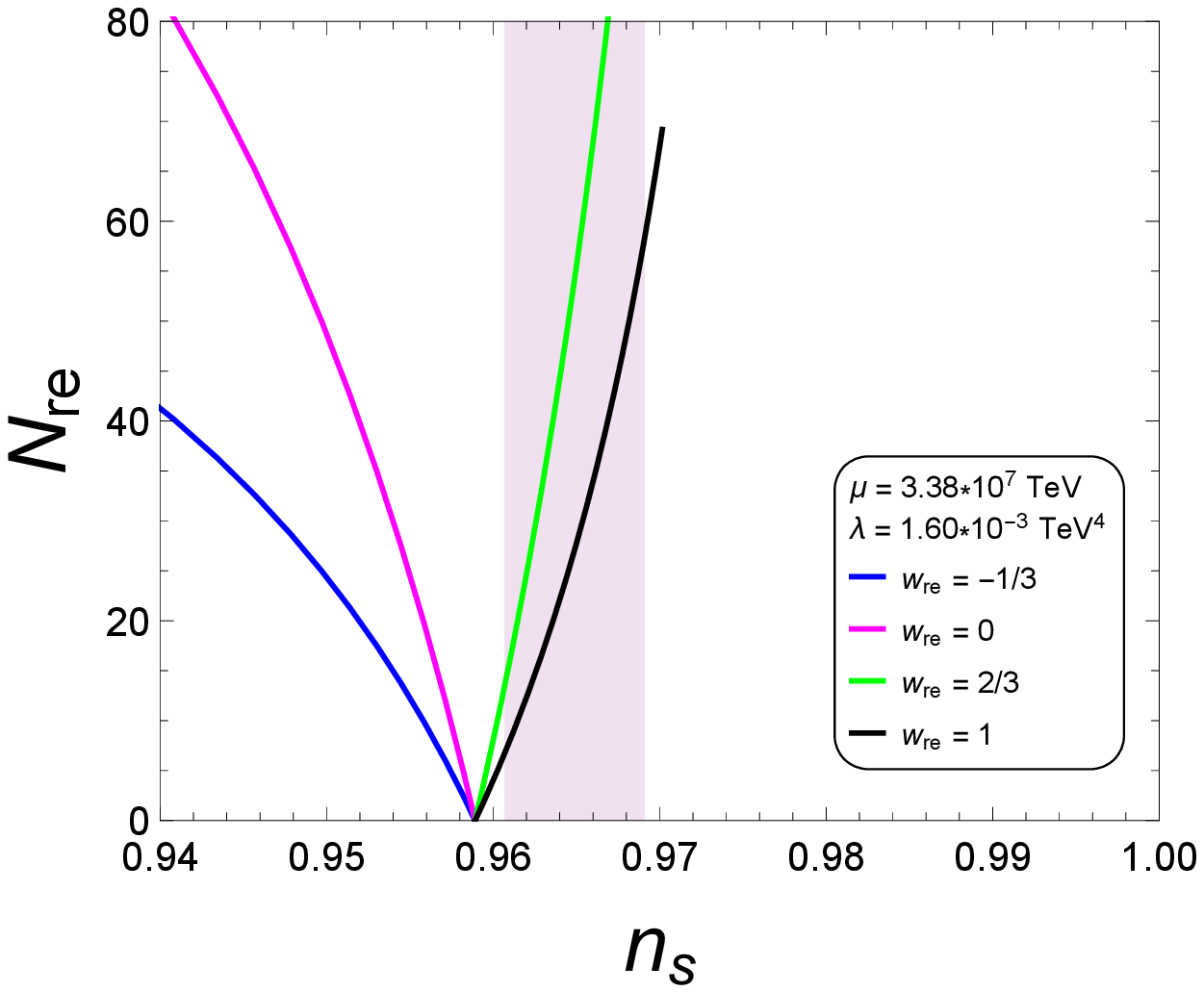}
\includegraphics[width=0.45\textwidth]{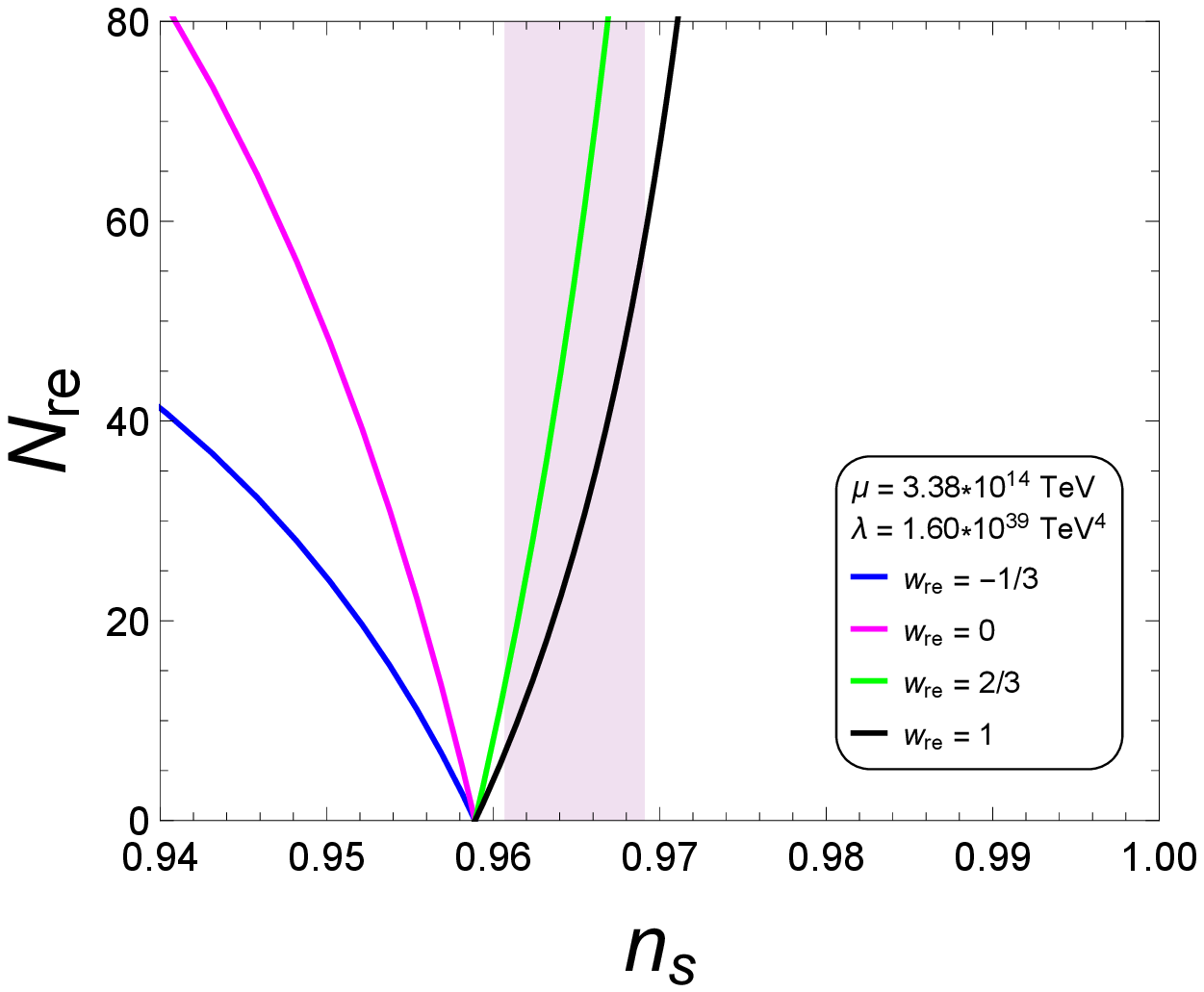}
\includegraphics[width=0.45\textwidth]{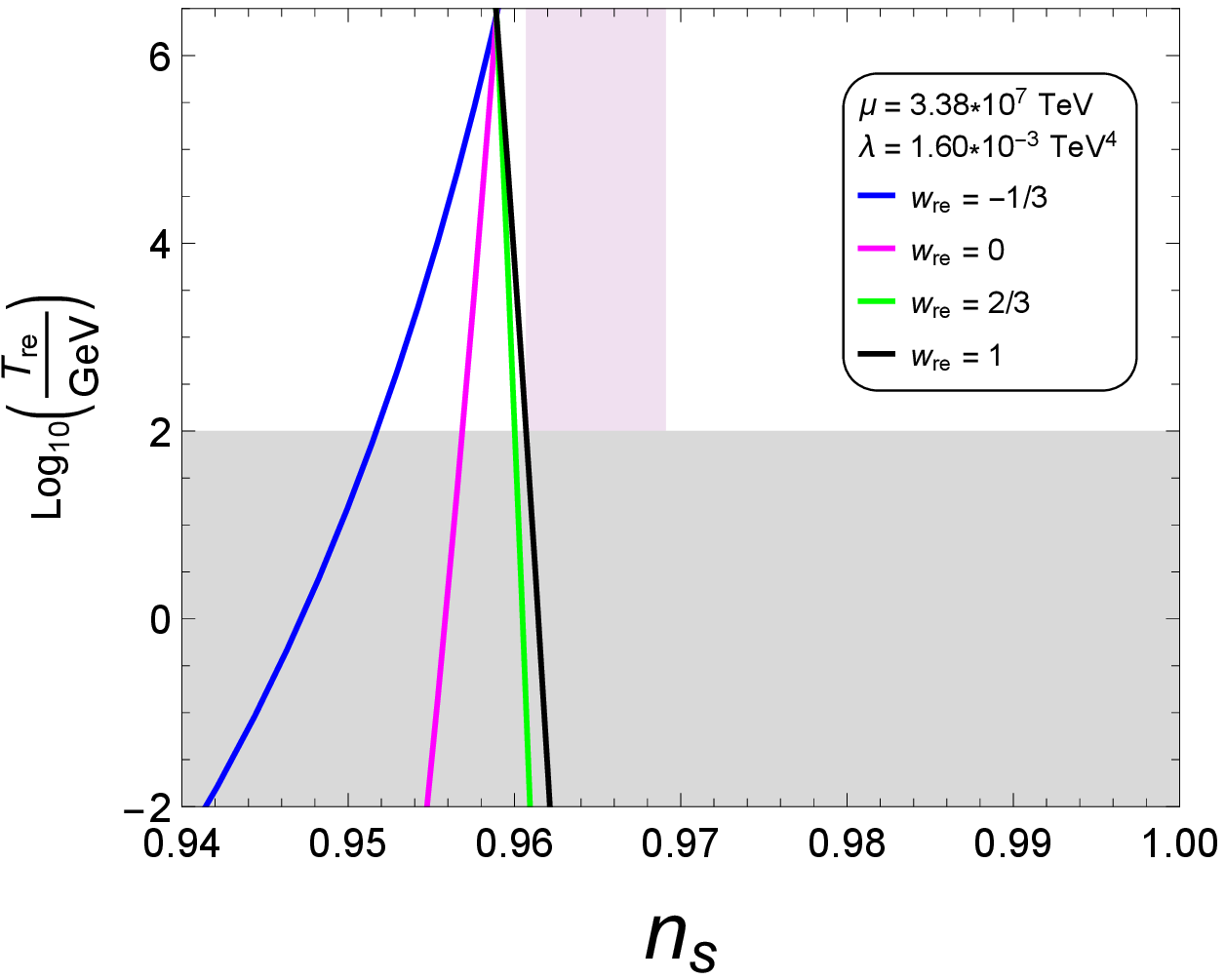}
\includegraphics[width=0.45\textwidth]{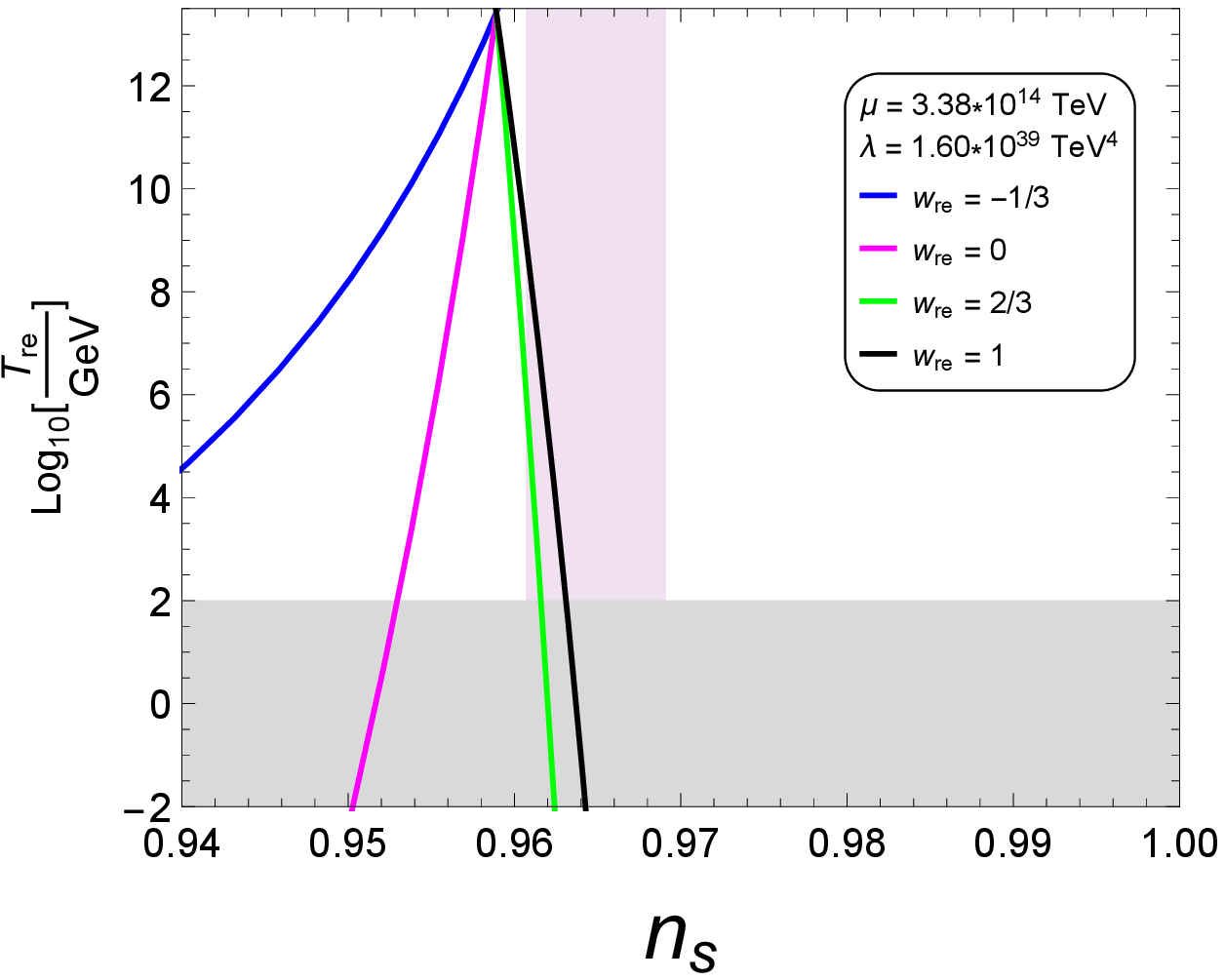}
\caption{Plots of $N_{re}$ and $T_{re}$ as functions of $n_s$ for Higgs-like inflation. The left panels shows the plots for $M_5=10^{5}$\,TeV while the right panels shows the plots for $M_5=10^{12}$\,TeV. The curves and the shading regions are the same as FIG. \ref{fig:NI5} and all plots corresponds to $\alpha=0.0115$.} \label{fig:rehiggs}
\end{figure}

Following the same method as previous sections, we can give predictions for reheating by means plotting parametrically Eqs. \eqref{nre} and \eqref{tre} with respect to $\alpha$ and $N_k$ over the range of the effective EoS $-\frac{1}{3}\leq w_{re}\leq1$. In Fig. \ref{fig:rehiggs} we show the plots for reheating using $M_5=10^5$\,TeV (left panels) and $M_5=10^{12}$\,TeV (right panels) for $\alpha=0.0115$ (corresponding to the constraints for $N_k=55$). Our analysis indicates that the behavior of the convergence point is the same as in Natural inflation and quadratic Hilltop inflation. As we can see, for $M_5=10^5$\,TeV the maximum reheating temperature is about $T_{re}\approx10^7$\,GeV and for $M_5=10^{12}$\,TeV is about $T_{re}\approx10^{14}$\,GeV.


Analyzing the curves for the reheating temperature, we found an allowed range for $N_k$ for each value of $w_{re}$ when 
$\alpha$ is fixed. The corresponding intervals are shown in Table \ref{whiggs}. For consistency, we only display the results for the plots of Fig. \ref{fig:rehiggs} for $M_5=10^{12}$\,TeV because the allowed range of $N_k$ for the lower limit of $M_5$ is too small. It should be noted that for $\alpha=0.0080$ and $\alpha=0.0068$ (plots not shown), all four curves enter to the purple region, while for $\alpha=0.0155$, $\alpha=0.0188$ and $\alpha=0.0200$ (plots not shown), none of the curves enter.


\begin{table}[H]
\begin{center}
\begin{tabular}{ |c|c|c| } \hline
$w_{re}$       &  $N_k$\\\hline
   \,2/3\,   & \,60 - 61\,  \\ 
   \,1\,     & \,60 - 65\,  \\ \hline
\end{tabular}
\caption{Summary of the allowed range for the number of $e$-folds for each EoS parameter $w_{re}$  when the dimensionless parameter $\alpha$ is fixed to be $\alpha=0.0115$ and $M_5=10^{12}$\,TeV.}\label{whiggs}
\end{center}
\end{table}


\begin{figure}[ht!]
\centering
\includegraphics[width=0.45\textwidth]{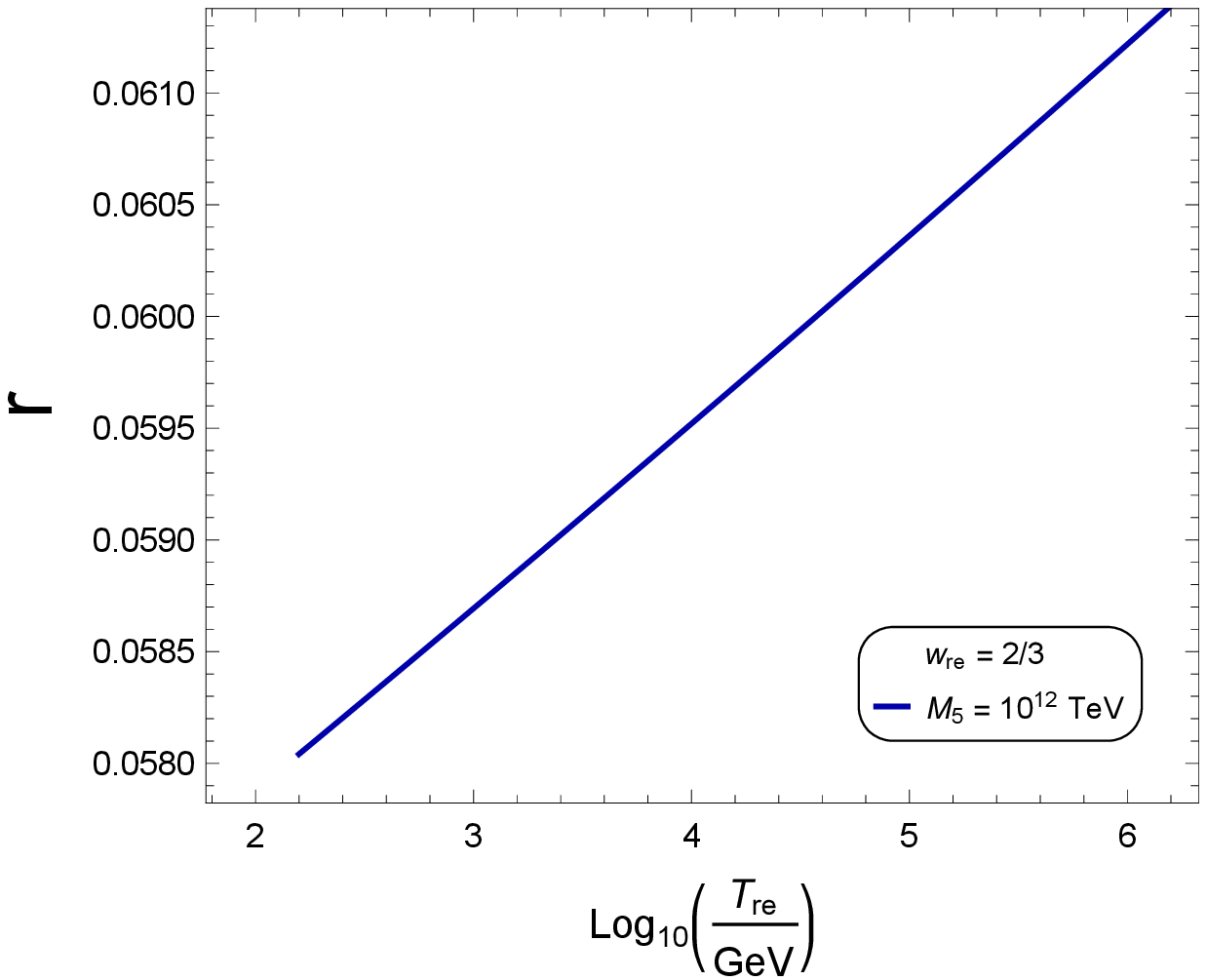}
\includegraphics[width=0.45\textwidth]{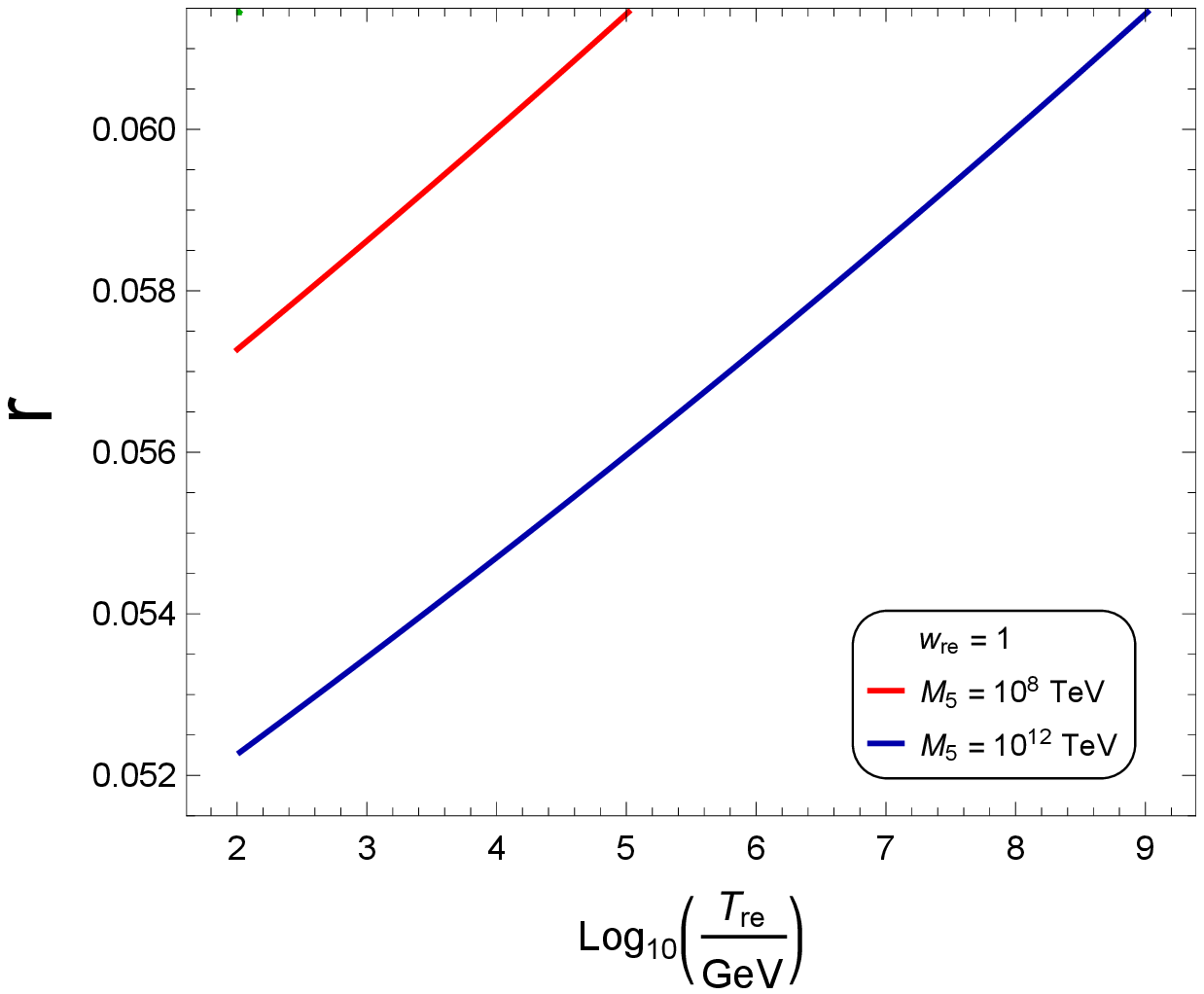}
\caption{Plots for the tensor-to-scalar ratio against the reheating temperature for Higgs-like inflation for $w_{re}=2/3,1$ and $\alpha=0.0115$. The red line corresponds to a mass of $M_5=10^{8}$\,TeV and the blue line corresponds to a mass of $M_5=10^{12}$\,TeV.} \label{fig:rhiggs}
\end{figure}


Evaluating Eqs. \eqref{rHE} and \eqref{tre} at the Hubble radius crossing, and plotting parametrically with respect to the number of $e$-folds, we can find the allowed values for the tensor-to-scalar ratio in terms of the reheating temperature.
The only values of $\alpha$ and $w_{re}$ consistent with the current bounds on the tensor-to-scalar ratio, correspond to $\alpha=0.0115$ and $w_{re}=2/3,\,1$ for values of $M_5$ greater than its lower limit, as it is shown in Fig. \ref{fig:rhiggs}. In this case, it is found that for $M_5=10^{12}$\,TeV and $w_{re}=2/3$ the reheating temperature must be in the range of
\begin{equation}
10^{2}\,\textup{GeV} \lesssim T_{re} \lesssim 10^{7} \,\textup{GeV},
\end{equation}
while for $w_{re}=1$, one finds that the allowed values for $T_{re}$ are found
within the ranges
\begin{eqnarray}
10^2 \,\textup{GeV} \lesssim T_{re} \lesssim 10^{5} \,\textup{GeV}, \\
10^2 \,\textup{GeV} \lesssim T_{re} \lesssim 10^{9} \,\textup{GeV},
\end{eqnarray}
when $M_5$ is fixed to $10^5$ TeV and $10^{12}$ TeV, respectively.





\section{Exponential SUSY Inflation on the brane}\label{SUSYbra}

\subsection{Dynamics of inflation}

The last potential we study in the present work is a well motivated one from SUGRA, namely Exponential SUSY inflation, 
given by Eq. \eqref{susy}
\begin{equation}
V(\phi)=\Lambda^4 \,(1-e^{\phi/f})\label{susypot}
\end{equation}

Replacing this potential into Eqs. \eqref{epHE} and \eqref{etHE} we obtain the set of slow-roll parameters in the high-energy regime as
\begin{eqnarray}
\epsilon &=&\frac{\alpha \,e^{-2x}}{(1-e^{-x})^3},\label{epsilons}\\
\eta &=&-\frac{\alpha\,e^{-x}}{(1-e^{-x})^2},\label{etas}
\end{eqnarray}
where the dimensionless parameter are defined by
\begin{eqnarray}
x &\equiv& \frac{\phi}{f},\\
\alpha &\equiv& \frac{M_4^2 \lambda}{4\,\pi^2\, f^2 \, \Lambda^4}.\label{alphas}
\end{eqnarray}

Similarly to the quadratic Hilltop and Higgs-like inflation models, we solve the expression for $x_k(N_k)$ numerically, and using as initial condition $x(N_k=0)=x_{end}$, where $x_{end}=\phi_{end}/f$ is obtained from the condition at the end of inflation, i.e. $\epsilon_{end}=1$.

\subsection{Cosmological perturbations}

\begin{figure}[ht!]
\centering
\includegraphics[scale=0.5]{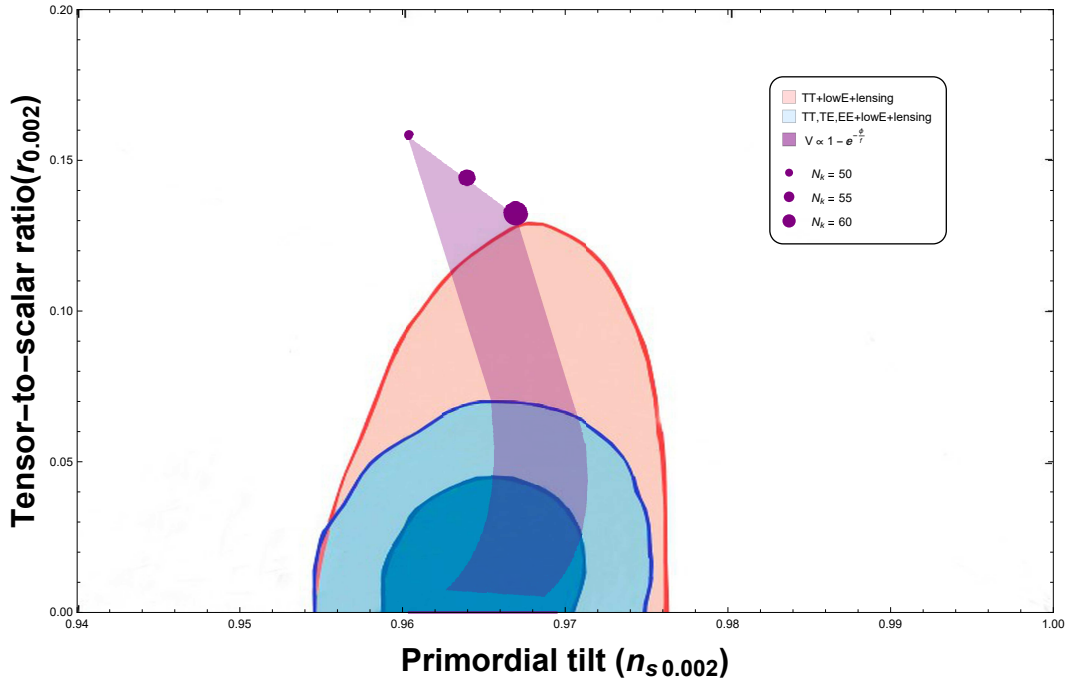}
\caption{Plot of the tensor-to-scalar ratio $r$ versus the scalar spectral index $n_s$ for Exponential SUSY inflation on the brane along with the two-dimensional marginalized joint confidence contours for ($n_s,r$) at the 68$\%$ (blue region) and 95$\%$ (light blue region) C.L., from the latest PLANCK 2018 results.}\label{contornoSUSY}
\end{figure}

Replacing the potential \eqref{susypot} into Eq. \eqref{ASHE}, we obtain the following expression for the scalar power spectrum 
\begin{equation}
P_S=\frac{\gamma^4\,e^{2x}\,(1-e^{-x})^6}{12\,\pi^2 \,\alpha^3},\label{perts}
\end{equation}
where $\gamma=\frac{\Lambda}{f}$. Evaluating $\epsilon$ and $\eta$ at the solution for $x_k$ and using Eqs. \eqref{ns} and \eqref{rr} to obtain $n_s$ and $r$, we plot the predictions on the $n_s$-$r$. In doing so, we vary simultaneously the dimensionless parameter $\alpha$ in a wide range and the number $e$-folds $N_k$ within the range $N_k=50-60$. Fig. \ref{contornoSUSY} shows the tensor-to-scalar ratio against the scalar spectral index plot using the two-dimensional marginalized joint confidence contours for ($n_s,r$) at the 68$\%$ (blue region) and 95$\%$ (light blue region) C.L., from the latest PLANCK 2018 results.

As we have seen already, the allowed values for $\alpha$ are found when a given curve, for a fixed $N_k$, enters and leaves the 2$\sigma$ region. We note that in this model, unlike previous potentials already studied, the trajectories never leave the 2$\sigma$ region, achieving a very small tensor-to-scalar ratio, which is well inside the $(68\%\,\textup{C.L.})$ contour for large values of $\alpha$. The latter implies that we only have a lower bound on $\alpha$ for each value of $N_k$. So, following the same method as before, one obtains that the predictions of the model are within the 95\% C.L. region from PLANCK data, for $N_k=50$, if $\alpha$ is such that $\alpha \gtrsim 1.85\times 10^{-2}$. Therefore, 
an upper bound for the scalar-to-tensor ratio is achieved, yielding $r \lesssim 0.070$. For $N_k=55$, the lower bound on 
$\alpha$ is $\alpha \gtrsim 1.29\times 10^{-2}$, while $r$ is found to be $r\lesssim 0.069$. Finally, for $N_k=60$, the corresponding constraint on $\alpha$ is found to be $\alpha \gtrsim 1.14\times 10^{-2}$, while the tensor-to-scalar ratio 
is such that $r\lesssim 0.064$. Taking the limit $\alpha \rightarrow \infty$, one finds the asymptotic limit of the tensor-to-scalar ratio, that yields $r\rightarrow 0$, whereas the asymptotic limit for the spectral index is found to be $n_s\rightarrow 0.960$ for $N_k=50$, $n_s \rightarrow 0.964$ for $N_k=55$ and $n_s \rightarrow 0.967$ for $N_k=60$.

Combining the previous constraints on $\alpha$ with Eq. \eqref{perts} and the amplitude of the scalar spectrum $P_S\simeq 2.2\times 10^{-9}$, we obtain the corresponding allowed ranges for the dimensionless parameter $\gamma$
\begin{eqnarray}
\gamma \gtrsim 8.70\times 10^{-4}, \\
\gamma \gtrsim 7.46\times 10^{-4}, \\
\gamma \gtrsim 6.91\times 10^{-4},
\end{eqnarray}
for $N_k=50$, $N_k=55$ and $N_k=60$, respectively. The allowed ranges for $\alpha$ and $\gamma$ are summarized in Table \ref{constalphas}.


\begin{table}[H]
\begin{center}
\begin{tabular}{| c | c | c |}\hline
 $N_k$   & \,Constraint on $\alpha$\,  & \,Constraint on $\gamma$\, \\\hline
  \,50\,  & \,$\alpha \gtrsim 0.0185$\,  & \,$\gamma \gtrsim 8.70\times 10^{-4}$\,  \\
  \,55\,  & \,$\alpha \gtrsim 0.0129$\,  & \,$\gamma \gtrsim 7.46\times 10^{-4}$\,  \\
  \,60\,  & \,$\alpha \gtrsim 0.0114$\,  & \,$\gamma \gtrsim 6.91\times 10^{-4}$\,  \\ \hline
\end{tabular}
\caption{Results of the constraints on the parameters $\alpha$ and $\gamma$ for Exponential SUSY inflation in the high-energy limit of Randall-Sundrum brane model, using the last data of PLANCK.} \label{constalphas}
\end{center}
\end{table}


Replacing Eq. \eqref{M5M4} into the definition of $\alpha$ (Eq. \eqref{alphas}) and using the fact that $\Lambda=\gamma\,f$, we found the expressions for the mass scales $f$ and $\Lambda$ as
\begin{equation}
f = \left ( \frac{3}{16 \pi^2 \alpha \gamma^4} \right )^{1/6} M_5 \label{fs}
\end{equation}
\begin{equation}
\Lambda = \gamma f=\gamma \left ( \frac{3}{16 \pi^2 \alpha \gamma^4} \right )^{1/6} M_5 \label{Ls}
\end{equation}

After evaluating these expressions at several values for $\alpha$ and $\gamma$ (Table \ref{constalphas}), we found that $\lambda=1.60 \times 10^{-3}$\,TeV$^4$ for $M_5=10^5$\,TeV and $\lambda=1.60 \times 10^{39}$\,TeV$^4$ for $M_5=10^{12}$\,TeV. The top panels of Table \ref{Tfls} show the corresponding values of the mass scales for the lower limit of $M_5$, while the bottom panels shows the values of the mass scales for the upper limit.


\begin{table}[H]
\centering
\begin{tabular}{| c | c | c |}\hline
 $N_k$   & \,Constraint on $f$\,[TeV]\,  & \,Constraint on $\Lambda$\,[TeV]\, \\\hline
  \,50\,  & \,$f \lesssim 1.10\times 10^7$\,  & \,$\Lambda \lesssim 9.59\times 10^{3}$\,  \\
  \,55\,  & \,$f \lesssim 1.30\times 10^7$\,  & \,$\Lambda \lesssim 9.67\times 10^{3}$\,  \\
  \,60\,  & \,$f \lesssim 1.39\times 10^7$\,  & \,$\Lambda \lesssim 9.63\times 10^{3}$\,  \\ \hline
\end{tabular}\\
\vspace{1cm}\begin{tabular}{| c | c | c |}\hline
 $N_k$   & \,Constraint on $f$\,[TeV]\,  & \,Constraint on $\Lambda$\,[TeV]\, \\\hline
  \,50\,  & \,$f \lesssim 1.10\times 10^{14}$\,  & \,$\Lambda \lesssim 9.59\times 10^{10}$\,  \\
  \,55\,  & \,$f \lesssim 1.30\times 10^{14}$\,  & \,$\Lambda \lesssim 9.67\times 10^{10}$\,  \\
  \,60\,  & \,$f \lesssim 1.39\times 10^{14}$\,  & \,$\Lambda \lesssim 9.63\times 10^{10}$\,  \\ \hline
\end{tabular}
\caption{Results for the constraints on the mass scales $f$ and $\Lambda$ for Exponential SUSY inflation in the high-energy limit of Randall-Sundrum brane model using the last data of PLANCK. The top table shows the results using $M_5=10^{5}$\,TeV while the bottom table shows the results using $M_5=10^{12}$\,TeV.} \label{Tfls}
\end{table}


Like previous models, we find numerically that the distance Swampland conjecture, $\overline{\Delta \phi}$ increases as both the number of $e$-folds and the 5-dimensional Planck mass increase, so this conjecture is fulfilled, while for the de Sitter conjecture, $\overline{\Delta V}$ decreases as the number of $e$-folds increases, and also as $M_5$ grows.

\subsection{Reheating}

\begin{figure}[ht!]
\centering
\includegraphics[width=0.45\textwidth]{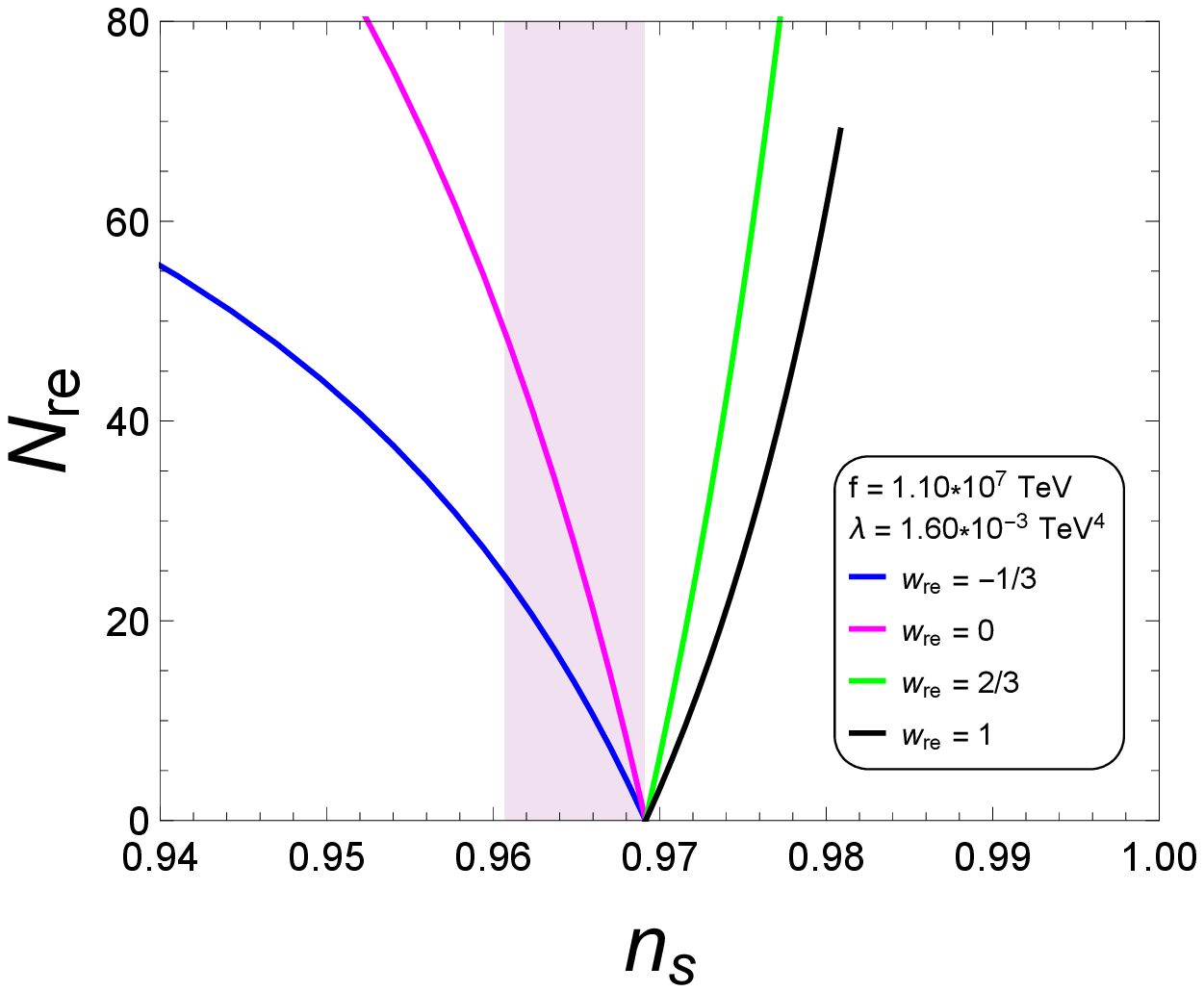}
\includegraphics[width=0.45\textwidth]{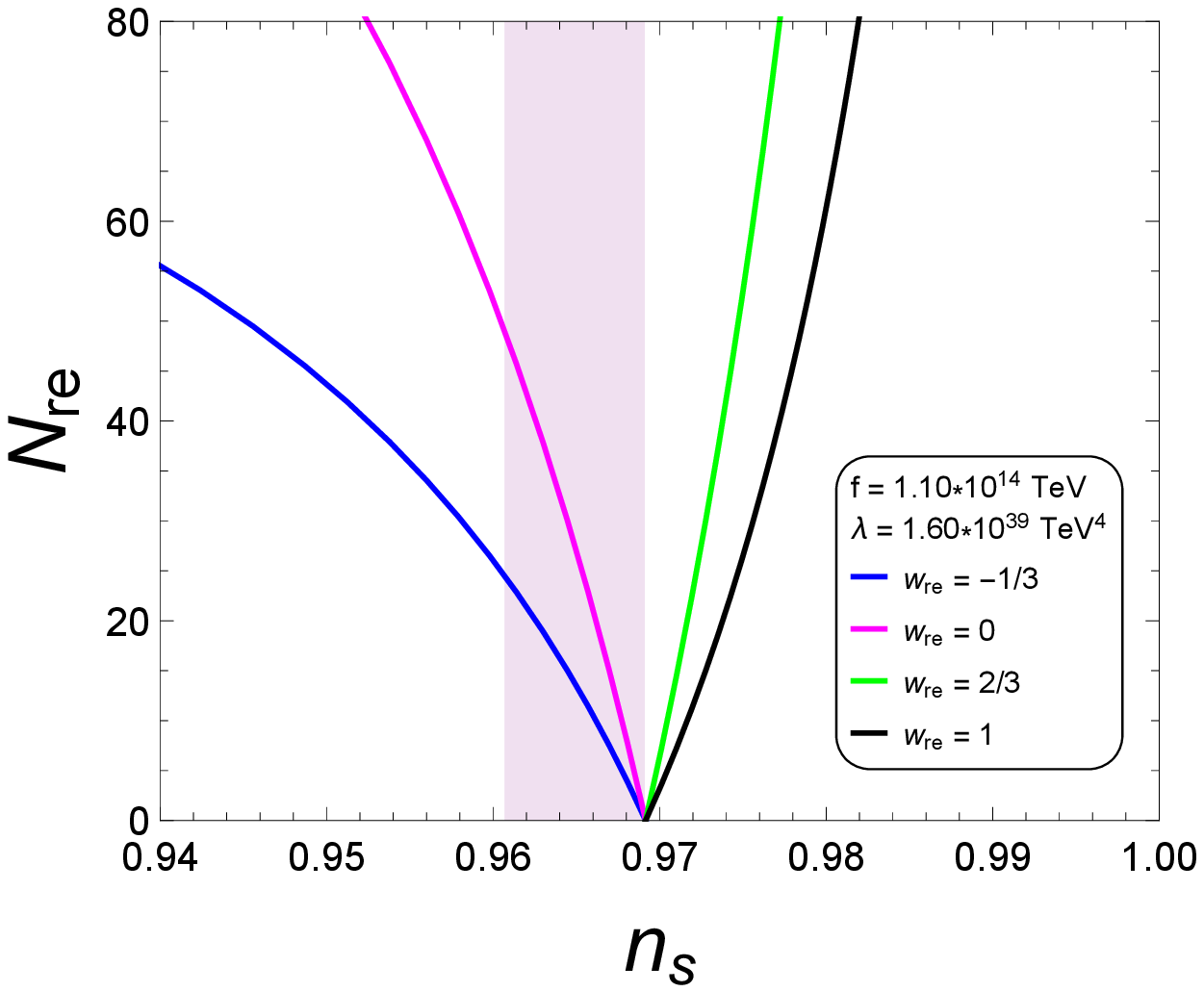}
\includegraphics[width=0.45\textwidth]{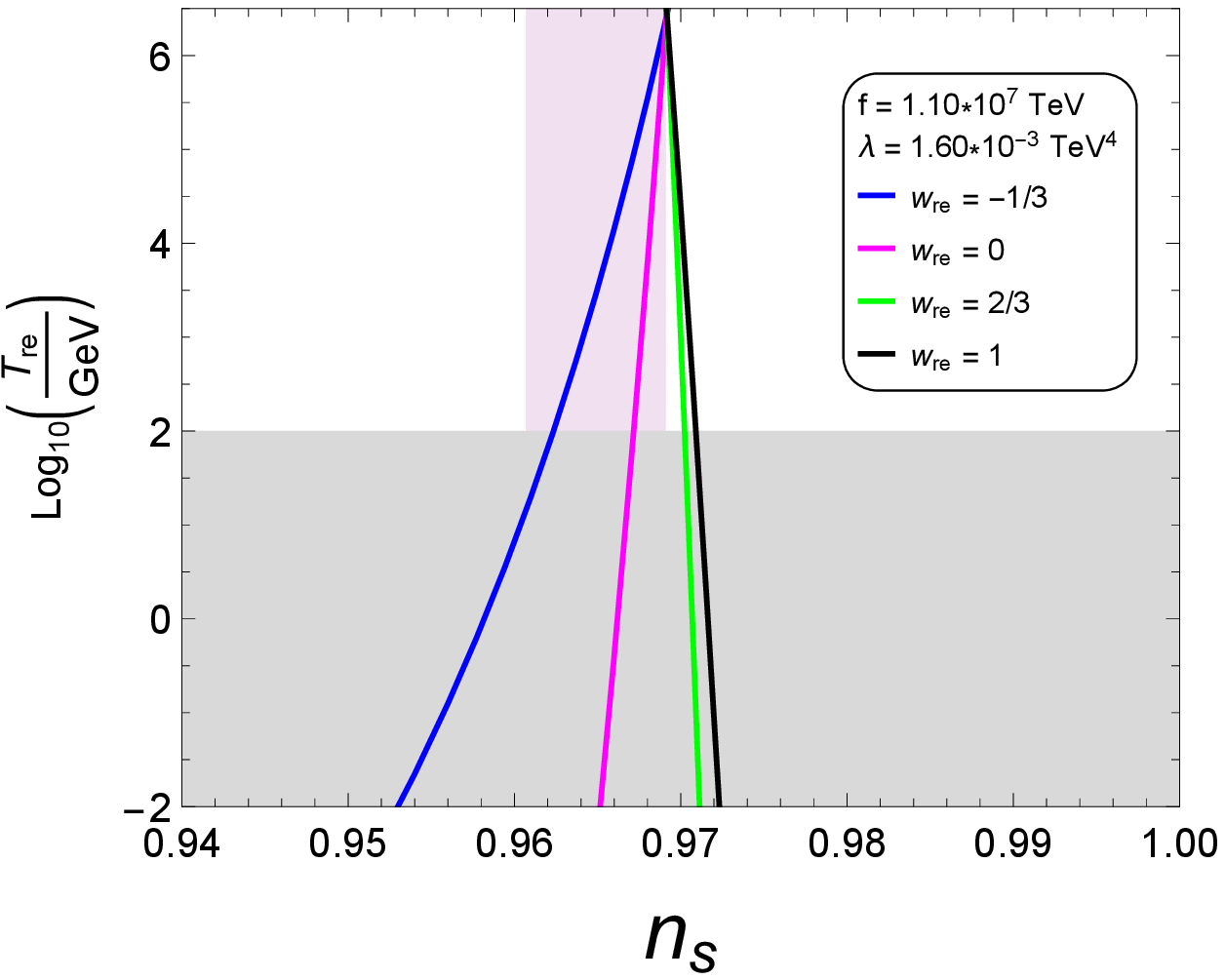}
\includegraphics[width=0.45\textwidth]{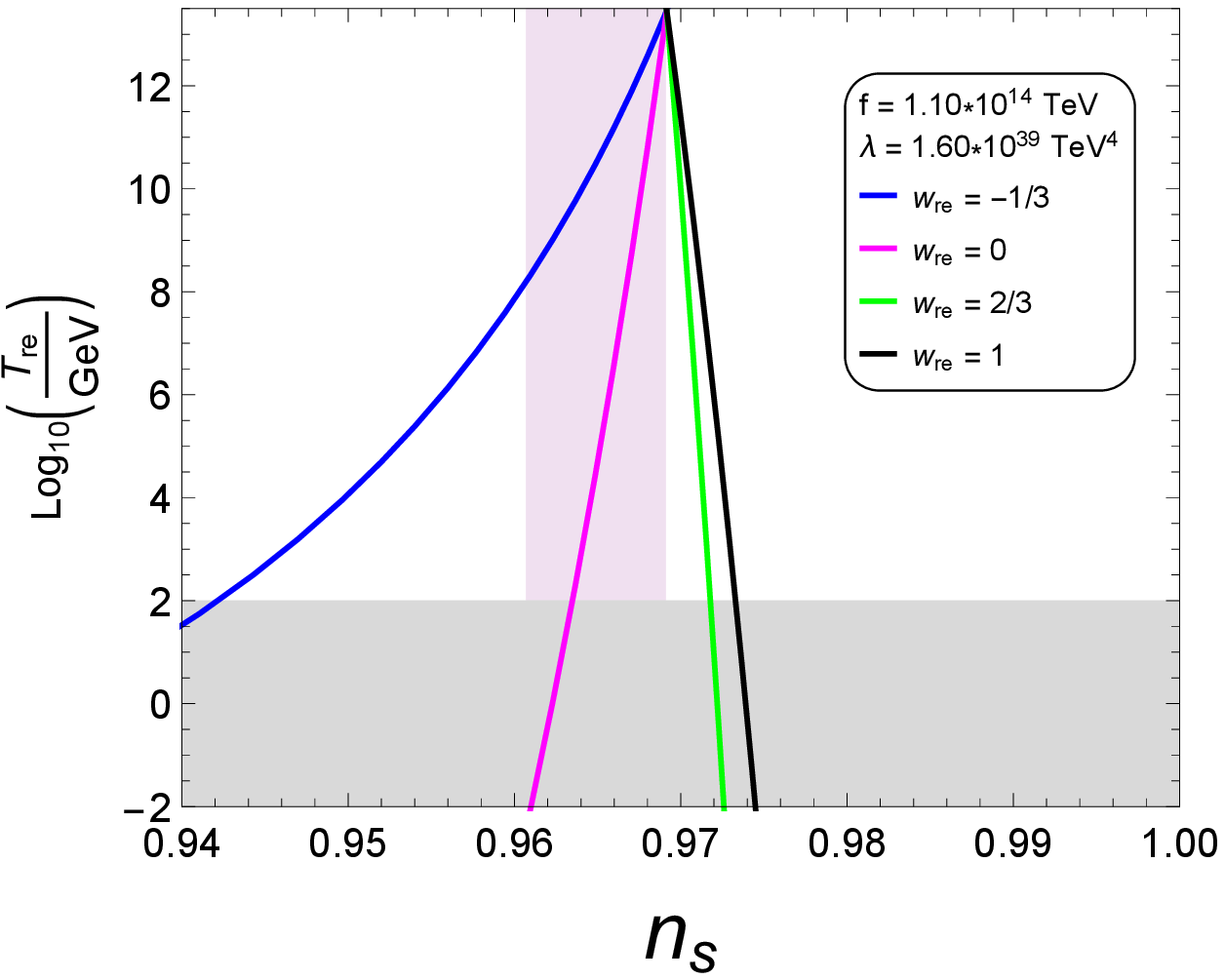}
\caption{Plots of $N_{re}$ and $T_{re}$ as functions of $n_s$ for Exponential SUSY inflation. The left panels  shows the plot for $M_5=10^{5}$\,TeV while the right panels  shows the plot for $M_5=10^{12}$\,TeV.The curves and the shading regions are the same as FIG. \ref{fig:NI5} and all plots corresponds to $\alpha=0.0185$.} \label{fig:S}
\end{figure}

If one follows the same procedure as in the previous sections, we can give predictions for reheating plotting parametrically Eqs. \eqref{nre} and \eqref{tre} with respect to $\alpha$ and $N_k$ over the range of the effective EoS $-\frac{1}{3}\leq w_{re}\leq1$. Unlike previous models, this kind of potential is derived from SUGRA, hence the corresponding degrees of freedom of relativistic particles at the end of reheating appearing in 
the expressions for $N_{re}$ and $T_{re}$ are $g_{re}={\mathcal{O}}(200)$. In Fig. \ref{fig:S} we show the plots of reheating using $M_5=10^5$\,TeV (left panels) and $M_5=10^{12}$\,TeV (right panels) for $\alpha=0.0185$ that corresponds to the constraints of $N_k=55$. As we can see, the maximum reheating temperature increases with the five-dimensional Planck mass, giving $T_{re}\approx10^7$\,GeV for $M_5=10^5$\,TeV and $T_{re}\approx10^{14}$\,GeV for $M_5=10^{12}$\,TeV.


Analyzing the curves of the plots for reheating, we found the allowed values for  number of $e$-folds $N_k$ when fixing $\alpha$ for a certain value of the EoS parameter $w_{re}$. For consistency, we display the results for the plots of Fig. \ref{fig:S} in Table \ref{ws}. It should be noted that for $\alpha=0.0129$ and $\alpha=0.0114$ (plots not shown) all the four curves enter to the purple region. 


\begin{table}[H]
\begin{center}
\begin{tabular}{ cc }   
\begin{tabular}{ |c|c|c| } \hline
 $w_{re}$    &  $N_k$\\\hline
   \,-1/3\,  & \,46 - 56\,  \\
   \,0\,     & \,53 - 56\,  \\ \hline
\end{tabular} &  
\hspace{1cm}
\begin{tabular}{ |c|c|c| } \hline
$w_{re}$     &  $N_k$\\\hline
   \,-1/3\,  & \,44 - 56\,  \\
   \,0\,     & \,47 - 56\,  \\ \hline 
\end{tabular} \\
\end{tabular}
\caption{Summary of the allowed range for the number of $e$-folds for each EoS parameter $w_{re}$  when the dimensionless parameter $\alpha$ is fixed to $\alpha=0.0185$. The left and right tables corresponds to a 5-dimensional Planck mass of $M_5=10^5$\,TeV and $M_5=10^{12}$\,TeV respectively.}\label{ws}
\end{center}
\end{table}


Plotting parametrically Eqs. \eqref{rHE} and \eqref{tre} with respect to the number of $e$-folds, we express the allowed values for the tensor-to-scalar ratio in terms of the reheating temperature. The only values for $\alpha$ and $w_{re}$ in agreement with current bounds on the tensor-to-scalar ratio, correspond to $\alpha=0.0185$ and $w_{re}=-1/3,0$, as it is depicted in Fig. \ref{fig:rS}. In particular, for $w_{re}=-1/3$, the reheating temperature must be in the ranges
\begin{eqnarray}
10^4\,\textup{GeV} \lesssim T_{re}\lesssim 10^{6}\,\textup{GeV}, \\
10^7\,\textup{GeV} \lesssim T_{re} \lesssim 10^{10}\,\textup{GeV}, \\
10^{11}\,\textup{GeV} \lesssim T_{re} \lesssim 10^{14}\,\textup{GeV},
\end{eqnarray}
when $M_5$ takes the values $10^5$ TeV, $10^8$ TeV, and $10^{12}$ TeV, respectively. On the other hand, for $w_{re}=0$, the allowed ranges for
$T_{re}$ are found to be
\begin{eqnarray}
10^2\,\textup{GeV} \lesssim T_{re} \lesssim 10^{6}\,\textup{GeV}, \\
10^2\,\textup{GeV} \lesssim T_{re} \lesssim 10^{10}\,\textup{GeV}, \\
10^{11}\,\textup{GeV} \lesssim T_{re} \lesssim 10^{14}\,\textup{GeV},
\end{eqnarray}
when fixing $M_5$ as $10^5$ TeV, $10^8$ TeV, and $10^{12}$ TeV, respectively. 

{The production of massive relics, such as gravitinos, is an important issue when discussing supersymmetric models, since their overproduction might spoil the success of BBN \cite{gravitino1,gravitino2,Okada:2004mh,gravitino3,gravitino4}. In the context of brane-world cosmology, the gravitino problem is avoided provided that the transition temperature, $T_t$, is bounded from above, $T_t \leq (10^6-10^7)$~GeV \cite{Okada:2004mh}. The transition temperature is the temperature at which the evolution of the Universe passes from the brane-world cosmology into the standard one, and it is given by \cite{TLRS1}
\begin{equation}
T_t = 1.6 \times 10^7 \left( \frac{100}{g_{re}} \right)^{1/4} \: \left( \frac{M_5}{10^{11} GeV} \right)^{3/2}
\end{equation}
Clearly, the upper bound on $T_t$ implies an upper bound on $M_5$, and therefore in the case of exponential SUSY inflation the five-dimensional Planck mass is finally forced to take values in the range
\begin{equation}
10^5~\textrm{TeV} \leq M_5 \leq 10^8~\textrm{TeV}.
\end{equation}



\begin{figure}[ht!]
\centering
\includegraphics[width=0.45\textwidth]{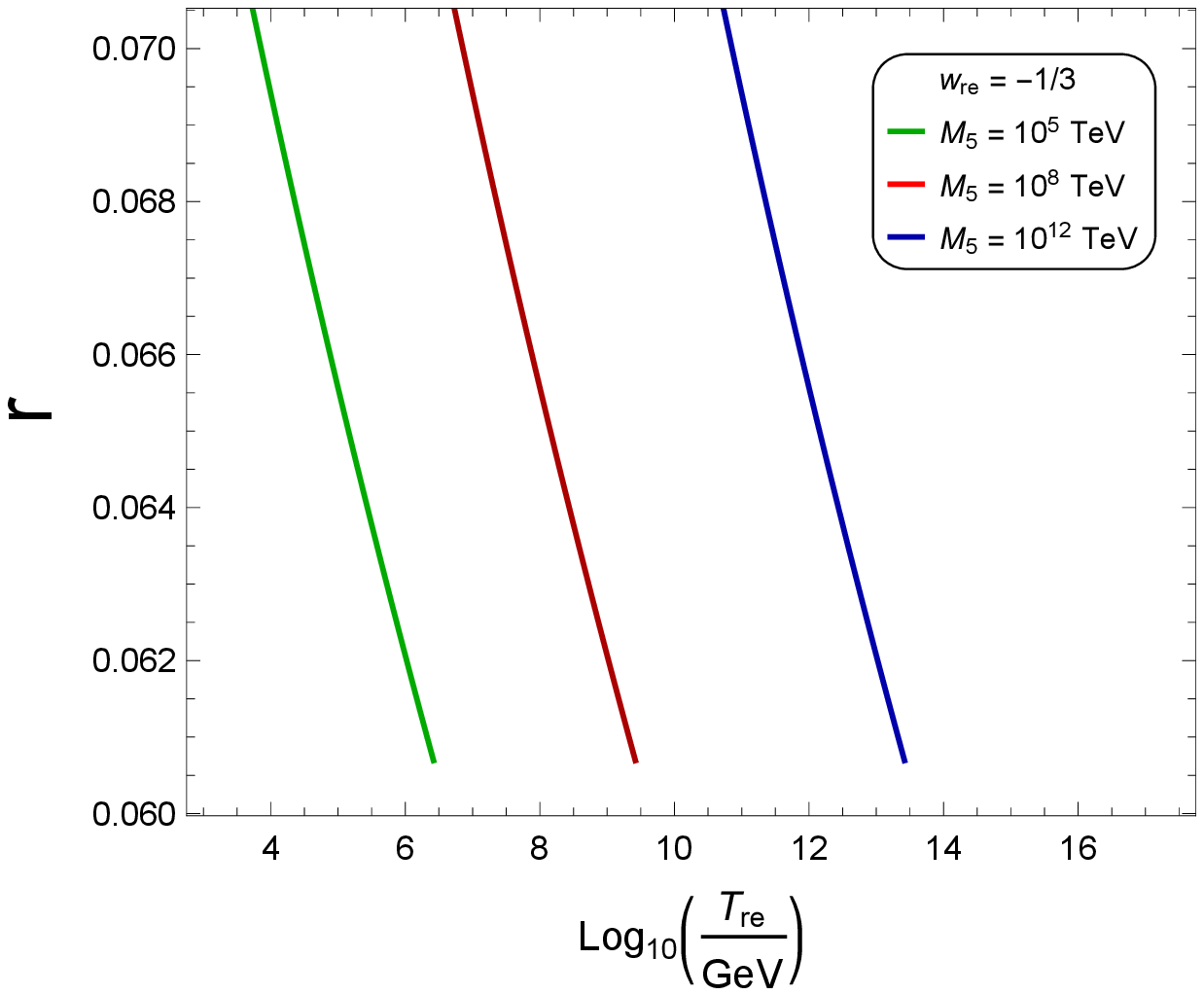}
\includegraphics[width=0.45\textwidth]{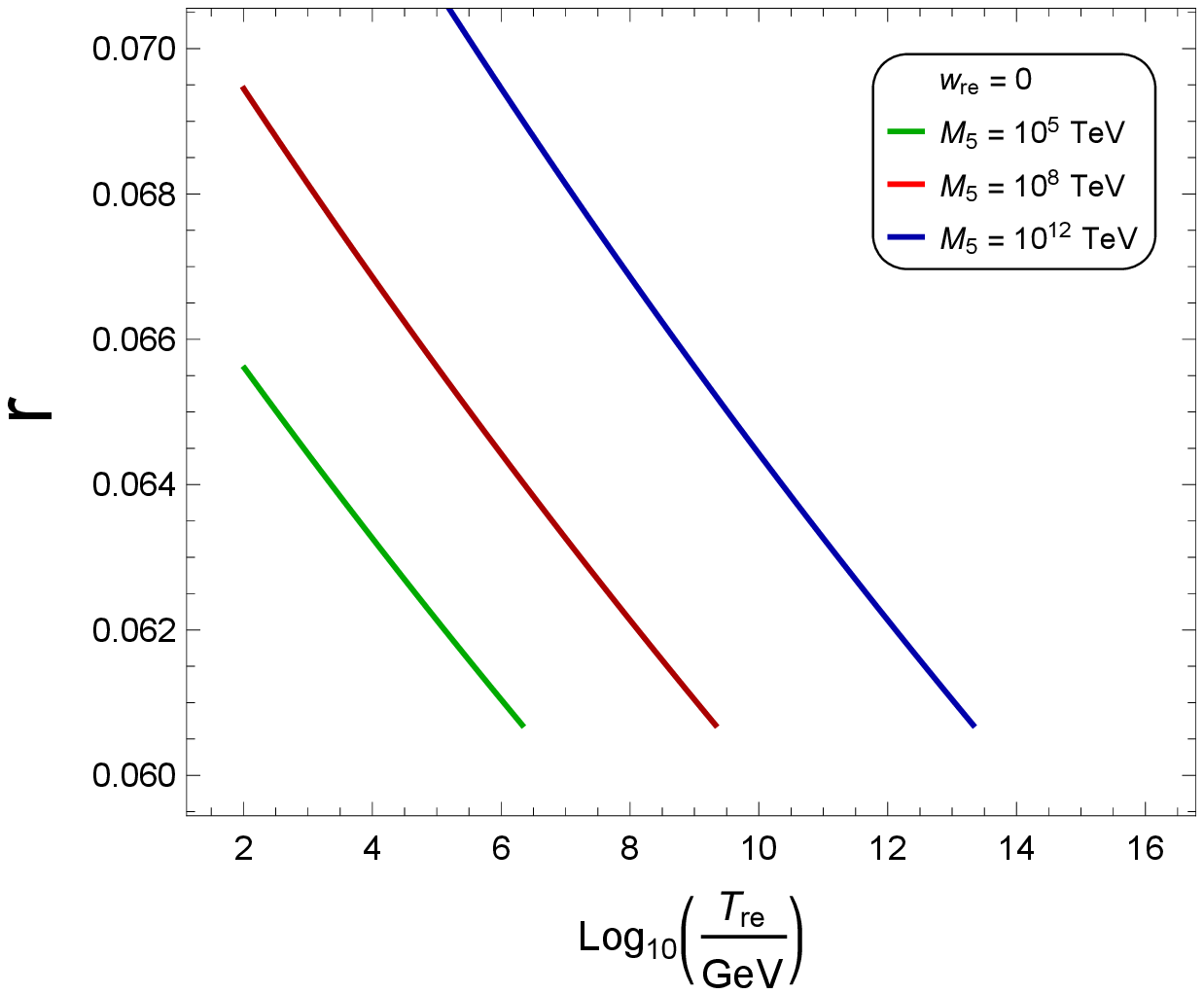}
\caption{Plots for the tensor-to-scalar ratio against the reheating temperature for Exponential SUSY inflation for $w_{re}=-1/3,\,0$ and $\alpha=0.0185$. The green, red and blue lines corresponds to a mass of $M_5=10^{5}$\,TeV, $M_5=10^8$\,TeV and $M_5=10^{12}$\,TeV respectively.} \label{fig:rS}
\end{figure}




\section{Baryogenesis via leptogenesis}

Finally, let us comment on the generation of baryon asymmetry in the Universe. Any viable and successful inflationary model must be capable of generating the baryon asymmetry, which comprises one of the biggest challenges in modern theoretical cosmology. Primordial Big Bang Nucleosynthesis \cite{bbn} as well as data from CMB temperature anisotropies \cite{wmap,Planck2015-1,Planck2015-2,Planck2018-1,Planck2018-2} indicate that the baryon-to-photon ratio is a very small but finite number, $\eta_B=6.19 \times 10^{-10}$ \cite{values}. This number must be calculable within the framework of the particle physics we know. Although as of today several mechanisms have been proposed and analysed, perhaps the most elegant one is leptogenesis \cite{leptogenesisn}. In this scenario a lepton asymmetry arising from the out-of-equilibrium decays of heavy right-handed neutrinos is generated first. Next, the lepton asymmetry is partially converted into baryon asymmetry via non-perturbative "sphaleron" effects \cite{sphalerons}. 

\smallskip

Of particular interest is the non-thermal leptogenesis scenario \cite{values,lepto10,lepto20,lepto30,lepto40,lepto50,lepto60,Panotopoulos-1,Panotopoulos-2,paperbase,Panotopoulos-3}, since the lepton asymmetry is computed to be proportional to the reheating temperature after inflation. Therefore, within non-thermal leptogenesis the baryon asymmetry and the reheating temperature, two key parameters of the Big Bang cosmology, are linked together. Furthermore, in supersymmetric models the gravitino problem \cite{linde1,linde2} puts an upper bound on the reheating temperature after inflation \cite{Kohri}, and therefore thermal leptogenesis \cite{DiBari,review}, which requires a high reheating temperature \cite{Strumia}, is much more difficult to be implemented. Moreover, contrary to thermal leptogenesis where one has to solve the complicated Boltzmann equations numerically, in the non-thermal leptogenesis scenario one can work with analytic expressions.

\smallskip

The initial lepton asymmetry, $Y_L=n_L/s$, is converted into baryon asymmetry $Y_B=n_B/s$ via sphaleron effects \cite{sphalerons}
\begin{equation}
Y_B = a Y_{B-L} 
\end{equation}
or
\begin{equation}
Y_B = \frac{a}{a-1} \: Y_L \equiv C \: Y_L 
\end{equation}
where $n$ is the number density of leptons or baryons, $s$ is the entropy density of radiation, $s=(2 \pi^2 h_* T^3)/45$, and the conversion factor $a$ is computed to be $a=(24+4N_H)/(66+13N_H)$ \cite{turner}, with $N_H$ being the number of Higgs doublets in the model. In SM with only one Higgs doublet, $N_H=1$, $a=28/79$ and $C=-28/51$, while in MSSM with two Higgs doublets, $N_H=2$, $a=8/23$ and $C=-8/15$.

\smallskip

In the scenario of non-thermal leptogenesis, the lepton asymmetry is computed to be
\begin{equation}
Y_L = \frac{3}{2} \frac{T_{re}}{M_I} \sum_i^3 BR(\phi \rightarrow N_i N_i) \epsilon_i
\end{equation}
where $\epsilon$ is the CP-violation asymmetry factor, and $BR(\phi \rightarrow N_i N_i)$ is the branching ratio of the inflaton decay channel into a pair of right-handed neutrinos $\phi \rightarrow N_i N_i$.

\smallskip

Moreover, lepton asymmetry is generated by the out-of-equilibrium decays of the heavy right-handed neutrinos into Higgs bosons and leptons
\begin{equation}
N \rightarrow H l, \quad N \rightarrow \bar{l} H^\dagger
\end{equation}
provided that $T_{re} < M_1$. The CP-violation asymmetry factor is defined by \cite{covi}
\begin{equation}
\epsilon = \frac{\Gamma-\bar{\Gamma}}{\Gamma+\bar{\Gamma}}
\end{equation}
where $\Gamma=\Gamma(N \rightarrow l H)$ and $\bar{\Gamma}=\Gamma(N \rightarrow \bar{l} H^\dagger)$, for any of the three right-handed neutrinos, and it arises from the interference of the one--loop diagrams with the tree level coupling \cite{covi}. In concrete SUSY GUT models based on the $SO(10)$ group it typically takes values $\epsilon \sim 10^{-5}$ \cite{Model-1,Model-2}.

\smallskip

Assuming the mass hierarchy $M_1 \ll M_{2,3}$, the inflaton is not sufficiently heavy to decay into $N_2,N_3$, and therefore the channels $\phi \rightarrow N_2 N_2$ and $\phi \rightarrow N_3 N_3$ are kinematically closed. Thus, we obtain for baryon asymmetry the final expression
\begin{equation} \label{asymmetry}
Y_B = \frac{3 C}{2} \; \frac{T_{re}}{M_I} \; \epsilon_1
\end{equation}

It thus becomes clear that the three relevant mass scales, namely $T_{re}, M_I, M_1$, must satisfy the following hierarchy
\begin{equation}
T_{re} < M_1 < M_I
\end{equation}
and therefore within non-thermal leptogenesis the inflaton mass must be always larger than the reheating temperature.

\smallskip

In the models discussed here the inflaton mass is given in terms of the two mass scales, $\mu,\Lambda$, as follows
\begin{equation} \label{inflatonmass}
M_I \sim \frac{\Lambda^2}{\mu}
\end{equation}
while for any given value of $M_5$ the allowed range for $\mu,\Lambda,T_{re}$ is known, according to the analysis presented in the previous sections. Given the numerical results already presented, it is easy to verify that for a given $M_5$, the inflaton mass is always lower than $T_{re}$. Hence, we conclude that in single-field inflationary models with a canonical scalar field in the RS-II brane model non-thermal leptogenesis cannot work, at least for the concrete inflationary potentials considered here.

\smallskip

There is another way to see that non-thermal leptogenesis cannot work here. Let us ignore for a moment the fact that the mass scales violate the required hierarchy mentioned before, and let us
show graphically how the CP-violation asymmetry factor depends on the reheating temperature after inflation. This is shown in the figures \ref{fig:Lepto1} and \ref{fig:Lepto2} for $M_5=10^5$ TeV and $M_5=10^{12}$ TeV, respectively. Clearly, it turns out that for the reheating temperature obtained before, the CP-violation asymmetry factor is many orders of magnitude lower than what typically concrete particle physics models predict, $~10^{-5}$, as already mentioned before.

\smallskip


\begin{figure}[ht!]
	\centering
	\includegraphics[scale=0.8]{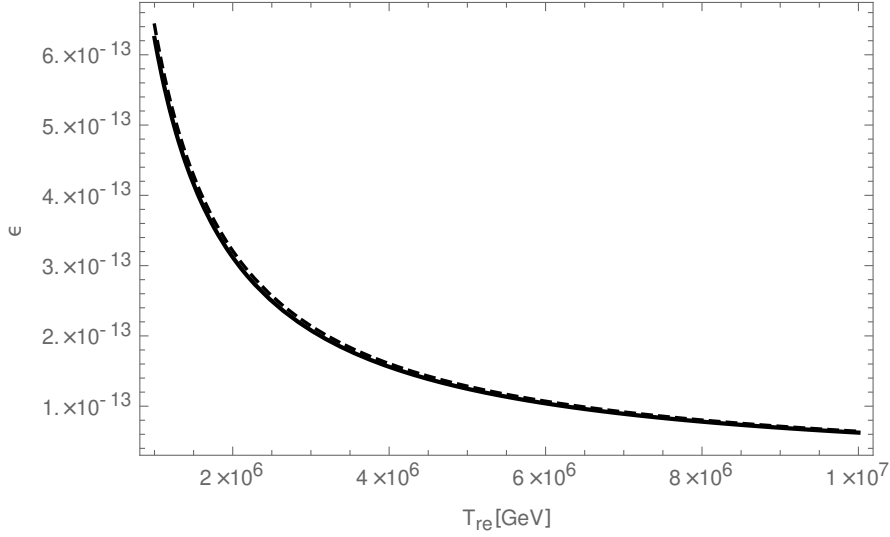}
	\caption{
	CP-violation asymmetry factor, $\epsilon$, as a function of the reheating temperature after inflation, $T_{re}$, 
	for the Higgs-like inflationary potential and $M_5=10^5$ TeV. The solid curve corresponds to the SM, while the dashed   curve to the MSSM.
	}
	\label{fig:Lepto1} 	
\end{figure}



\begin{figure}[ht!]
	\centering
	\includegraphics[scale=0.8]{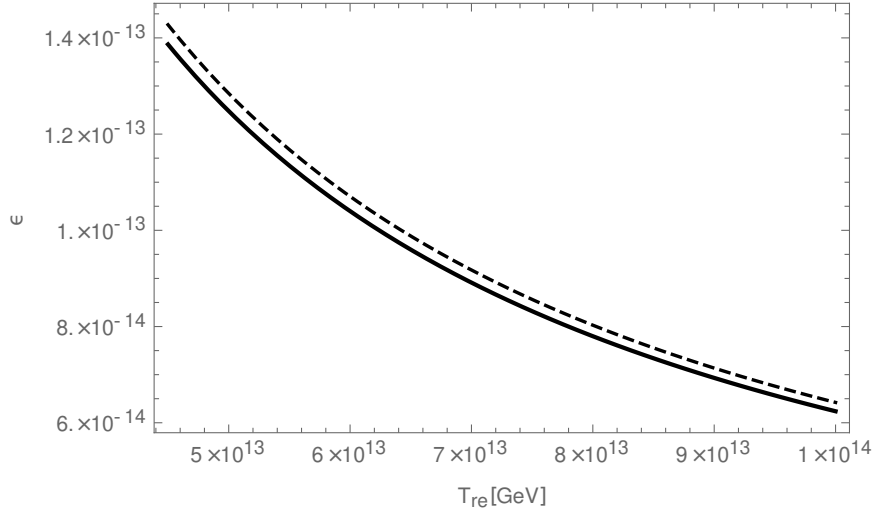}
	\caption{
	Same as previous figure, but for $M_5=10^{12}$ TeV.
	}
	\label{fig:Lepto2} 	
\end{figure}


Therefore, for those two independent reasons we conclude that non-thermal leptogenesis cannot work within the framework of RS-II brane cosmology, at least for the inflationary potentials considered here. Consequently, one must rely on the mechanism of thermal leptogenesis, which in the framework of RS brane cosmology has been analysed in \cite{TLRS1,TLRS2}, and it requires a sufficiently high $M_5$. In particular, in the high-energy regime of brane cosmology, it is found that $M_5$ must take values in the range $10^{12}\,\textup{GeV} < M_5 < 10^{16}\,\textup{GeV}$ \cite{TLRS2}.

As a final remark, we have not included the TCC in our analysis. We hope to be able to address this point in a future work.

\smallskip

{\bf Note added:} As our work was coming to its end, another work similar to ours appeared \cite{Mohammadi:2020ake}. There, too, the authors have studied different types of inflationary potentials in the framework of five-dimensional RS brane model, and they could determine models that satisfy both data and swampland criteria at the same time. We find the following differences compared to our analysis: i) the allowed range for the free parameters of each model is not determined, ii) nothing is mentioned about baryon asymmetry, and iii) the tensor-to-scalar ratio has been overlooked.

\section{Conclusions}

We have studied the dynamics of four concrete small-field inflationary models based on a single, canonical scalar field in the framework of the high-energy regime of the Randall-Sundrum II brane model. In particular, we have considered 
i) an axion-like potential for the inflaton (Natural Inflation), ii) Hilltop potential with a quadratic term (quadratic Hilltop inflation), iii) a potential arising in the context of dynamical symmetry breaking (Higgs-like inflation), and iv) a SUGRA-motivated potential (Exponential SUSY inflation). Adopting the Randall-Sundrum fine-tuning, all the models are
characterized by 3 free parameters in total, namely the 5-dimensional Planck
mass, $M_5$, and the two mass scales of the inflaton potential. We have shown in the $n_s-r$ plane the theoretical predictions of the models together with the allowed contour plots from the PLANCK Collaboration, and we have determined the allowed range of the parameters for which a viable inflationary Universe emerges. The mass scales of the inflaton potential have been expressed in terms of the five-dimensional Planck mass, which remains unconstrained using the PLANCK results only. However, on the one hand current tests for deviation from Newton's gravitational law at millimeter scales, and on the other hand the assumption that inflation takes place in the high-energy limit of the RS-II brane model force the five-dimensional Planck mass to lie in the range $10^5\,\textup{TeV}\lesssim M_5\lesssim 10^{12}\,\textup{TeV}$, and therefore all parameters are finally known. After that, we have shown that for those types of potentials the inflation incursion is sub-Planckian, then the distance Swampland conjecture is satisfied. Nevertheless, the de Sitter Swampland Criteria and its refined version may be evaded for these potentials in the high-energy regime of the RS-II brane model instead. Finally, we have computed the reheating temperature $T_{re}$ as well as the duration of reheating, $N_{re}$, versus the scalar spectral index $n_s$ assuming four different values of the EoS parameter $w_{re}=-1/3, 0, 2/3, 1$ of the fluid into which the inflaton decays.  
Our results show that the reheating temperature depends on the five-dimensional Planck mass, and particularly that the maximum reheating temperature increases with $M_5$. Then, by applying the constraint on $M_5$ already found, an allowed range for the reheating temperature as well as for the tensor-to-scalar ratio could be obtained for each model. Furthermore, we have shown that non-thermal leptogenesis cannot work within the framework of RS-II brane cosmology, at least for the inflationary potentials considered here. Consequently, one must rely on the mechanism of thermal leptogenesis, which in the high-energy regime of the RS brane cosmology requires a sufficiently high five-dimensional Planck mass, $M_5 > 10^{12}\,\textup{GeV}$.


\section*{Acknowlegements}

C.~O. is supported by CONICYT Chile, scholarship ANID-PFCHA/Doctorado Nacional/2018-21181476 and N.~V. is supported by FONDECYT Grant N$^{\textup{o}}$ 11170162. G.~P. thanks the Funda\c c\~ao para a Ci\^encia e Tecnologia (FCT), Portugal, for the financial support to the Center for Astrophysics and Gravitation-CENTRA, Instituto Superior T{\'e}cnico,  Universidade de Lisboa, through the Project No.~UIDB/00099/2020.

\end{document}